\documentclass{amsart}
\usepackage{verbatim,amsmath,latexsym,amssymb}

\theoremstyle{plain}
\newtheorem{thm}{Theorem}[section]
\newtheorem{claim}[thm]{Claim}

\def\qed{\hfill$\Box$}

\input epsf

\begin{document}

\title{Numerically Invariant Signature Curves}
\author{Mireille Boutin}
\begin{abstract}
Corrected versions of the numerically invariant expressions for the affine and
Euclidean signature of a planar curve given in \cite{COSTH} are presented. The new formulas are valid for fine but otherwise arbitrary partitions of the curve.  We also give numerically invariant expressions for the four differential invariants parametrizing the three dimensional version of the Euclidean signature curve, namely the curvature, the torsion and their derivatives with respect to arc length.      
\end{abstract}
\maketitle

\section{Introduction}
The concept of signature was introduced by Calabi et al.\ \cite{COSTH} as ``a new paradigm for the invariant recognition of visual objects.'' For the case of planar curves, the signature is defined as follows: if $\alpha(s)$ is a smooth curve in ${\mathbb R}^2$ parametrized by arc length $s$ and if $G$ is any finite dimensional transformation group acting transitively on ${\mathbb R}^2$, then the signature curve $S$ of $\alpha(s)$ with respect to $G$ is given parametrically as $(\kappa(s),\kappa_s(s))$, where $\kappa$ is the $G$-invariant curvature and $\kappa_s$ its derivative with respect to arc length. One of the consequences of a theorem proved by Cartan \cite{C} is that $S$ fully determines the given curve $\alpha$ modulo $G$, provided that $\kappa$ is never zero.

The signature curve can therefore be used to program a computer for recognizing curves modulo certain group transformations. However differential invariants are high order derivatives and hence very sensitive to round off errors and noise. The idea of writing a numerical scheme in terms of joint invariants was also introduced by Calabi et al.\  \cite{COSTH}. Hoping to obtain less sensitive approximations, they suggested to find  numerical expressions for $\kappa$ and $\kappa_s$ in terms of joint invariants. A joint invariant  of the action of a group on a manifold is a real valued function J  which depends on a finite number of points $x_1,...,x_n$ of the manifold and which remains unchanged under the simultaneous action of the group G on the point configuration, i.e. $J(x_1,...,x_n)=J(g\cdot x_1,...,g\cdot x_n)$ $\forall g\in G$. For example, the Euclidean distance is a joint invariant of the action of the Euclidean group on ${\mathbb R}^2$. Expressing differential invariants in terms of joints invariants results in a $G$-invariant numerical approximation. In their paper, Calabi et al.\ proposed numerically invariant expressions for $\kappa$ and $\kappa_s$ for two specific group actions, namely the proper Euclidean group and the equi-affine group. 

But contrary to their claim, the expressions given for $\kappa_s$ are not convergent for arbitrary partitions of the curve. In the next section, we give correct formulas for approximating $\kappa_s$ and explain why the old ones do not work in generaL. In order to prove our claims, we also compute the resulting numerical signatures in a practical example and with different partitions. We then compare with the signatures obtained with the old formulas.

Cartan's theorem also provides us with a way to characterize curves in ${\mathbb R}^3$ modulo group transformation. The generalization of the signature curve is a curve in ${\mathbb R}^4$ determined by four differential invariants: the $G$-invariant curvature $\kappa$, its derivative $\kappa_s$ with respect to $G$-invariant arc length $s$, the $G$-invariant torsion $\tau$ and its derivative $\tau_s$ with respect to $s$. For practical applications, we are interested in the case where $G$ is the proper Euclidean group.  Following the example of \cite{COSTH}, we have found approximations for $\kappa$, $\kappa_s$, $\tau$ and $\tau_s$ in terms of the simplest joint invariant of the action of the Euclidean group on ${\mathbb R}^3$: the Euclidean distance. The results are given in section 3. That section also contains the results of numerical tests performed on a space curve with different parametrizations.

\section{Corrections for the Case of a Planar Curve}

In principle, one must keep track of the order of the approximation when manipulating an approximation. This is especially true when trying to approximate a derivative using an approximate expression. For example, if $\tilde{c}(t)=c(t)+O(t)$ is a first order approximation for $c(t)$, then for $0<h,t\ll1$ 
\[
\frac{\tilde{c}(t)-\tilde{c}(0)}{t}=\frac{c(t)+O(t)-c(0)}{t}\approx c'(0)+O(t^0)
\] 
is in general {\em not} an approximation for $c'(0)$.

By keeping track of the order of approximation throughout the computations performed in \cite{COSTH}, one finds out that the expressions given for $\kappa_s$ will converge only for a $2^{nd}$ order approximation of $\kappa$. In the case of a regular partition of the curve, the approximations used for the curvature are of second order, as one can see from equation (\ref{1}) below. This fortunate fact explains the correct numerical results presented in \cite{COSTH}. However for a generic partition there is no guarantee of success and failure is likely as we show by example.

In the following we present the corrected formulas together with their justification for the Euclidean and affine group action..

\subsection{Euclidean Group Action}

Let $P_{i-2}$, $P_{i-1}$, $P_{i}$, $P_{i+1}$, and $P_{i+2}$ be 5 consecutive points on a smooth planar curve. Denote by $d(P_j,P_k)$ the Euclidean distance between the points $P_j$ and $P_k$. As illustrated in figure 1, let\par
\begin{tabular}{llll}$a:=d(P_{i-1},P_{i})$,& $c:=d(P_{i-1},P_{i+1})$, & $e:=d(P_{i},P_{i+2})$,& $g:=d(P_{i-2},P_{i-1})$\\
 $b:=d(P_{i},P_{i+1}$), & $d:=d(P_{i+1},P_{i+2}$)& $f:=d(P_{i-1},P_{i+2})$ &      
\end{tabular}

\begin{figure}[here]
\caption{}
\vspace{0.5cm}
\centerline{
\hbox{
\epsfysize=6.0cm
\epsfxsize=7.0cm
{\leavevmode \epsfbox{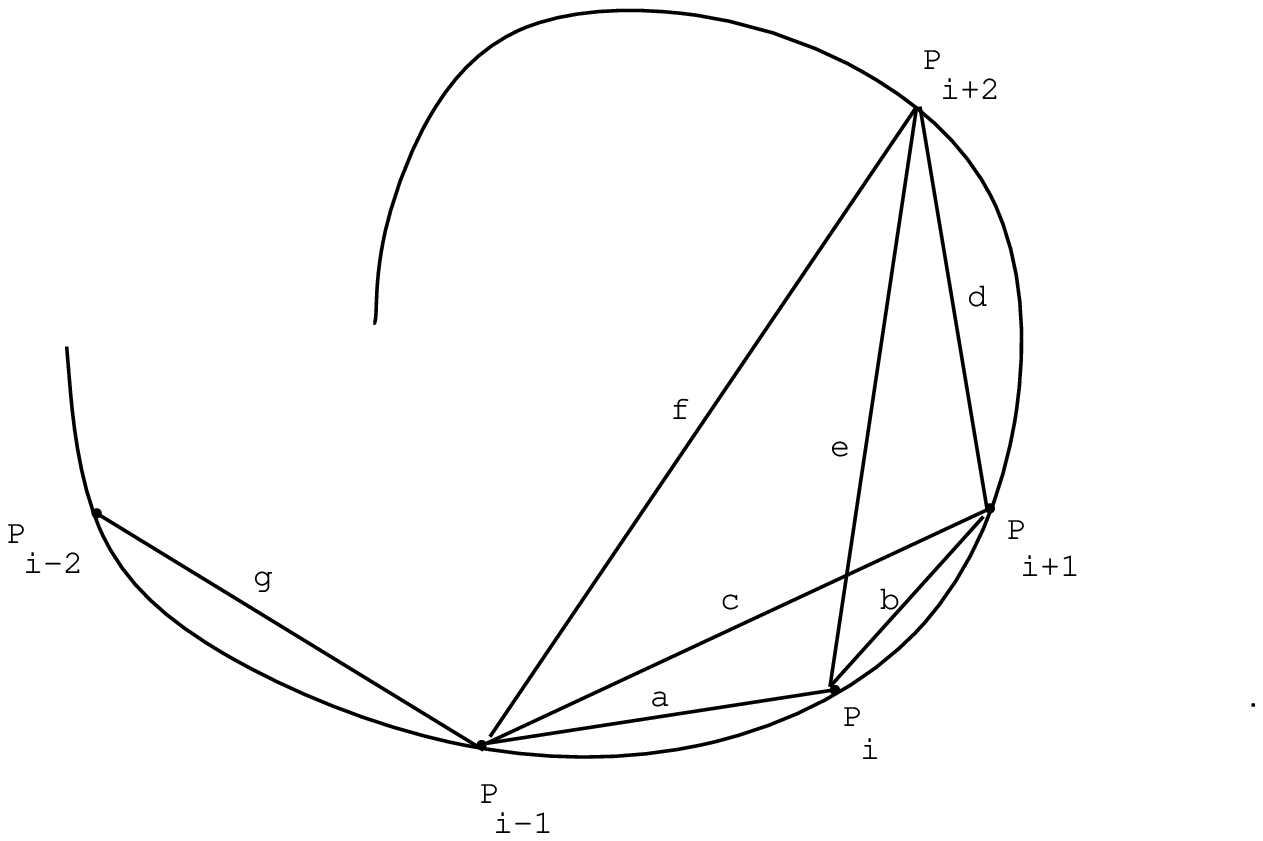}}
}
}
\end{figure}

It was shown in \cite{COSTH} that
$\tilde{\kappa}(P_{i})=\pm\frac{4\Delta}{abc}$, where $\Delta$ denotes the area of the triangle with sides of length $a$, $b$ and $c$, is a good approximation for
the Euclidean curvature at $P_{i}$ . In fact, the following expansion has
been proven to be valid for small $a$ and $b$. 
\begin{eqnarray}
\label{1}
\tilde{\kappa}(P_{i})&=& \kappa (P_{i})+\frac{1}{3}(b-a)\kappa_s(P_{i})+O(2).
\end{eqnarray}
Observe that $\tilde{\kappa}$ depends only on the distances between the three points. Since the distance is an Euclidean invariant, we thus have a Euclidean  numerically invariant expression for the curvature. 

When $a \not= b$, the approximation obtained is only of first order. Therefore care must be taken when approximating the derivative $\kappa_s(P_{i})$. The formulas used in \cite{COSTH} are
\begin{eqnarray}
\label{initial}
\tilde{\kappa}_{s,1}(P_{i})&:=&\frac{\tilde{\kappa}(P_{i+1})-\tilde{\kappa}(P_{i})}{b},\\
\label{initial_symmetric}
\tilde{\kappa}_{s,1}(P_{i})&:=&\frac{\tilde{\kappa}(P_{i+1})-\tilde{\kappa}(P_{i-1})}{c}.
\end{eqnarray}

We would like to have expressions that do not assume a priori conditions on the parametrizations. Hence we propose the following numerically Euclidean invariant expression for $\kappa_s(B)$:

\begin{eqnarray}
\label{2}
\tilde{\kappa}_{s,3}(P_{i})&:=& 3\cdot\frac{\tilde{\kappa}(P_{i+1})-\tilde{\kappa}(P_{i})}{a+b+d},\\
\label{3}
\tilde{\kappa}_{s,4}(P_{i})&:=& \frac{3}{2}\cdot\frac{\tilde{\kappa}(P_{i+1})-\tilde{\kappa}(P_{i})}{a+b+d}+\frac{3}{2}\cdot\frac{\tilde{\kappa}(P_{i})-\tilde{\kappa}(P_{i-1})}{a+b+g}, \\
\label{4}
\tilde{\kappa}_{s,5}(P_{i})&:=&3\cdot\frac{\tilde{\kappa}(P_{i+1})-\tilde{\kappa}(P_{i-1})}{2a+2b+d+g}. 
\end{eqnarray}

\begin{claim}
Formulas (\ref{2}), (\ref{3}) and (\ref{4})  converge to $\kappa_{s}(P_{i})$ as $a$, $b$, $d$, $g$ $\to 0$

\proof By (\ref{1}) we have
\[
\tilde{\kappa}(P_{i})=\kappa(P_{i})+\frac{1}{3}(b-a)\kappa_s(P_{i})+O(2)
\]
and
\[
\tilde{\kappa}(P_{i+1})=\kappa(P_{i+1})+\frac{1}{3}(d-b)\kappa_s(P_{i+1})+O(2).
\]
Assume that $\alpha$=$\alpha (s)$ is a smooth curve parametrized by arc length, that $P_{i}=\alpha (0)$ and that $P_{i+1}=\alpha (h)$. We can write

 \[\kappa(P_{i+1})=\kappa(P_{i})+h\kappa_s(P_{i})+O(2)\approx\kappa(P_{i})+b\kappa_s(P_{i})+O(2)  \]
and   
\[\kappa_s(P_{i+1})=\kappa_s(P_{i})+O(1).\]

\noindent Thus
\[
  \tilde{\kappa}(P_{i+1})-\tilde{\kappa}(P_{i})=\frac{1}{3}(d-2b+a)\kappa_s(P_{i})+b\kappa_s(P_{i})+O(2).
\]
Rearranging, we obtain

\[
 3\cdot\frac{\tilde{\kappa}(P_{i+1})-\tilde{\kappa}(P_{i})}{a+b+d} =\kappa_s(P_{i})+O(1)
\]
Therefore (\ref{2}) is valid. 

Equation (\ref{3}) is just a symmetric formula obtained by averaging (\ref{2}) with
its reflected version. We obtain (\ref{4}) by expanding
$\tilde{\kappa}(P_{i+1})-\tilde{\kappa}(P_{i-1})$  about $P_i$ in a similar way. \qed 
\end{claim}

Figures 2-8 show that our formulas are just as good as (\ref{initial}) and (\ref{initial_symmetric}) when the parametrization is very regular. But when the parametrization is not, we see that (\ref{initial}) and (\ref{initial_symmetric}) diverge from the exact solution, while (\ref{2}), (\ref{3}) and (\ref{4}) remain closer to it. Thus we believe that our formulas are better for a generic partition of $\alpha$.  Observe that formula (\ref{2}) suffers from a small bias due to its asymmetry. This phenomenon was also observed for formula (\ref{initial}) in \cite{COSTH}.

\subsection{Affine Group Action}

Let $\alpha$ be a convex smooth curve. Let $P_{0}$, $P_{1}$,
$P_{2}$, $P_{3}$, $P_{4}$ be five consecutive points on
$\alpha$. Denote by $[ijkl]$ the signed area of the parallelogram whose
sides are $P_i - P_j$ and $P_k - P_l$ and by $[ijk]$ the signed area of the parallelogram whose
sides are $P_i - P_j$ and $P_i - P_k$. Define $T$ and $S$  at $P_{2}$ by 

\begin{equation*}
4\cdot T(P_2)=\prod_{0\leq l<m<n\leq 4} [lmn],
\end{equation*}

\begin{eqnarray*}
4\cdot S(P_2)&=&[013]^2[024]^2[1234]^2+[012]^2[034]^2[1324]^2\\
&&\phantom{xxx}-2[012][034][013][024]([123][234]+[124][134]).
\end{eqnarray*}
Observe that the expressions for $T$ and $S$ are equi-affine
invariant.
The following is a numerically invariant approximation for the affine curvature at $P_i$:
\begin{equation}
\tilde{\kappa}(P_i)=\frac{S(P_i)}{T(P_i)^{2/3}}.
\end{equation} 
Indeed the following expansion has been shown to be valid for sufficiently close
points \cite{COSTH}

\begin{equation}
\tilde{\kappa}_i=\kappa+\frac{1}{5}\left( \sum_{j=i-2}^{i+2}L_j \right)\frac{d\kappa}{ds}+...
\end{equation} 
where $L_j$ denotes the signed affine arc length of the conic from
$P_i$ to $P_j$ and the higher order terms are quadratic in the $L_j$'s.
Again, since this is a first order approximation, care must be taken
when approximating the derivative of $\kappa$ with respect to affine arc
length $s$. 

The formula used in \cite{COSTH} is

\begin{equation}
\label{initial_affine}
\tilde{\kappa}_{s}=\frac{\tilde{\kappa}(P_{i+1})-\tilde{\kappa}(P_{i-1})}{\tilde{S}_{i}+\tilde{S}_{i-1}},
\end{equation}
where $\tilde{S}_{j}$ is the triangular approximation to the affine arc length from $P_j$ to $P_{j+1}$. This formula would be valid only if $\tilde{\kappa}$ were a second order approximation of $\kappa$. However, this is not the case unless $\sum_{j=i-2}^{i+2}L_j$ is small compared to all of the $L_j$'s. We would like a more general expression. Hence we propose the following numerically affine invariant expression for $\kappa_s$ which does not assume a priori conditions on the parametrizations:

\begin{equation}
\label{affine1}
\tilde{\kappa}_{s}^1=5\cdot\frac{\tilde{\kappa}(P_{i+1})-\tilde{\kappa}(P_{i-1})}{\tilde{S}_{i-3}+2\tilde{S}_{i-2}+2\tilde{S}_{i-1}+2\tilde{S}_{i}+2\tilde{S}_{i+1}+\tilde{S}_{i-2}}
\end{equation}
The convergence can be proved by an argument similar to the one used in the Euclidean case. 

Figures 9-12 show the results obtained with our formula using different partitions of a curve. Comparison is made with the signature curve obtained with (\ref{initial_affine}). Just like in the Euclidean case, we see that equations (\ref{initial_affine}) and (\ref{affine1}) seem just as good when the partition of the curve is regular. But when the partition is not regular, (\ref{initial_affine}) diverges from the exact solution while (\ref{affine1}) remains valid. 

\section{Numerically Invariant Euclidean Signature for a Space Curve}

The signature curve $S$ of a space curve is given by $S=(\kappa, \kappa_s, \tau, \tau_s)$. In the following section, we present some numerically Euclidean invariant approximations for $S$ for the case of the Euclidean group action on ${\mathbb R}^3$.
This is achieved by expressing each of the differential invariants constituting $S$ in terms of Euclidean distances, which is the simplest joint invariant of the Euclidean group action on ${\mathbb R}^3$. 

\subsection{Numerically Invariant Expressions for $\kappa$ and $\kappa_s$.}
Let $P_{i-1}$, $P_i$ and $P_i+1$ be three consecutive points on a space curve and let $a$, $b$ and $c$ be their mutual distances as illustrated in figure 1. Denote by $\Delta$ the area of the triangle with sides $a$, $b$ and $c$. The curvature of a space curve is defined exactly the same way as the curvature of a planar curve: it is the inverse of the radius of the oscullating circle at $p$. So we expect that $\tilde{\kappa}=\pm\frac{4\Delta}{abc}$ is a good approximation for $\kappa$ at $P_i$. Indeed, a Mathematica routine using the local canonical form of a curve in ${\mathbb R}^3$ gave us the following Taylor series expansion.

\begin{eqnarray}
\tilde{\kappa}&=&\kappa+\frac{(b-a)}{3}\kappa_{s}+(a^2+b^2)(\frac{\kappa_{ss}}{12}-\frac{\kappa \tau^2}{36})-ab(\frac{\kappa_{ss}}{12}+\frac{\kappa\tau^2}{36})\nonumber\\
&+&(b^3-a^3)(\frac{\kappa^2 \kappa_{s}}{40}-\frac{7\kappa_{s}\tau^2}{570}-\frac{\kappa\tau\tau_{s}}{45}+\frac{\kappa_{sss}}{60})\nonumber\\
&+&(a^2 b-a b^2)(-\frac{\kappa^2 \kappa_{s}}{60}+\frac{\kappa_{s}\tau^2}{180}+\frac{\kappa\tau\tau_{s}}{180}+\frac{\kappa_{sss}}{60})+...\nonumber
\end{eqnarray}
Note that if we set $\tau=0$, then our result agrees with the 2-dimensional case. Also observe that $\tau$ does not appear in the first two terms of the expansion. Therefore the expressions for $\kappa_s$ given in the planar case ((\ref{2}), (\ref{3}) and (\ref{4})) are valid numerically invariant expressions for $\kappa_s$ in the three dimensional case. 

\subsection{Numerically Invariant Expressions for $\tau$ and $\tau_s$}

The Euclidean invariant torsion $\tau$ is defined as the derivative of the angle of the oscullating plane with respect to arc-length. Let $-1\ll-\delta<0<\epsilon_1<\epsilon_2\ll1$. Suppose $P_{i-1}=\alpha (-\delta)$, $P_i=\alpha (0)$, $P_{i+1}=\alpha (\epsilon_1)$ and $P_{i+2}=\alpha (\epsilon_2)$ are four consecutive points on a space curve $\alpha$. Let $a$, $b$, $c$, $d$, $e$ and $f$ be their mutual distances as illustrated in figure 1. Denote by $H$ the height of the tetrahedron with sides $a$, $b$, $c$, $d$, $e$ and $f$ with respect to $P_{i+2}$ and write $\Delta_{l_1l_2l_3}$ for the area of a triangle with sides of length $l_1$, $l_2$ and $l_3$.

We propose the two following numerically invariant expressions for the torsion at $P_i$:
\begin{equation}
\label{torsion1}
\tilde{\tau}_1=6\cdot\frac{H}{def\tilde{\kappa}}
\end{equation}

\begin{equation}
\label{torsion2}
\tilde{\tau}_2=\frac{3}{2}\cdot\frac{Hb}{f\Delta_{ebd}}
\end{equation}

Observe that (\ref{torsion1}) and (\ref{torsion2}) both use the minimal number of points (i.e. four) for approximating $\tau$. One might argue that this induces an asymmetry in the numerical results. However, this is something easy to fix in a practical situation (for example by averaging the results obtained in the two directions).

We justify (\ref{torsion1}) as follows: It is known that $\tau=\frac{-\alpha_s\times\alpha_{ss}\cdot\alpha_{sss}}{\kappa^2}$, \cite{DC}. In order words, $|\tau|$ is the component of $\alpha_{sss}$ which is perpendicular to the oscullating plane. We start by writing down a finite difference scheme for $\alpha_{sss}$ (for justifications, see \cite{FK} \S 5.4). 
\begin{equation}
\alpha_{sss}(0)\approx 6\cdot\frac{\displaystyle\frac{\displaystyle\frac{\alpha(\epsilon_2)-\alpha(\epsilon_1)}{\epsilon_2-\epsilon_1}-\frac{\alpha(\epsilon_1)-\alpha(0)}{\epsilon_1}}{\epsilon_2}-\displaystyle\frac{\displaystyle\frac{\alpha(\epsilon_1)-\alpha(0)}{\epsilon_1}-\frac{\alpha(0)-\alpha(-\delta)}{\delta}}{\delta+\epsilon_1}}{\delta+\epsilon_2}.
\end{equation}
When $\epsilon_1$ and $\delta$ approach zero, the plane defined by $P_{i-1}$, $P_i$ and $P_{i+1}$ approaches the oscullating plane. More precisely, if $\delta$ and $\epsilon_1$ are small enough, we find (from the local canonical form of the curve) that the equation of the plane passing through $\alpha (-\delta)$, $\alpha (0)$ and $\alpha (\epsilon)$ can be written as 
\[Ax+By+z=0
\]
with $A=\frac{\kappa\tau}{6}\epsilon_1\delta +O(3)$ and $B=-\frac{\tau}{3}(\epsilon_1-\delta)+O(2)$. Here,  $x$, $y$ and $z$ are the Frenet frame coordinates. The normal vector to this plane is $\frac{(A,B,1)}{\sqrt{A^2+B^2+1}}$ which is a first order approximation of $(0,0,1)$, the normal vector to the oscullating plane. We have

\begin{align}
\kappa\tau=&\alpha_{sss}\cdot (0,0,1)\nonumber\\
          =&\alpha_{sss}\cdot \frac{(A,B,1)}{\sqrt{A^2+B^2+1}}+O(1).\nonumber
\end{align} 
But \[
\alpha(-\delta)\cdot (A,B,1)=\alpha(0)\cdot (A,B,1)=\alpha(\epsilon_1)\cdot (A,B,1)=0
\]
and
\[
\alpha(\epsilon_2)\cdot\frac{(A,B,1)}{\sqrt{A^2+B^2+1}}=\pm H.
\]
Therefore
\begin{align}
\kappa\tau\approx& \frac{\pm 6H}{(\epsilon_2-\epsilon_1)\epsilon_2(\delta+\epsilon_2)}\nonumber\\
          \approx& \pm \frac{6H}{def}.\nonumber 
\end{align}
From the previous section we know that $\kappa\approx\pm \tilde{\kappa}=\frac{4\Delta_{abc}}{abc}$ and therefore $\kappa\approx\pm \frac{6H}{def\tilde{\kappa}}$.

There is an interesting similarity between (\ref{torsion1}) and $\tilde{\kappa}$. This is easily seen if we write $\tilde{\kappa}$ in a slightly different way. By basic geometry argument we can say that $\tilde{\kappa}=\pm\frac{2h}{ab}$ where $h$ is the height of the triangle with sides $a$, $b$ and $c$. Thus $\tilde{\kappa}\tilde{\tau}$ can be looked at as a three dimensional version of $\tilde{\kappa}$. This is one reason why we like (\ref{torsion1}). Another reason is that it provides us with an easy visual understanding of the torsion similar to the understanding we already have of the curvature. It is indeed easier to evaluate the height and sides of a tetrahedron together with the radius of the oscullating circle than a derivative. It makes the evaluation of the torsion from a picture of the curve very intuitive.

For finding (\ref{torsion2}), we started from the following definition: if $P=\alpha(t)$ and $Q=\alpha(0)$, then  $\tau=\lim_{P\rightarrow Q}\frac{\theta}{t}$, where $\theta$ is the angle between the oscullating planes at $P$ and $Q$ respectively \cite{E}. We approximated $\theta$ by $\sin \tilde{\theta}$ where $\tilde{\theta}$ denotes the angle between the plane $P_{i}P_{i+1}P_{i+2}$ and the plane $P_{i-1}P_iP_{i+1}$. If we assign to the plane $P_{i-1}P_iP_{i+1}$ the equation $z=0$ and if we let the segment $P_iP_{i+1}$ represent the x-axis, then the plane passing through $BCD$ has equation $H y-D_{y} z=0$ where $D_{y}$ is the projection of the point $D$ on the y-axis.

 Given the equations of two planes $A_{1}x+B_{1}y+C_{1}z+D_{1}=0$ and $A_{2}x+B_{2}y+C_{2}z+D_{2}=0$, the angle between the two planes is given by 
\[\cos\theta=\frac{A_{1}A_{2}+B_{1}B_{2}+C_{1}C_{2}}{\sqrt{A_{1}^{2}+B_{1}^2+C_{1}^2}\sqrt{A_{2}^2+B_{2}^{2}+C_{2}^{2}}}. 
\]
A straightforward computation gives us
\begin{align}
\sin \tilde{\theta}&=\pm \frac{H}{\sqrt{H^2+D_{y}^2}}\nonumber\\
           &=\pm \frac{H}{\sqrt{e^2-D_{x}^2}}\nonumber\\
           &=\pm \frac{H}{\sqrt{e^2-(\frac{e^2+b^2-d^2}{2b})^2}}\nonumber\\
           &=\pm \frac{2Hb}{\sqrt{4e^2 b^2-(e^2+b^2-d^2)^2}}\nonumber\\
           &=\pm \frac{2Hb}{4\Delta_{ebd}}\nonumber\\
\label{temporaire}
           &=\pm \frac{Hb}{2\Delta_{ebd}}.
\end{align}
On the other hand, we can use the local canonical form to approximate the equations of the plane $P_{i-1}P_iP_{i+1}$ and $P_{i}P_{i+1}P_{i+2}$ in the Frenet frame coordinates. Respectively, we have approximately
\[(\frac{\kappa\tau\epsilon_1\delta}{6})x+(\frac{\tau}{3})(\epsilon_1-\delta)y+z=0
\]
and
\[(\frac{-\kappa\tau\epsilon_1\epsilon_2}{6})x+(\frac{\tau}{3})(\epsilon_1+\epsilon_2)y+z=0.
\]
So
\[\cos^2\tilde{\theta}\approx\frac{(1+(\frac{\tau}{3})^2(\epsilon_1-\delta)(\epsilon_1+\epsilon_2))^2}{1+(\frac{\tau}{3}(\epsilon_1-\delta))^2+(\frac{\tau}{3}(\epsilon_1+\epsilon_2))^2}
\]
and therefore
\[
\sin^2\tilde{\theta}\approx\frac{\tau^2}{9}(\epsilon_2+\delta)^2\approx \frac{\tau^2}{9}f^2
\]
i.e. $\frac{3\sin \tilde{\theta}}{f}\approx \pm\tau$. Combining this result with (\ref{temporaire}), we get (\ref{torsion2}).

With the help of the symbolic computation software Mathematica and using the local canonical form of a curve, we computed the Taylor series expansion for both expressions. For (\ref{torsion1}), we obtained

\begin{equation}
\label{taylortorsion1}
\tilde{\tau}_1=\tau+\frac{\tau\kappa_s}{6\kappa}(a-b+3e)+\frac{\tau_s}{4}(b-a+e)+O(2).
\end{equation}
For (\ref{torsion2})  we obtained 
\begin{equation}
\label{taylortorsion2}
\tilde{\tau}_2=\tau+\frac{\tau\kappa_s}{6\kappa}(a+b+e)+\frac{\tau_s}{4}(b-a+e)+O(2).
\end{equation}
These expansions prove that (\ref{torsion1}) and (\ref{torsion2}) are not equivalent.

There are many ways to approximate $\tau_s$. For the reasons mentioned above, we chose to use (\ref{torsion1}). In a similar way as for $\kappa_s$, we used (\ref{taylortorsion1}) to obtain the following numerically invariant expression for $\tau_s$. For symmetry reasons and in view of the results obtained for $\kappa_s$, we decided on using a centered formula. The result is the following five point approximation.

\begin{equation}
\tilde{\tau}_s(P_i)=4\cdot\frac{\tilde{\tau}^1(P_{i+1})-\tilde{\tau}^1(P_{i-1})+(2a+2b-2d-3h+g)\frac{\tilde{\tau}^1(P_i)\tilde{\kappa}_s(P_i)}{6\tilde{\kappa}(P_i)}}{2a+2b+2d+h+g}
\end{equation}
Where $g=d(P_{i-2},P_{i-1})$ as illustrated on figure 1 and $\tilde{\kappa}_s$ can be taken to be either (\ref{2}), (\ref{3}) or (\ref{4}).      

\subsection{Numerical test}

We considered the curve $\alpha (t)=(\cos t, \sin t, \sqrt{t})$ whose graph is given in figure 12. Note that $\alpha$ is not parametrized by arc length. The signature of this curve is given by the four following quantities.
\begin{eqnarray}
\kappa&=&|\alpha_{ss}|=\frac{|\alpha_t\times\alpha_{tt}|}{|\alpha_t|^{3}}=\frac{2\sqrt{16t^3+4t^2+1}}{(1+4t)^{3/2}},\nonumber\\
\kappa_s&=&\frac{d\kappa}{dt}\frac{dt}{ds}=\frac{\frac{d\kappa}{dt}}{\frac{ds}{dt}}=\frac{8(8t^2+2t-3)\sqrt{t}}{\sqrt{16t^3+4t^2+1}(1+4t)^3},\nonumber\\
\tau&=&-\frac{\alpha_t\times\alpha_{tt}\cdot\alpha_{ttt}}{|\alpha_t\times\alpha_{tt}|^2}=\frac{-2\sqrt{t}(3+4t^2)}{16t^3+4t^2+1},\nonumber\\
\tau_s&=&\frac{d\tau}{dt}\frac{dt}{st}=\frac{\frac{d\tau}{dt}}{\frac{ds}{dt}}=\frac{2(64t^5-16t^4+240t^3+16t^2-3)}{\sqrt{4t+1}(16t^3+4t^2+1)^2}.\nonumber
\end{eqnarray}
We tested our numerical expressions on the portion of the curve given by $\frac{\pi}{2}\leq t\leq \frac{3\pi}{2}$ for different partition size. We used $\tilde{\kappa}_s$ given by equation (\ref{4}) for evaluating $\tilde{\tau}_s$. Care was taken to partition the curve very irregularly. In the example presented on figures 13-22, we computed the signature for a partition $\{t_i\}_{i=0}^N$ built the following way
\begin{eqnarray}
t_0 &=&o\nonumber\\
t_1 &=&t_0+\Delta t\nonumber\\
t_2 &=&t_1+\frac{\Delta t}{2}\nonumber\\
t_3 &=&t_2+\frac{\Delta t}{3}\nonumber\\
t_4 &=&t_3+\Delta t\nonumber\\
t_5 &=&t_4+\frac{\Delta t}{2}\nonumber\\
t_6 &=&t_5+\frac{\Delta t}{3}\text{ etc. }\nonumber
\end{eqnarray}

The resulting graphs are presented for $\Delta t=0.1$, $0.05$, $0.025$ and $0.0125$.
For simplicity, we graphed two projections of the signature curve. Although a lot of the information is lost this way, we believe that it gives a good measure of the effectivity of the method. We can see that although the partition is not regular, the graphs seem to converge to the exact graphs  when  $\Delta t$ becomes small. Therefore we believe that our formulas work properly for generic small partitions.

\section{Conclusion}

We now have in our hands good approximations for the Euclidean and affine signature of a planar curve. The formulas we obtained are invariant under the action of the Euclidean and affine group respectively. Another important property is that they are valid for any fine partition of the given curve. In a near future, we hope to use them for the recognition of planar curves and perhaps also for some other applications of the signature curve among the numerous possible ones. In particular, we wish to improve the numerical results obtained with noisy images and prove that the signature can be a practical tool of object recognition.

We also have good approximations for the four differential invariants which parametrize the Euclidean signature of a space curve, namely the curvature, the torsion and their derivatives with respect to arc length. The formulas we obtained are Euclidean invariants and valid for any fine partition of the given curve, although better results are obtained with equidistant partitions. In particular, we showed in section 3 that $\tilde{\tau}=\pm 6\cdot\frac{H}{def\kappa}$ is a good approximation for the torsion. We believe that this expression has a value on its own as it gives an easy visual understanding of the concept of torsion. Space curve recognition (for example blood vessels and trajectory of particles) is an interesting computer vision problem where our formulas could find applications.

\newpage

\begin{figure}[here]
\caption{Exact Data Used for Testing the Planar Euclidean Case.
} 
\vspace{1cm}
\[
\begin{array}{ll}
\epsfysize=6.0cm
\epsfxsize=6.0cm
{ \epsfbox{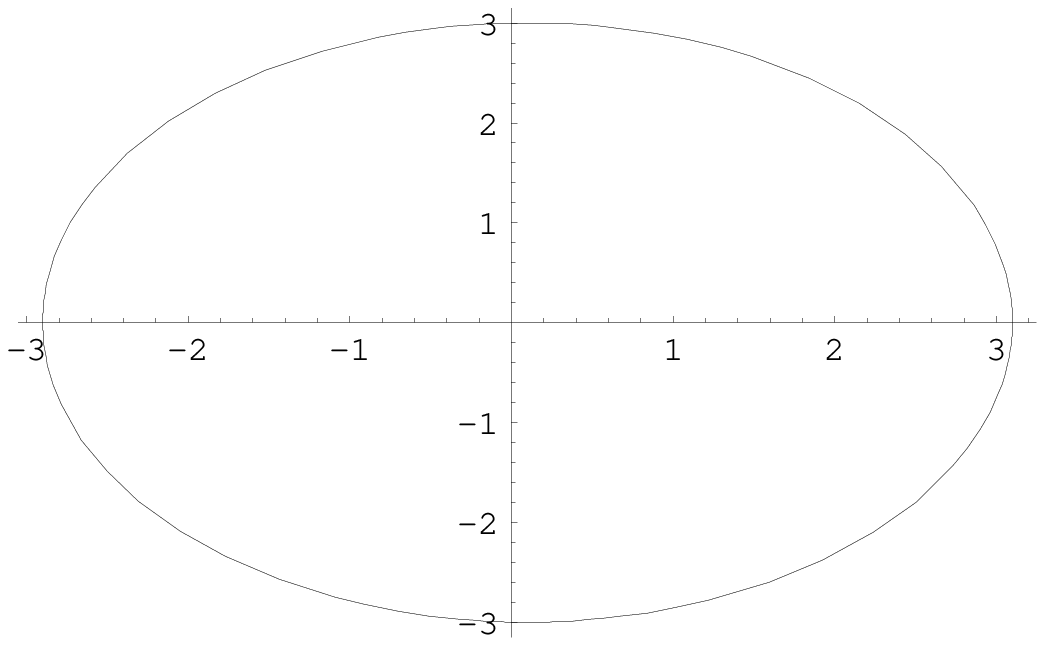}}\hspace{1cm}&
\epsfysize=6cm
\epsfxsize=6cm
{ \epsfbox{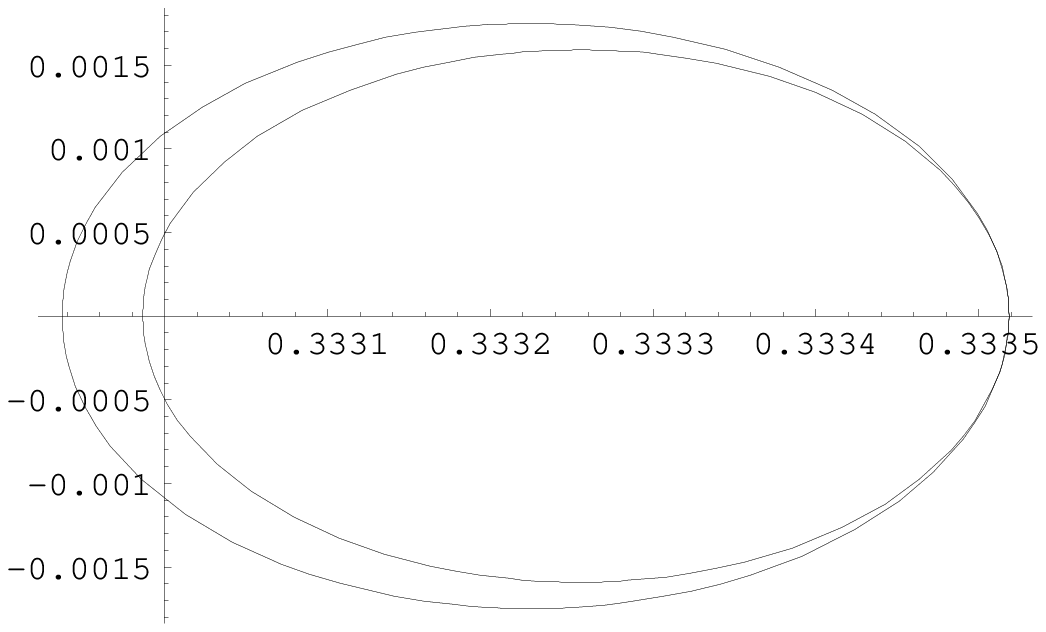}}
\\
\text{\small The Curve $r=1+\frac{1}{10}\cos t$ }&\text{\small Corresponding Euclidean signature. }
\end{array}
\]
\end{figure}
\vspace{3cm}

\begin{figure}[here]
\caption{Three Different Partitions of the Initial Curve Used in the Planar Euclidean Case.}
\vspace{1cm}
$\begin{array}{lll}
\epsfysize=5cm
\epsfxsize=5cm
{ \epsfbox{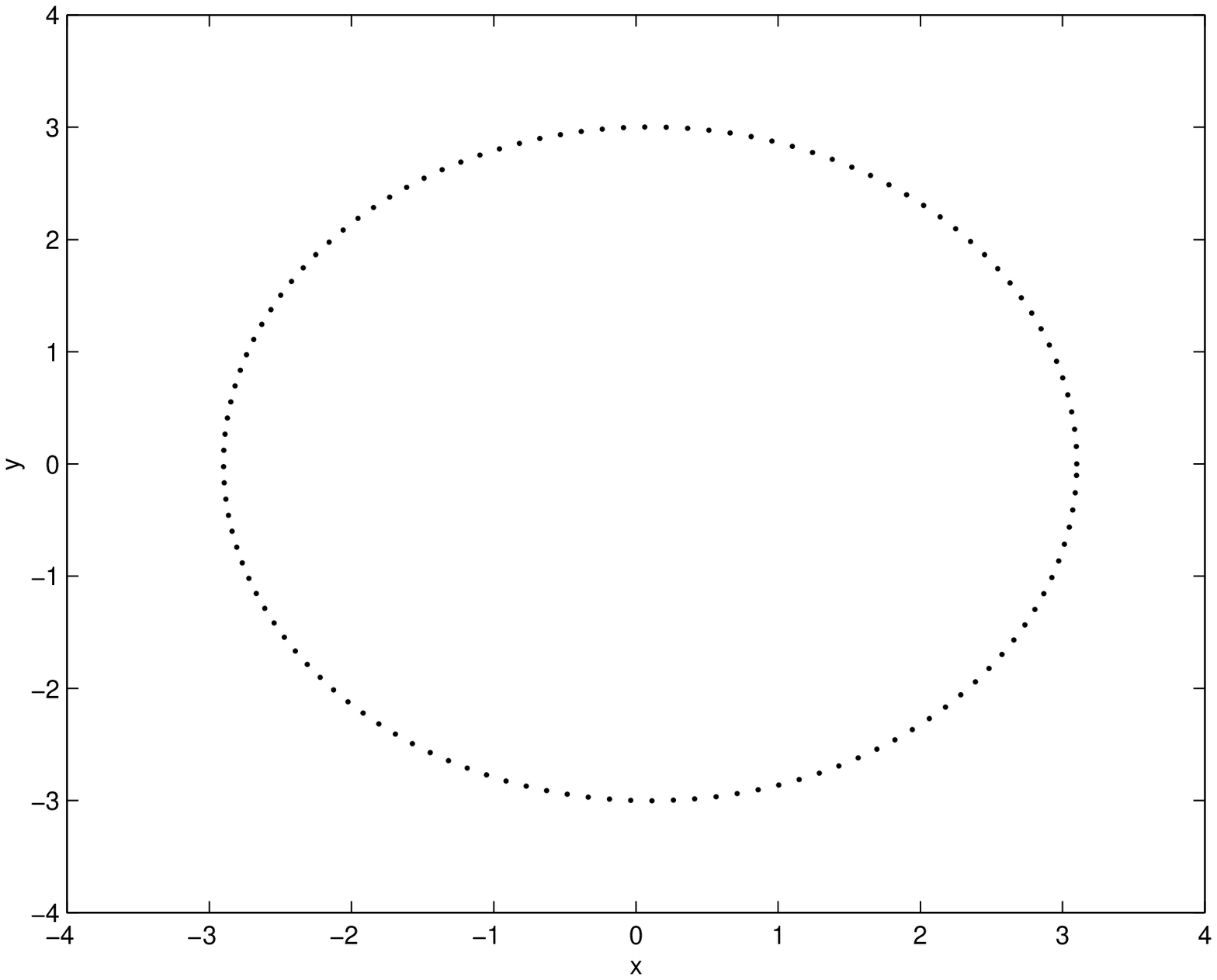} }&
\epsfysize=5cm
\epsfxsize=5cm
{ \epsfbox{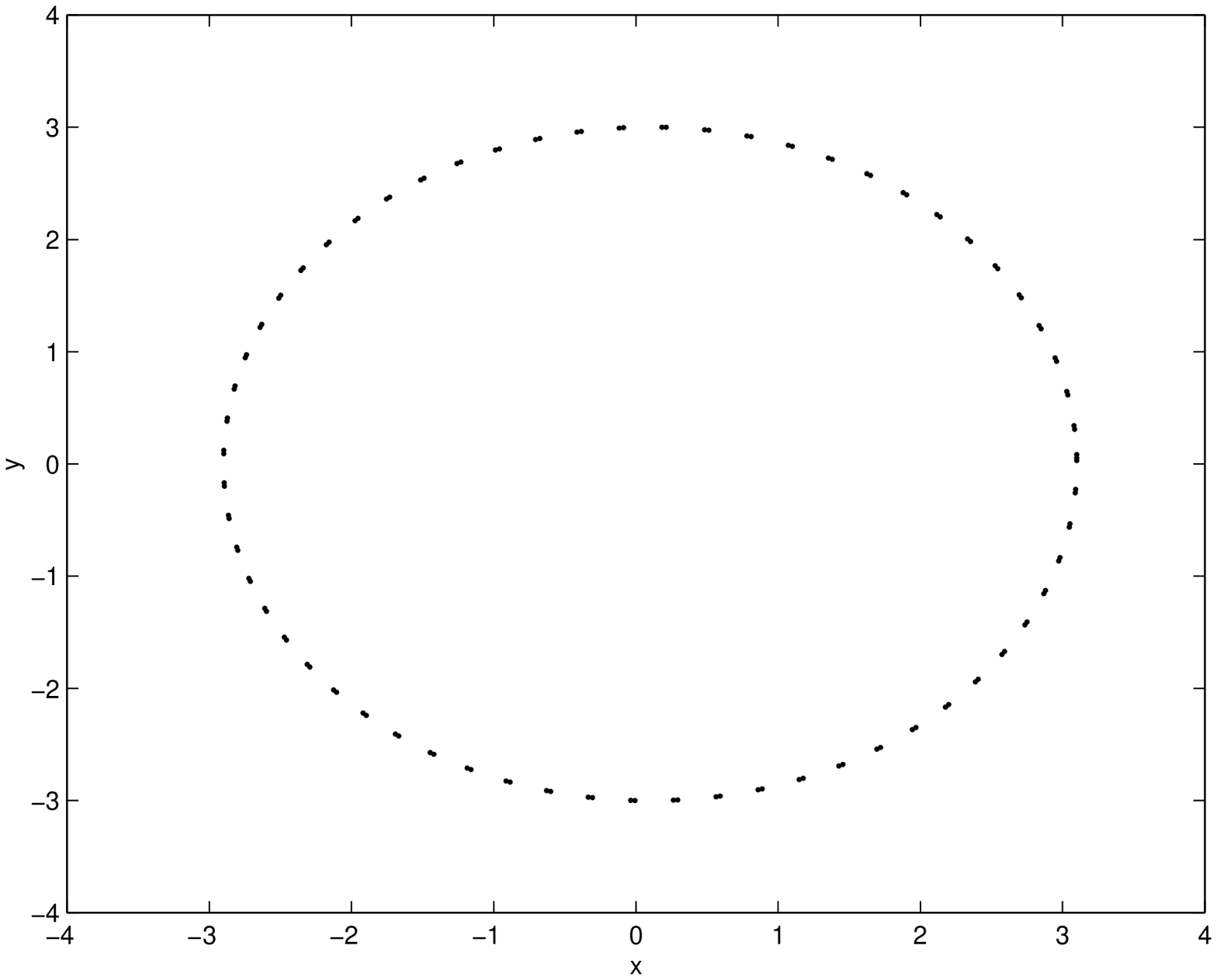}  }&
\epsfysize=5cm
\epsfxsize=5cm
{ \epsfbox{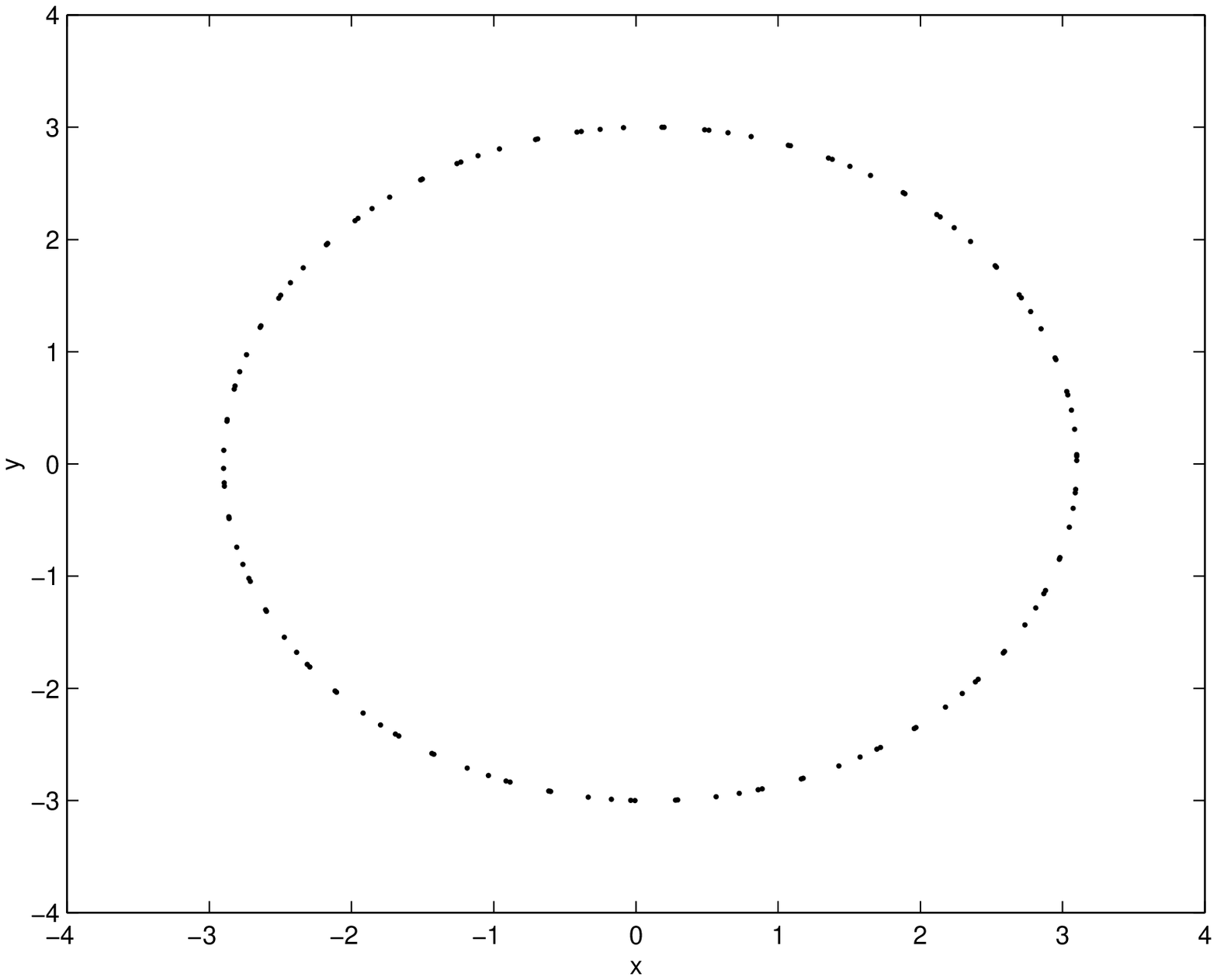}  }
\\
\text{\small a) Regular Partition.} & \text{\small b) Irregular Partition.} & \text{\small c) Very Irregular Partition.}
\end{array}$
\end{figure}

\newpage

\begin{figure}[here]
\caption{Approximations of the Euclidean Signature Curve Obtained with (\ref{initial}).}
\vspace{1cm} 
$
\begin{array}{lll}
\epsfysize=5cm
\epsfxsize=5cm
{ \epsfbox{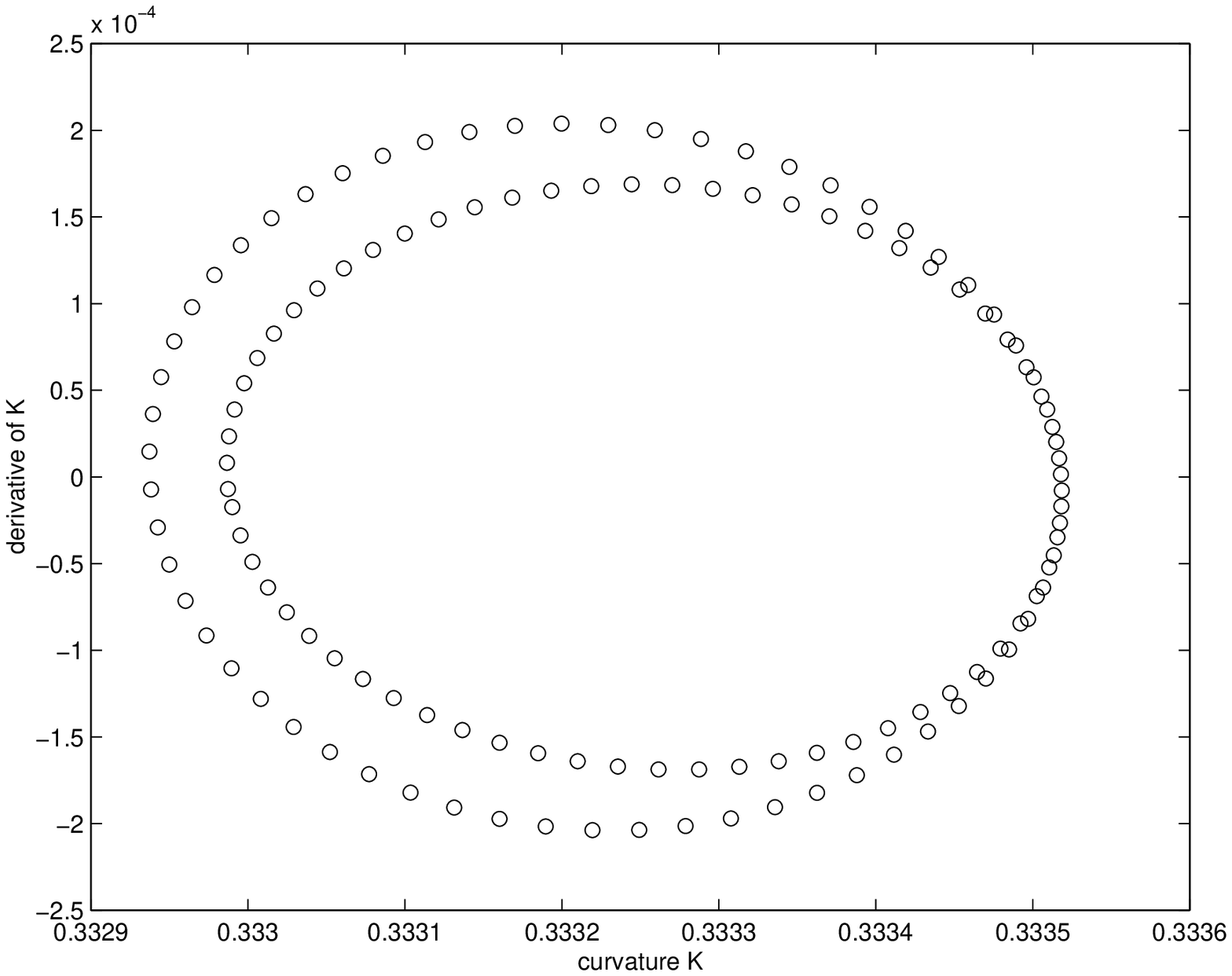}}&

\epsfysize=5cm
\epsfxsize=5cm
{ \epsfbox{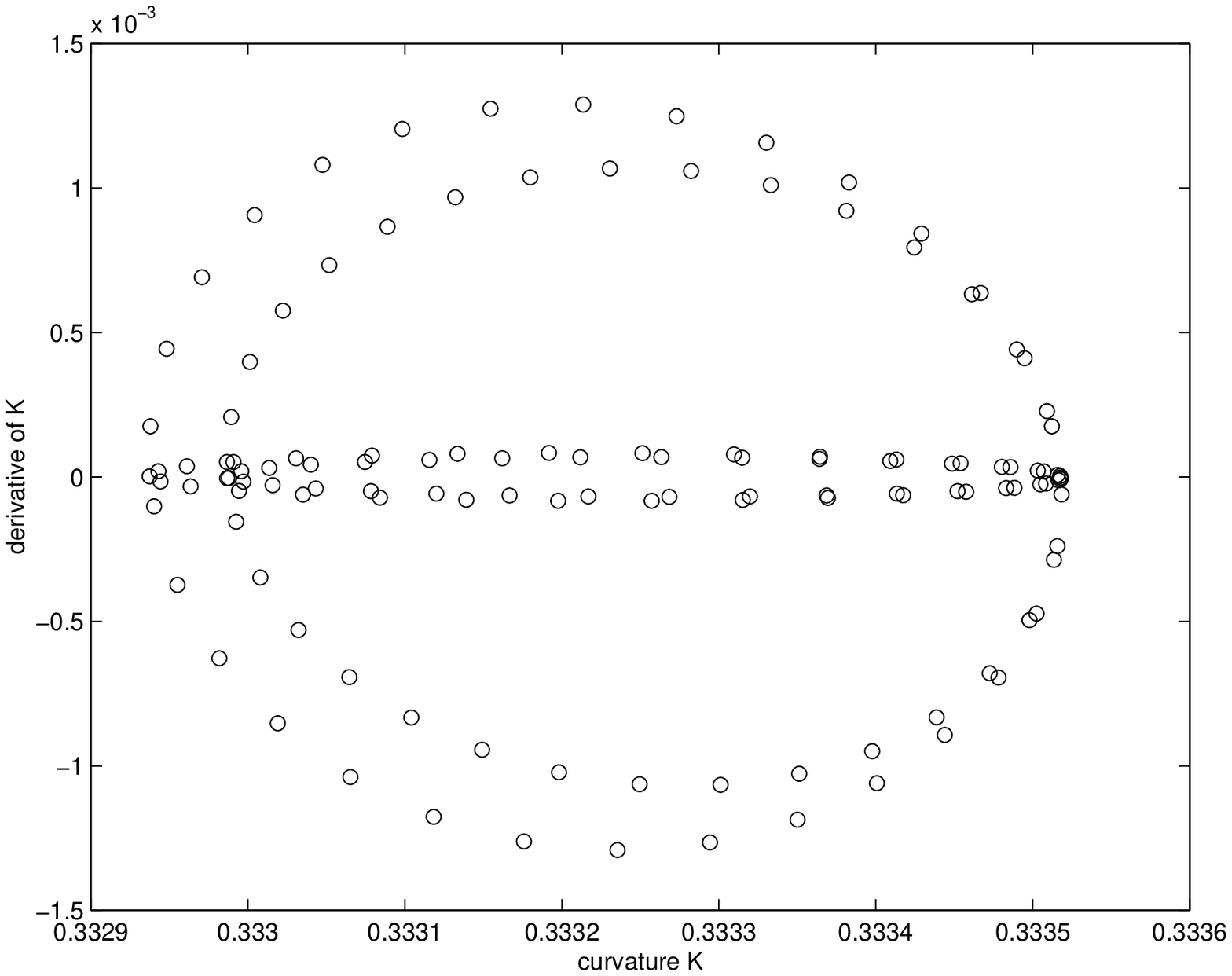}  }&

\epsfysize=5cm
\epsfxsize=5cm
{ \epsfbox{2Knice_init_form.ps}}
\\
\text{\small With Partition 3-a.} & \text{\small With Partition 3-b.} & \text{\small With Partition 3-c.}
\end{array}
$
\end{figure}

\vspace{0.5cm}

\begin{figure}[here]
\caption{Approximations of the Euclidean Signature Curve Obtained with (\ref{initial_symmetric}).} 
\vspace{1cm}
$ 
\begin{array}{lll}
\epsfysize=5cm
\epsfxsize=5cm
{ \epsfbox{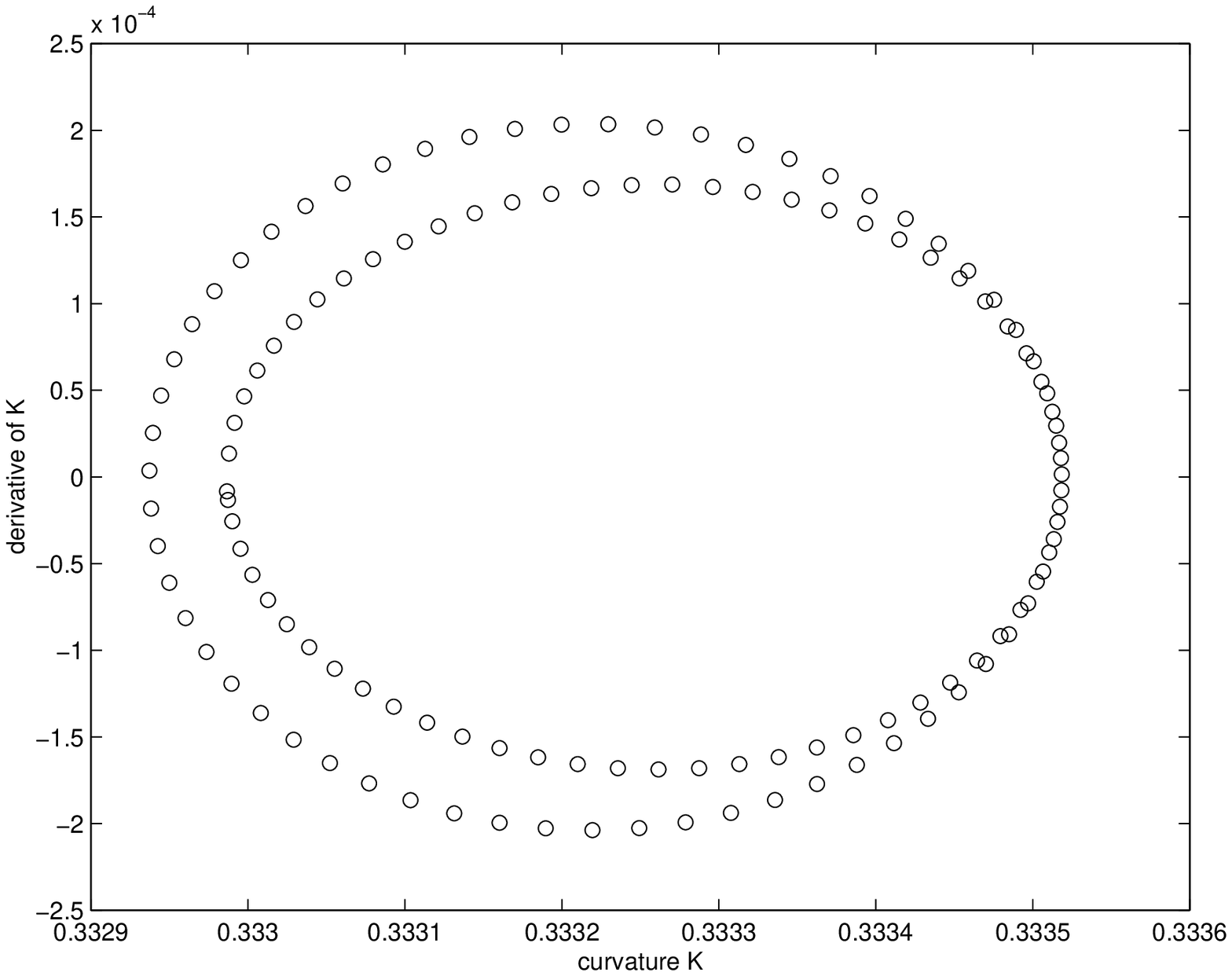}} &

\epsfysize=5cm
\epsfxsize=5cm
{ \epsfbox{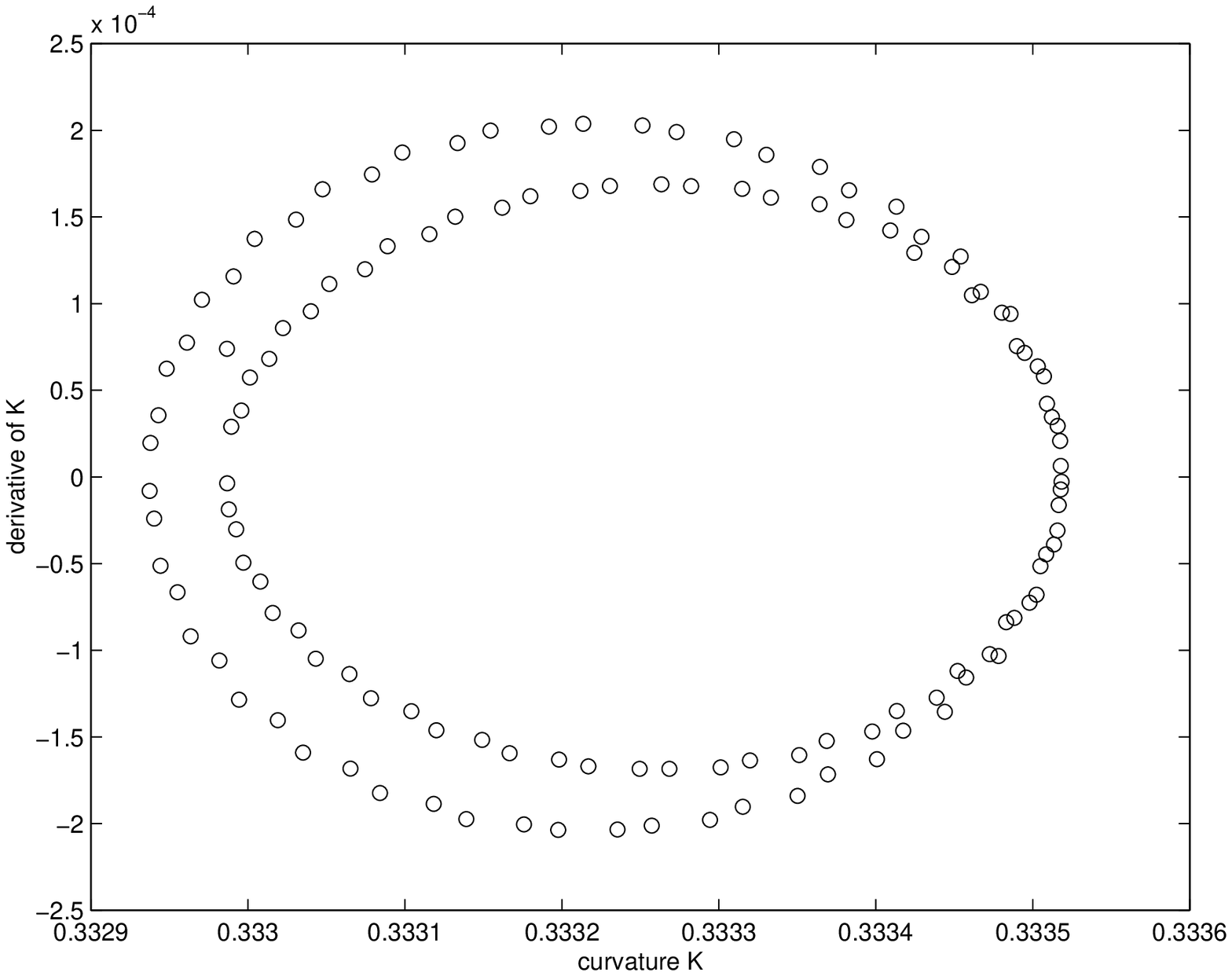}  } &

\epsfysize=5cm
\epsfxsize=5cm
{ \epsfbox{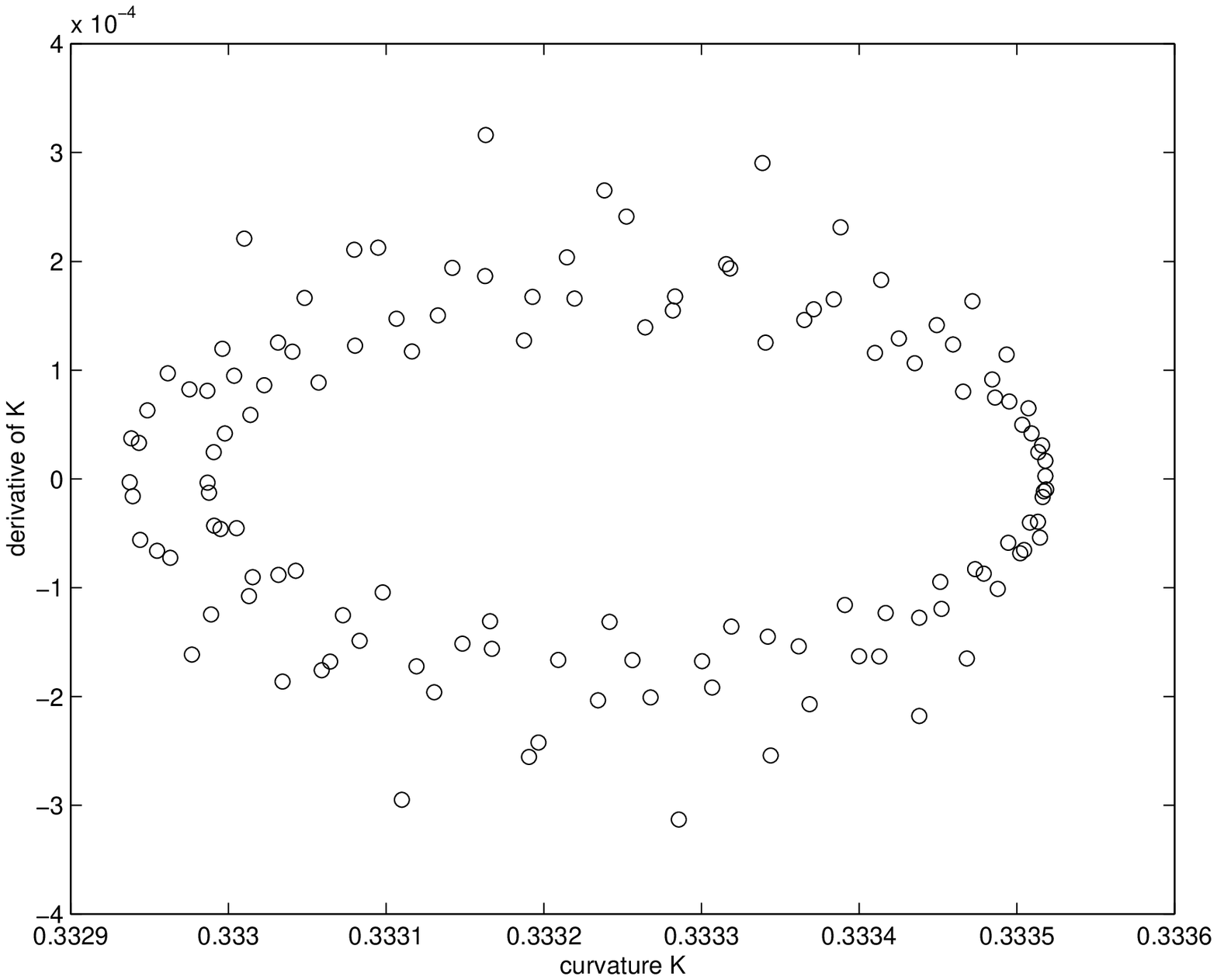}}
\\
\text{\small With Partition 3-a.} & \text{\small With Partition 3-b.} & \text{\small With Partition 3-c.}
\end{array}
$
\end{figure}

\newpage

\begin{figure}[here]
\caption{Approximations of the Euclidean Signature Curve Obtained with (\ref{2}).} 
$
\begin{array}{lll}

\epsfysize=4cm
\epsfxsize=4cm
{ \epsfbox{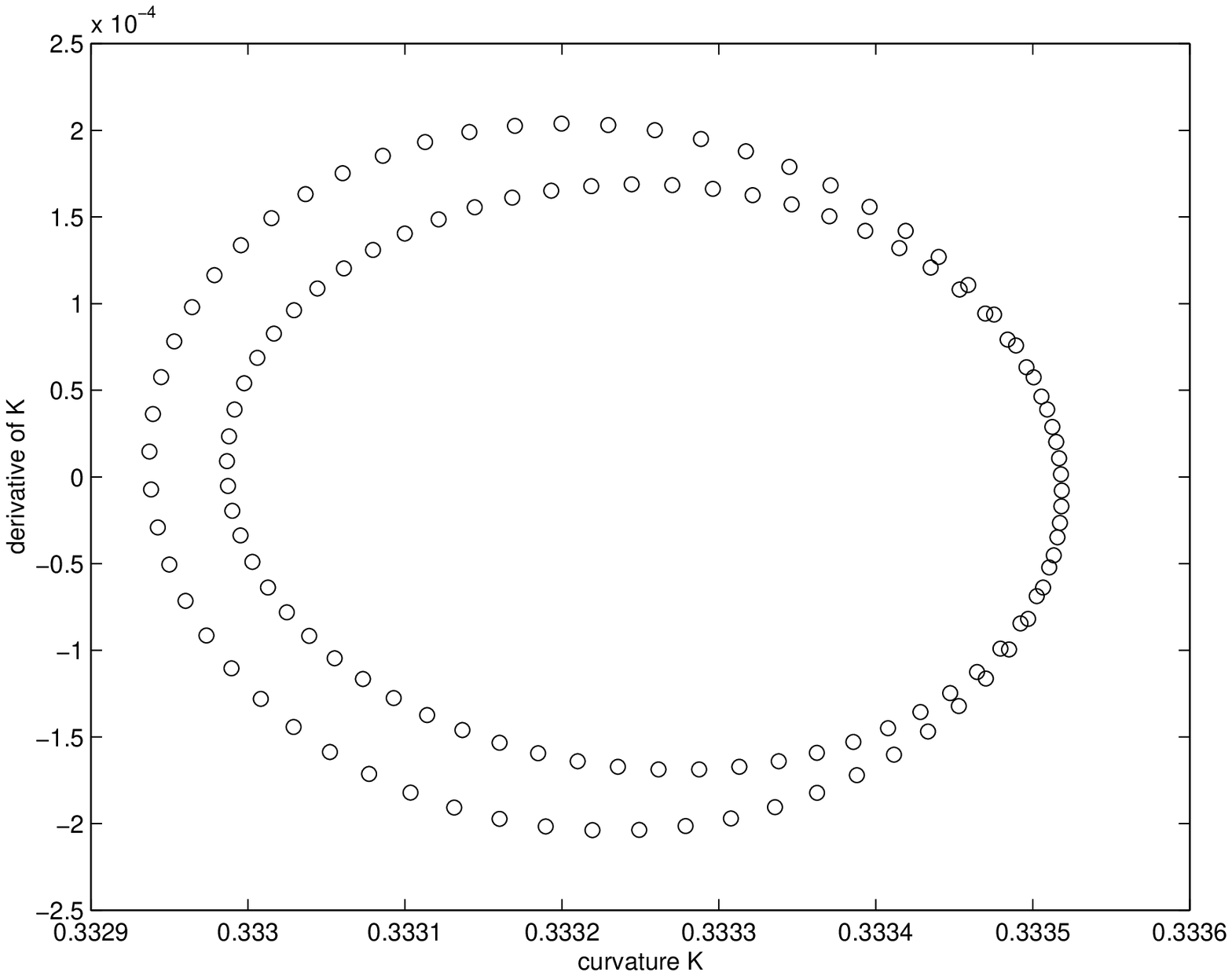}} &

\epsfysize=4cm
\epsfxsize=4cm
{ \epsfbox{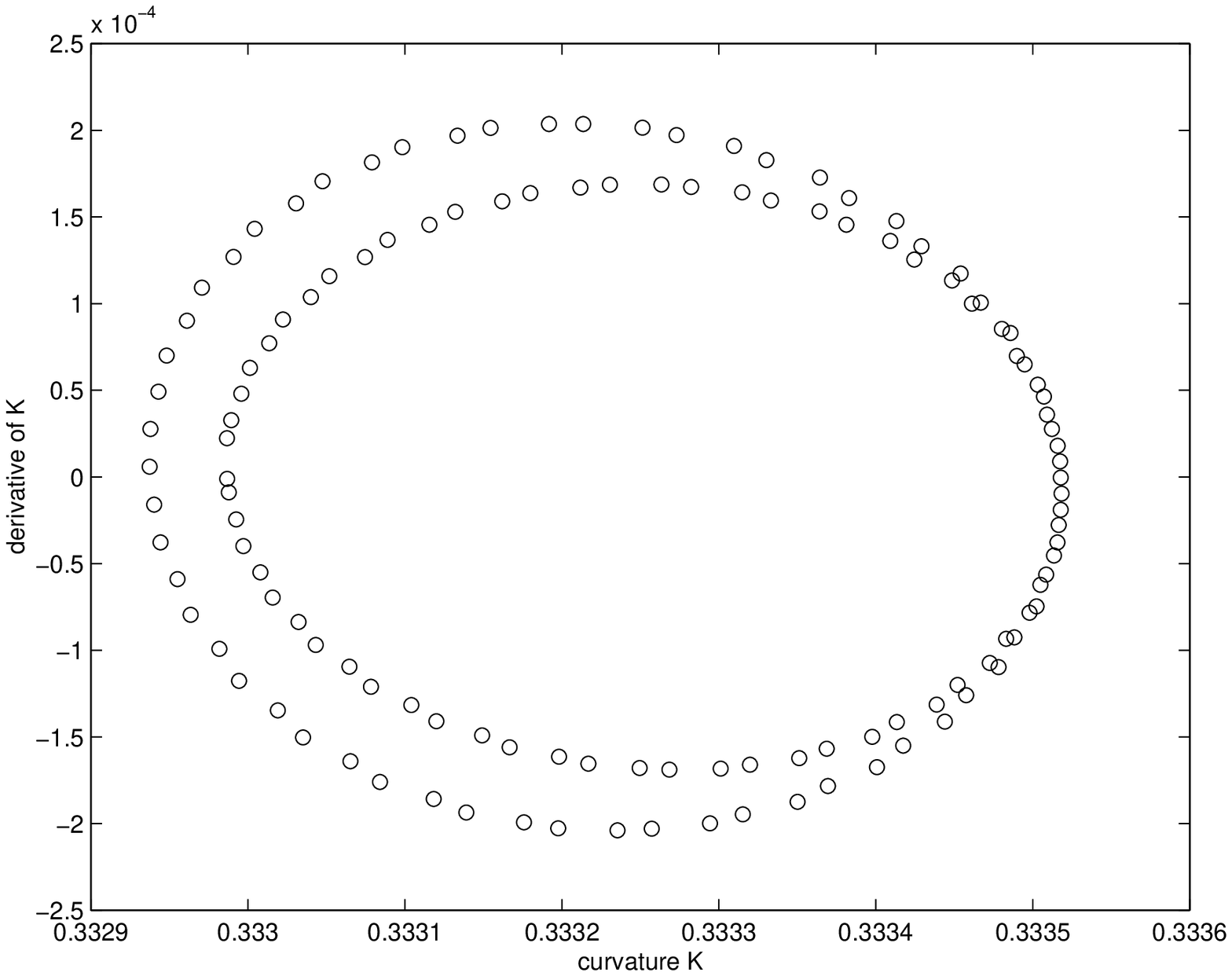}  }&

\epsfysize=4cm
\epsfxsize=4cm
{ \epsfbox{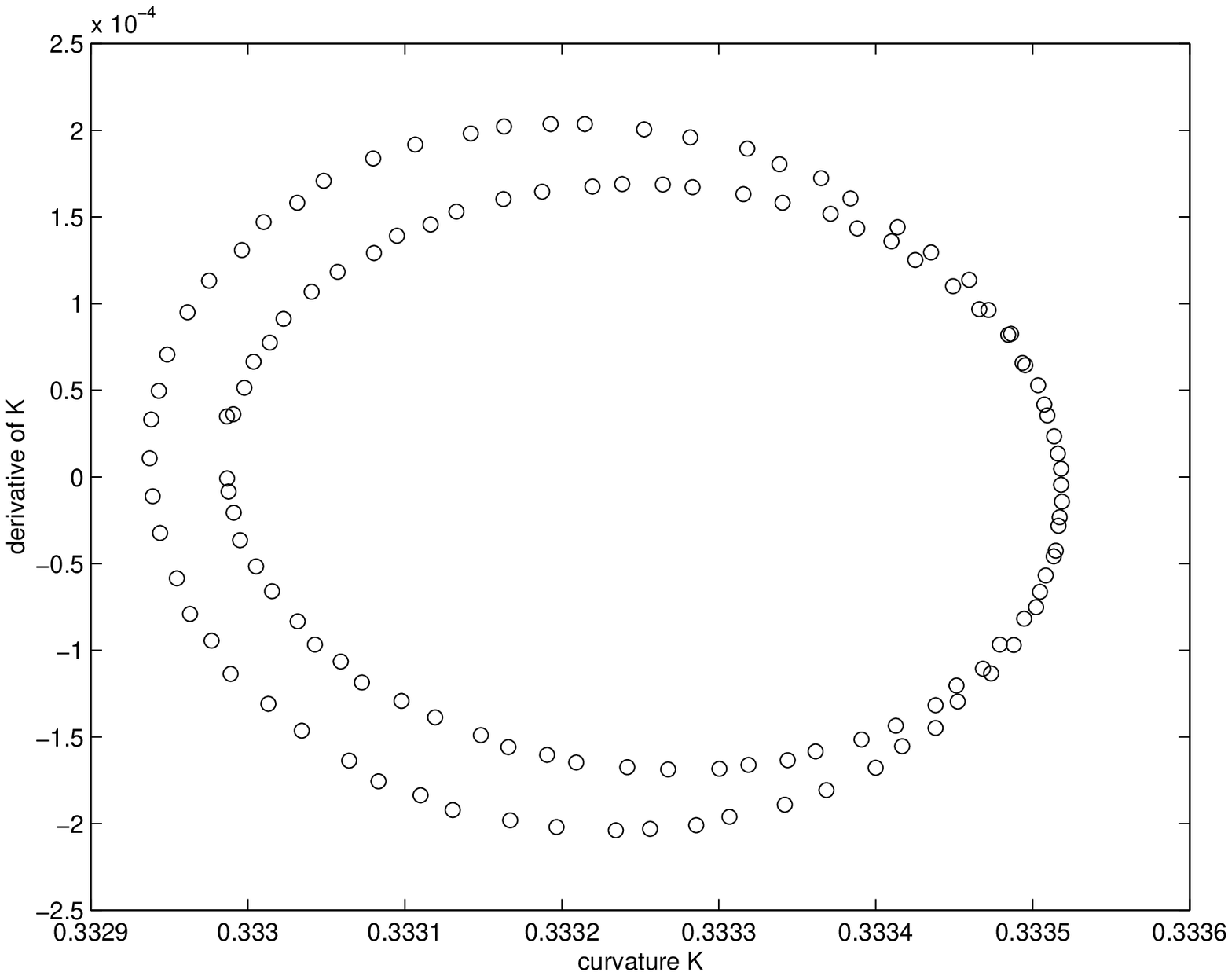}}
\\
\text{\small With Partition 3-a.} & \text{\small With Partition 3-b.} & \text{\small With Partition 3-c.}
\end{array}
$
\end{figure}

\begin{figure}[here]
\caption{Approximations of the Euclidean Signature Curve Obtained with (\ref{3}).} 
$
\begin{array}{lll}

\epsfysize=4cm
\epsfxsize=4cm
{ \epsfbox{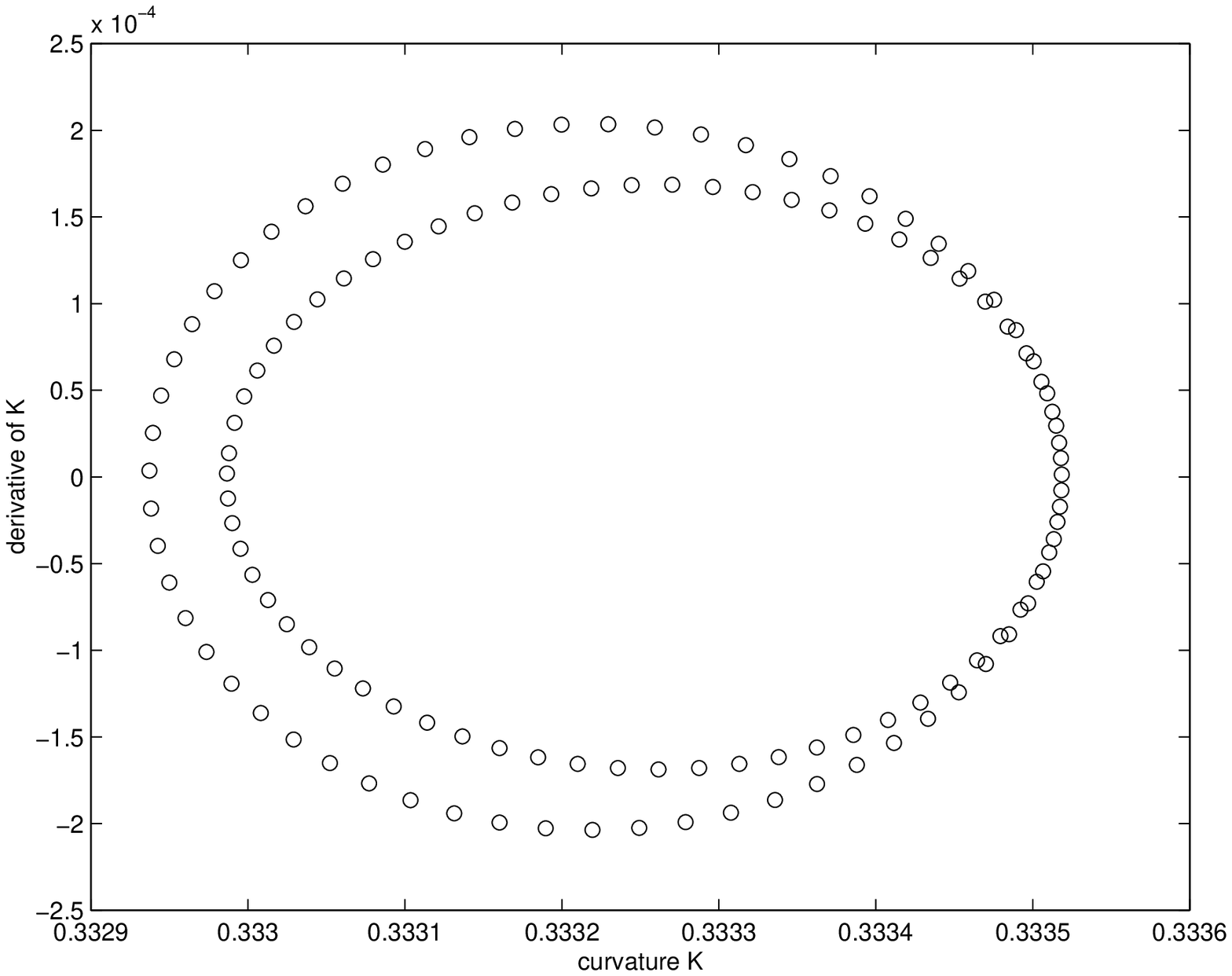}}&
\epsfysize=4cm
\epsfxsize=4cm
{ \epsfbox{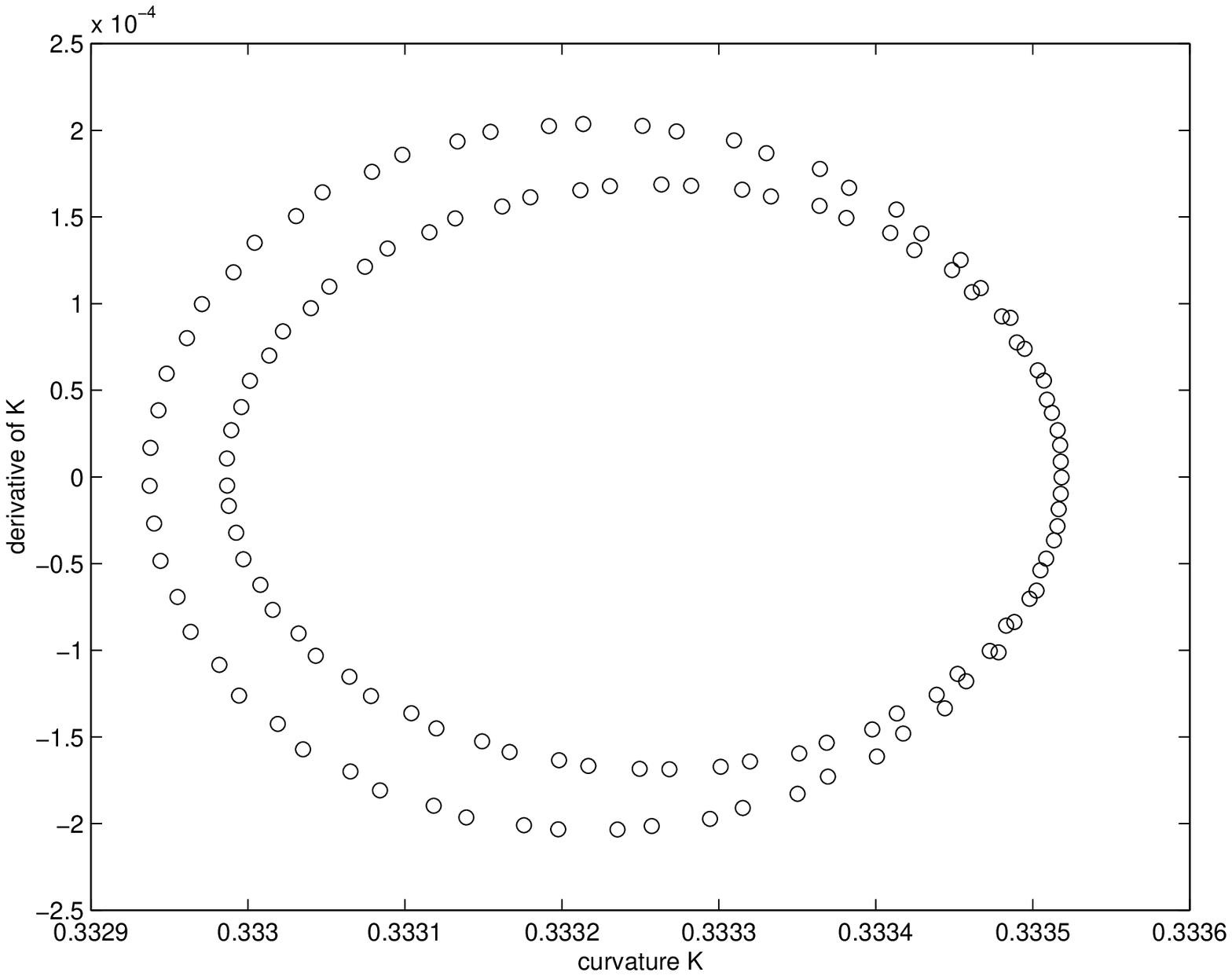}  }&
\epsfysize=4cm
\epsfxsize=4cm
{ \epsfbox{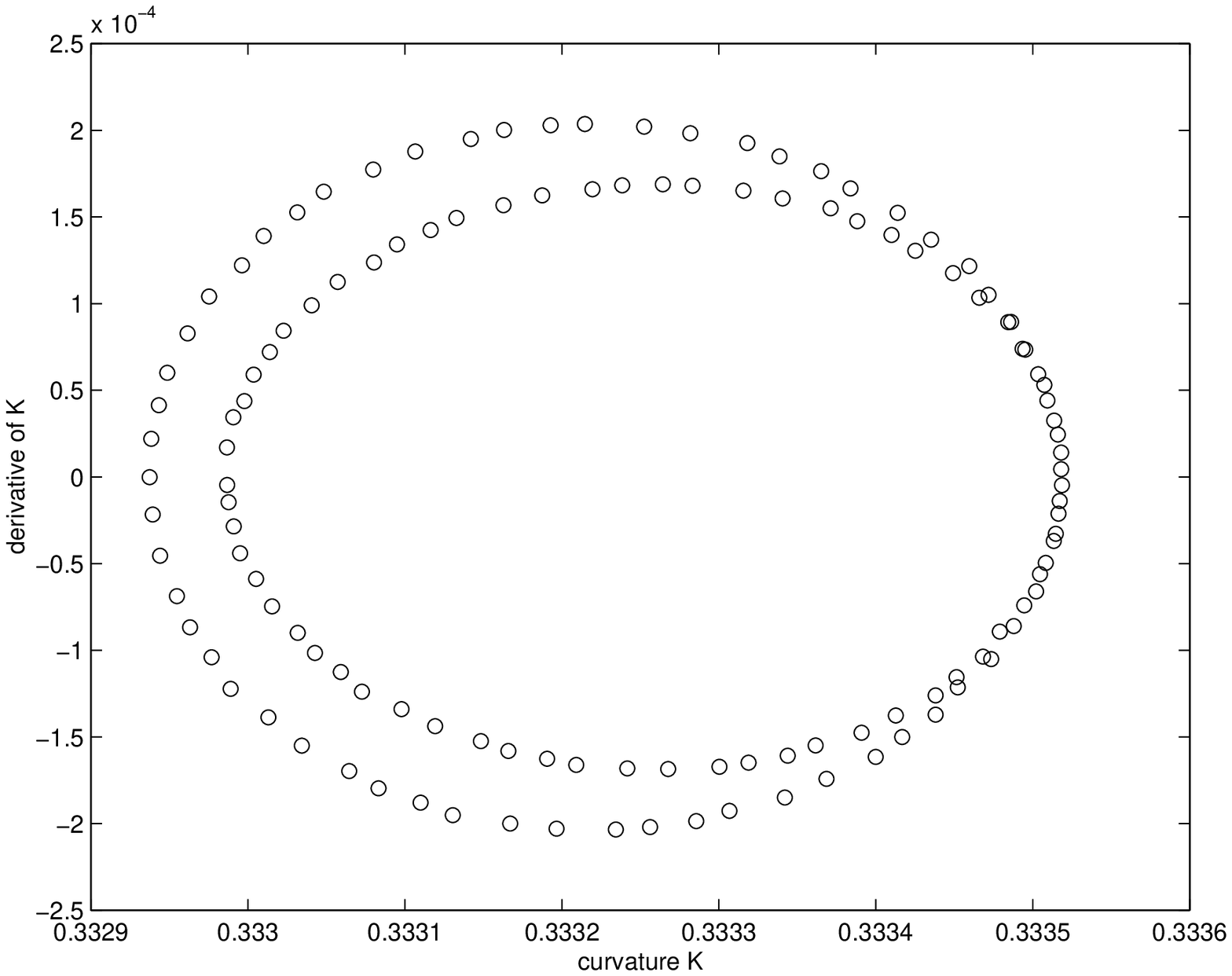}}
\\
\text{\small With Partition 3-a.} &\text{\small With Partition 3-b.} & \text{\small With Partition 3-c.}
\end{array}
$
\end{figure}

\begin{figure}[here]
\caption{ Approximation of the Euclidean Signature Curve Obtained with (\ref{4}).}\
$
\begin{array}{lll}

\epsfysize=4cm
\epsfxsize=4cm
{ \epsfbox{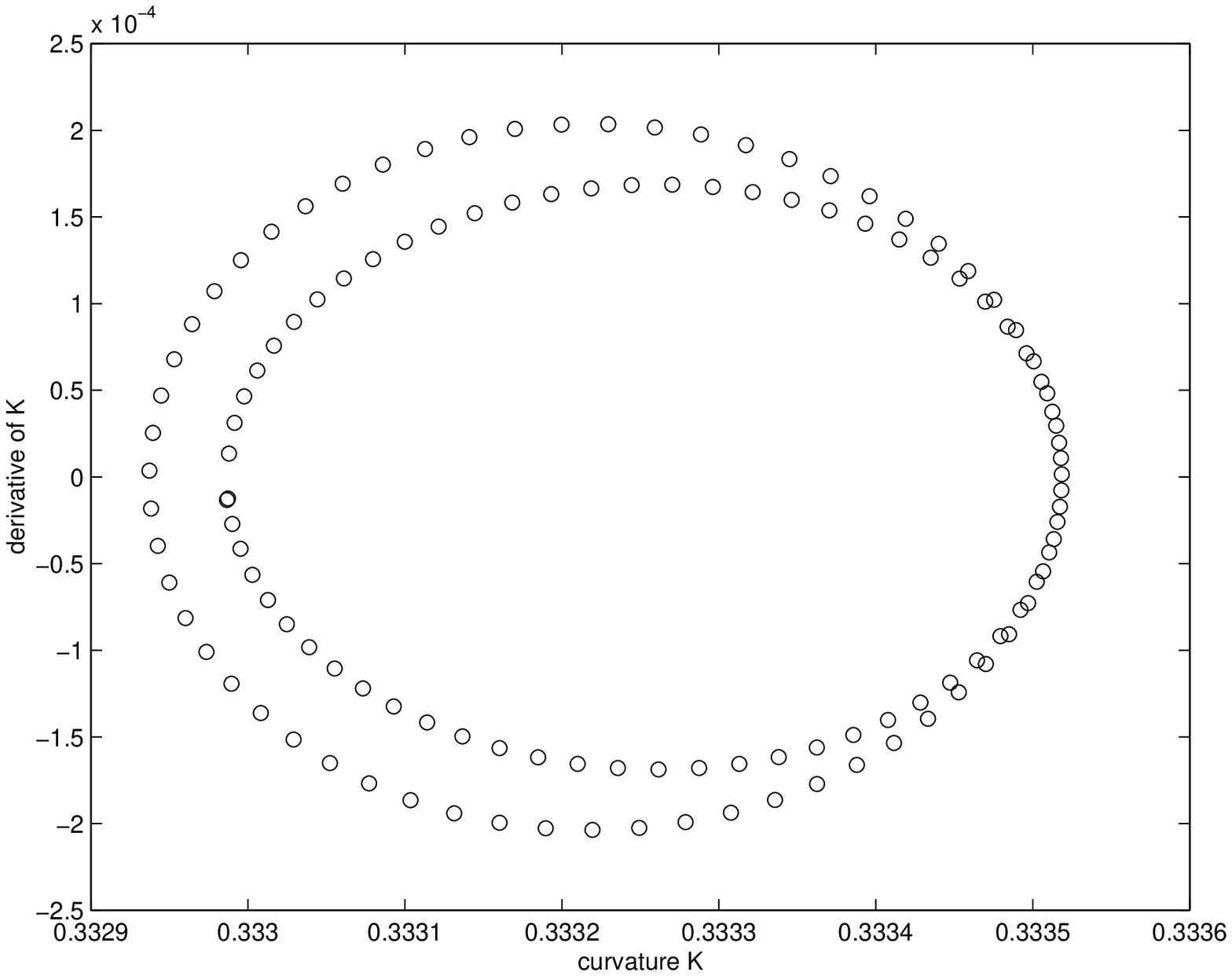}}&

\epsfysize=4cm
\epsfxsize=4cm
{ \epsfbox{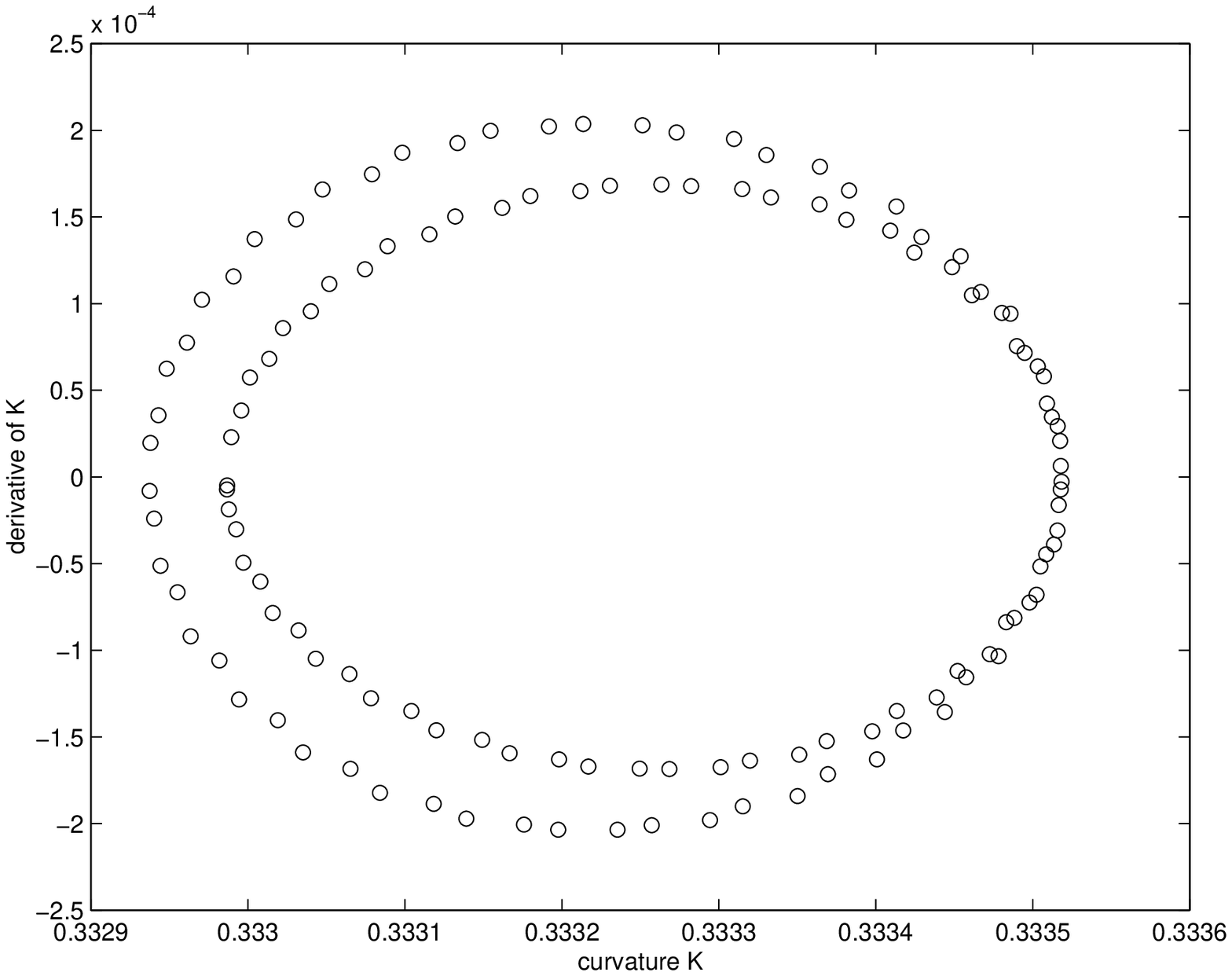}  }&

\epsfysize=4cm
\epsfxsize=4cm
{ \epsfbox{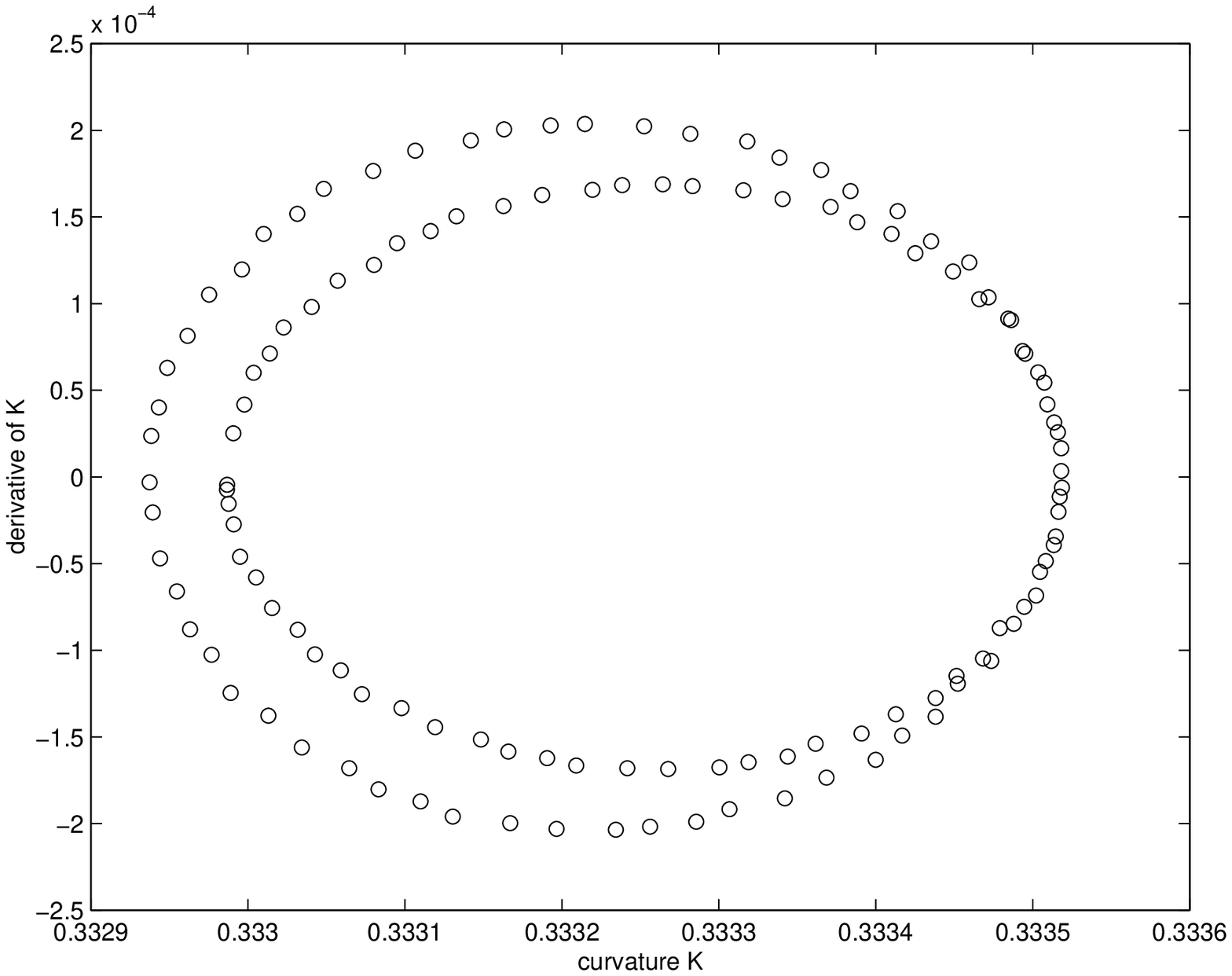}}
\\
\text{\small With Partition 3-a.} &\text{\small With Partition 3-b.} & \text{\small With Partition 3-c.}
\end{array}
$
\end{figure}

\newpage
\begin{figure}[here]
\caption{Exact Data Used for Testing the Planar Affine Case.} 
\vspace{1cm}
$
\begin{array}{ll}
\epsfysize=6.0cm
\epsfxsize=6.0cm
{\epsfbox{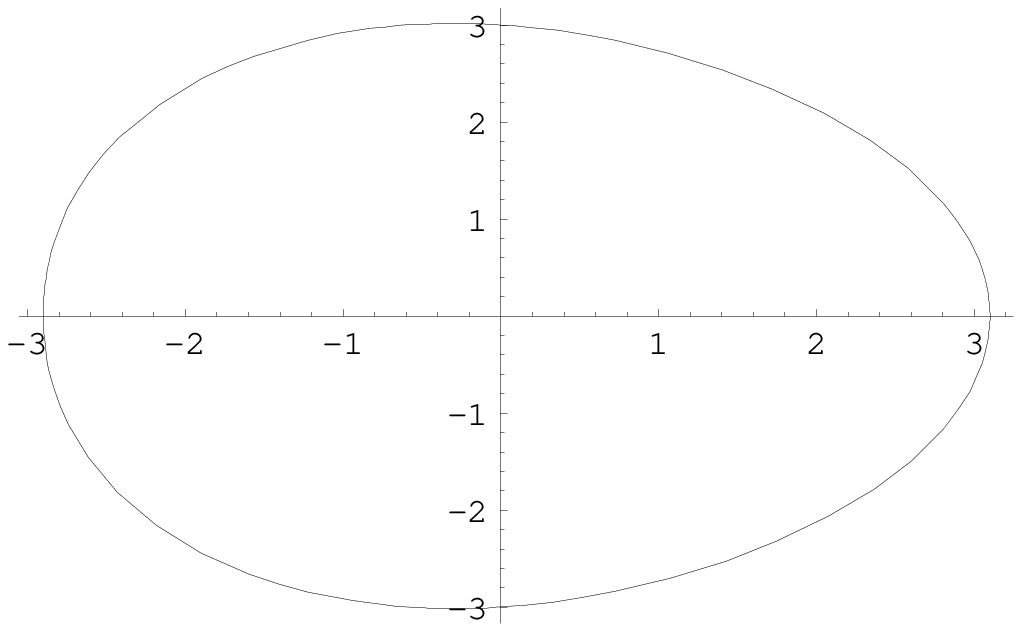} }\hspace{1cm}&
 \epsfysize=6.0cm
\epsfxsize=6.0cm
{ \epsfbox{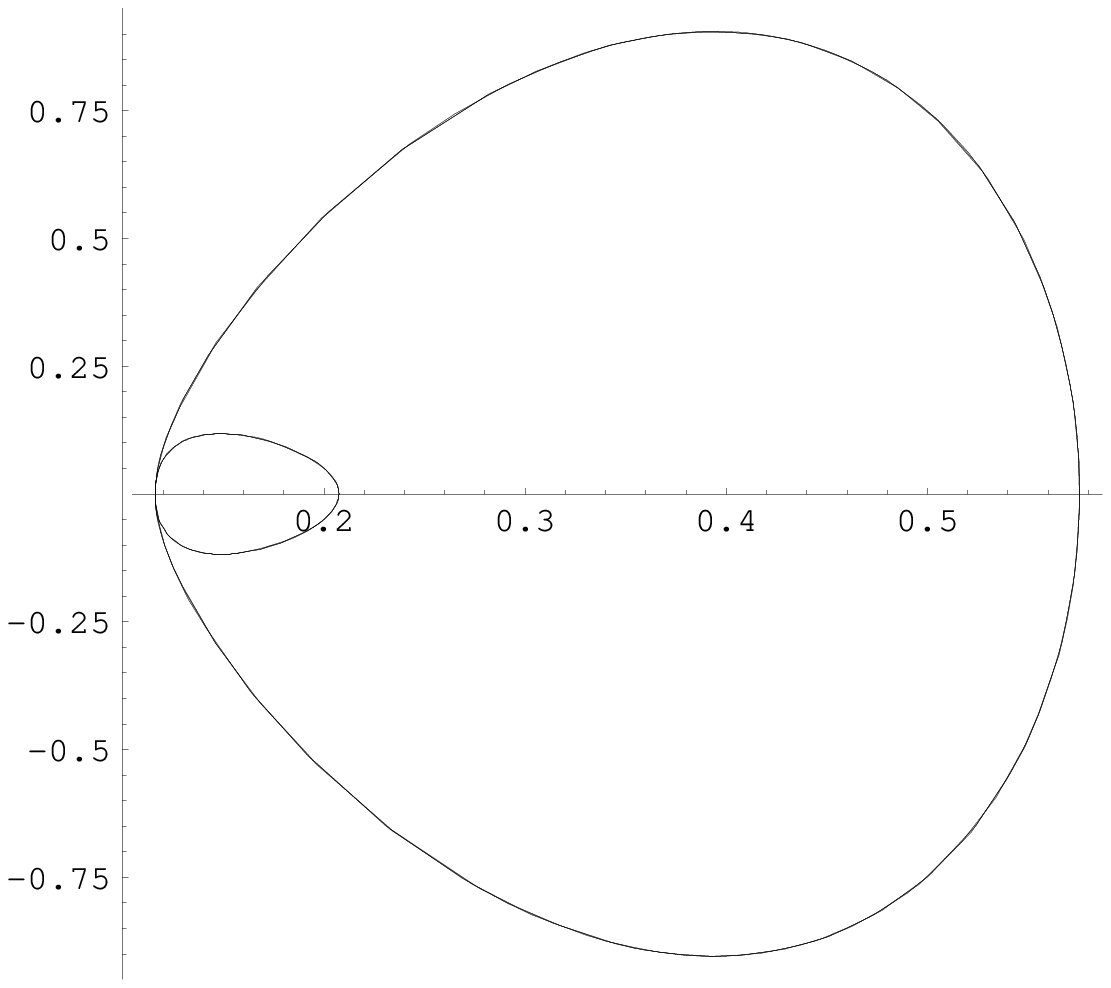}}
\\
\text{\small The Curve $r=1+\frac{1}{10}\cos 3t$} & \text{\small Corresponding affine signature.}
\end{array}
$
\end{figure}
\vspace{2cm}

\begin{figure}[here]
\caption{Three Different Partitions of the Initial Curve Used in the Planar Affine Case.}
\vspace{1cm}
$
\begin{array}{lll}
\epsfysize=5cm
\epsfxsize=5cm
{ \epsfbox{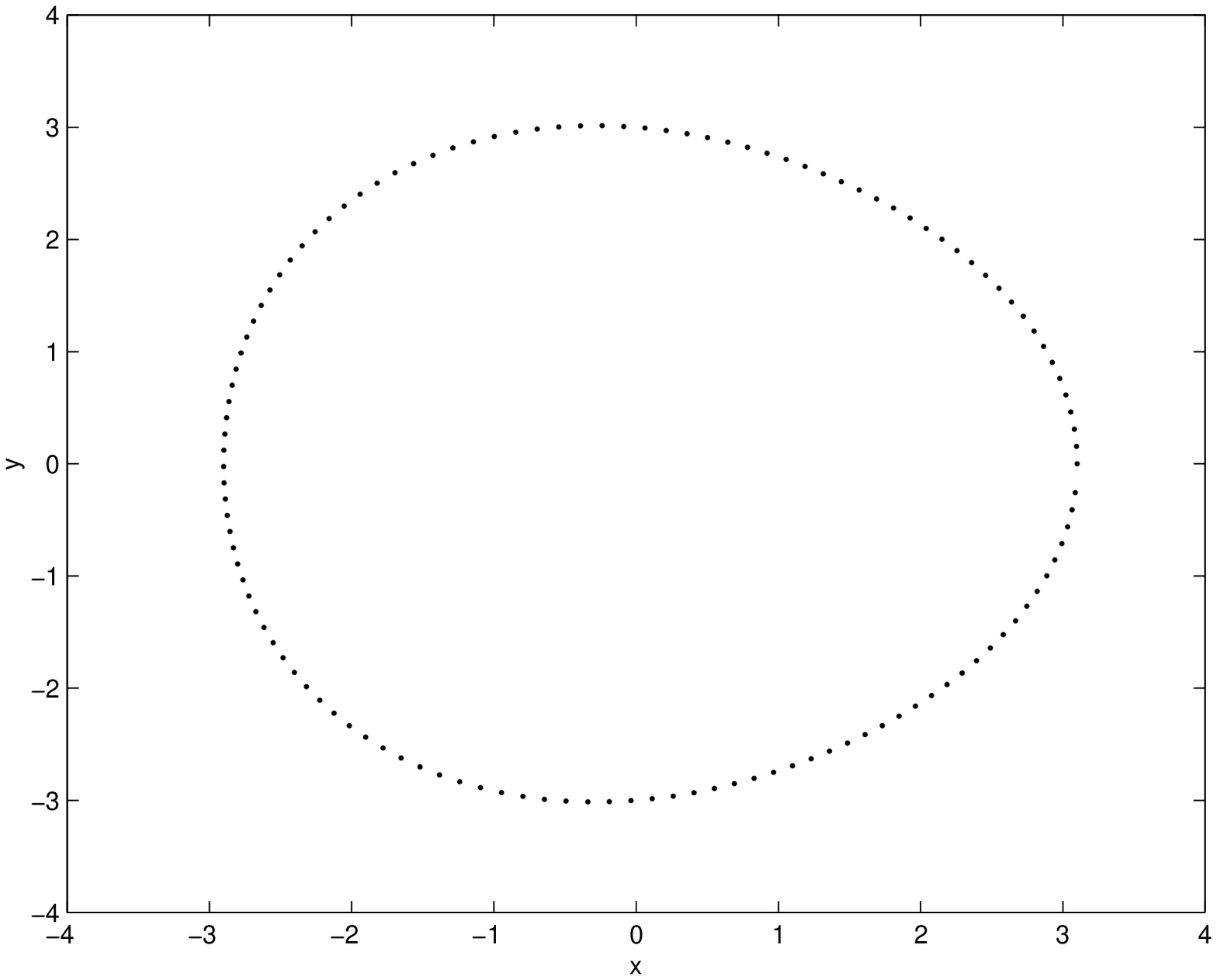} } &

\epsfysize=5cm
\epsfxsize=5cm
{\epsfbox{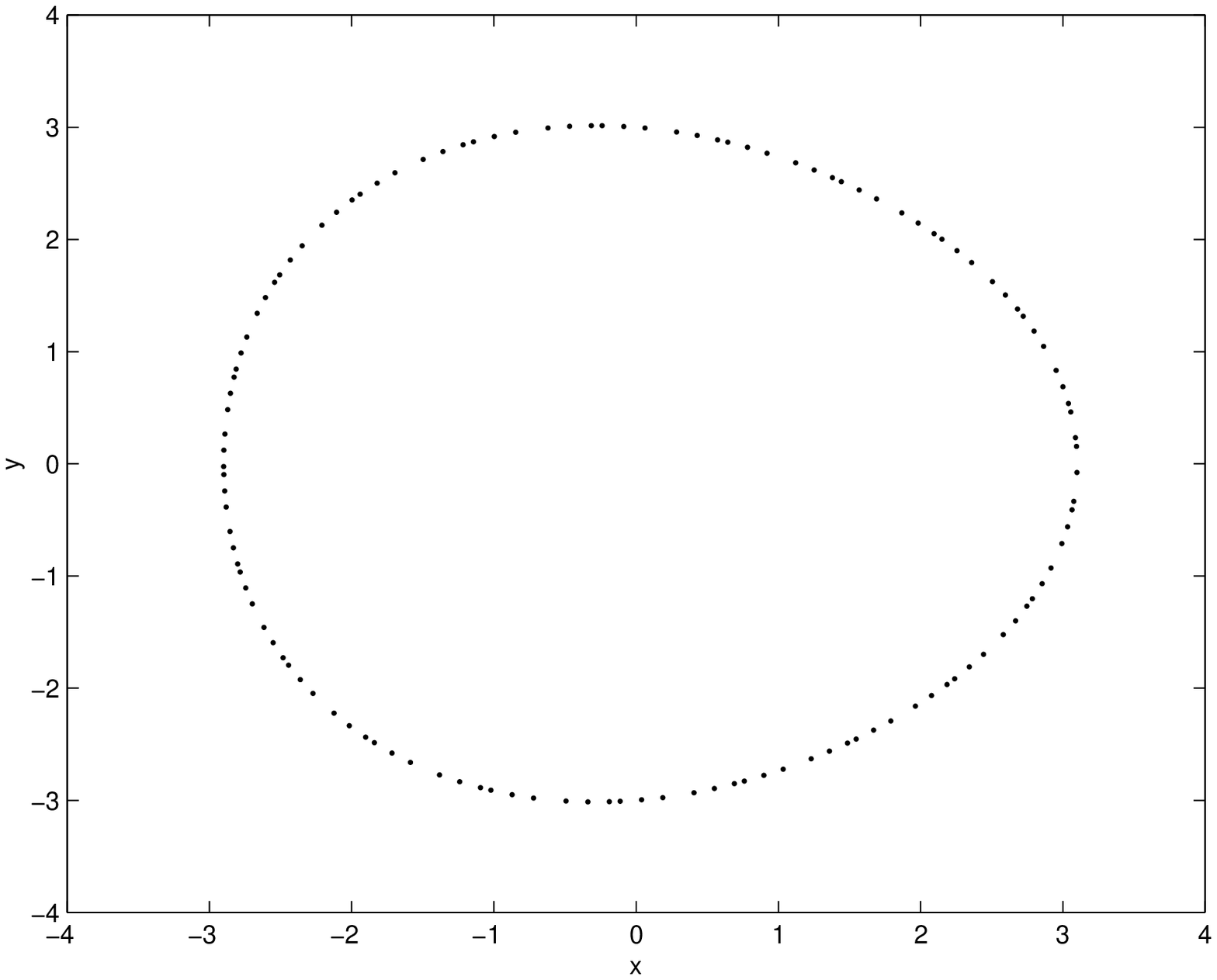}  }&

\epsfysize=5cm
\epsfxsize=5cm
{ \epsfbox{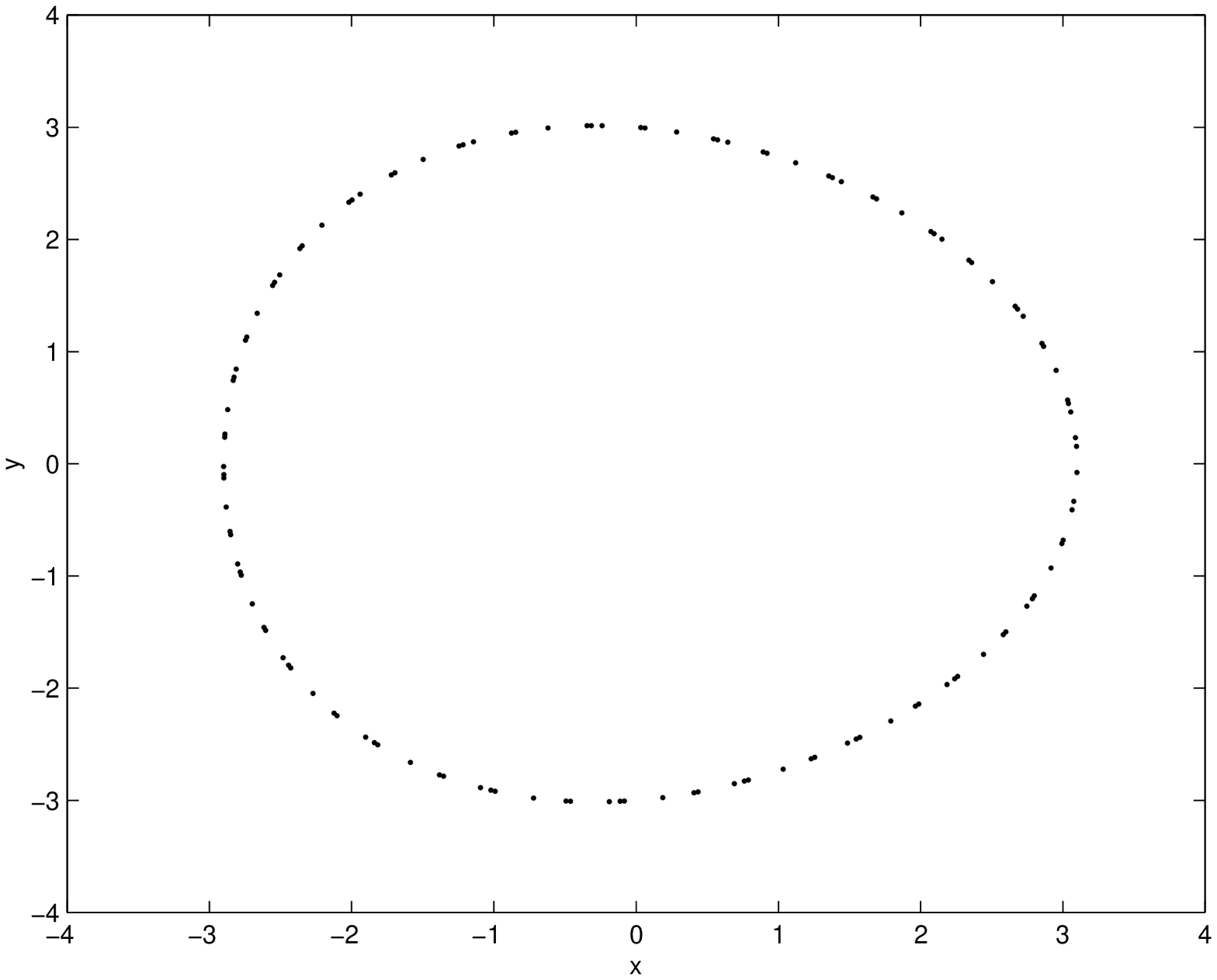}  }
\\

\text{\small a) Regular Partition.} & \text{\small b) Irregular Partition.} & \text{\small c) Very Irregular.}
\end{array}
$
\end{figure}

\newpage
\begin{figure}[here]
\caption{Approximations of the Affine Signature Curve Obtained with (\ref{initial_affine}).} 
\vspace{1cm} 
$
\begin{array}{lll}
\epsfysize=5cm
\epsfxsize=5cm
{ \epsfbox{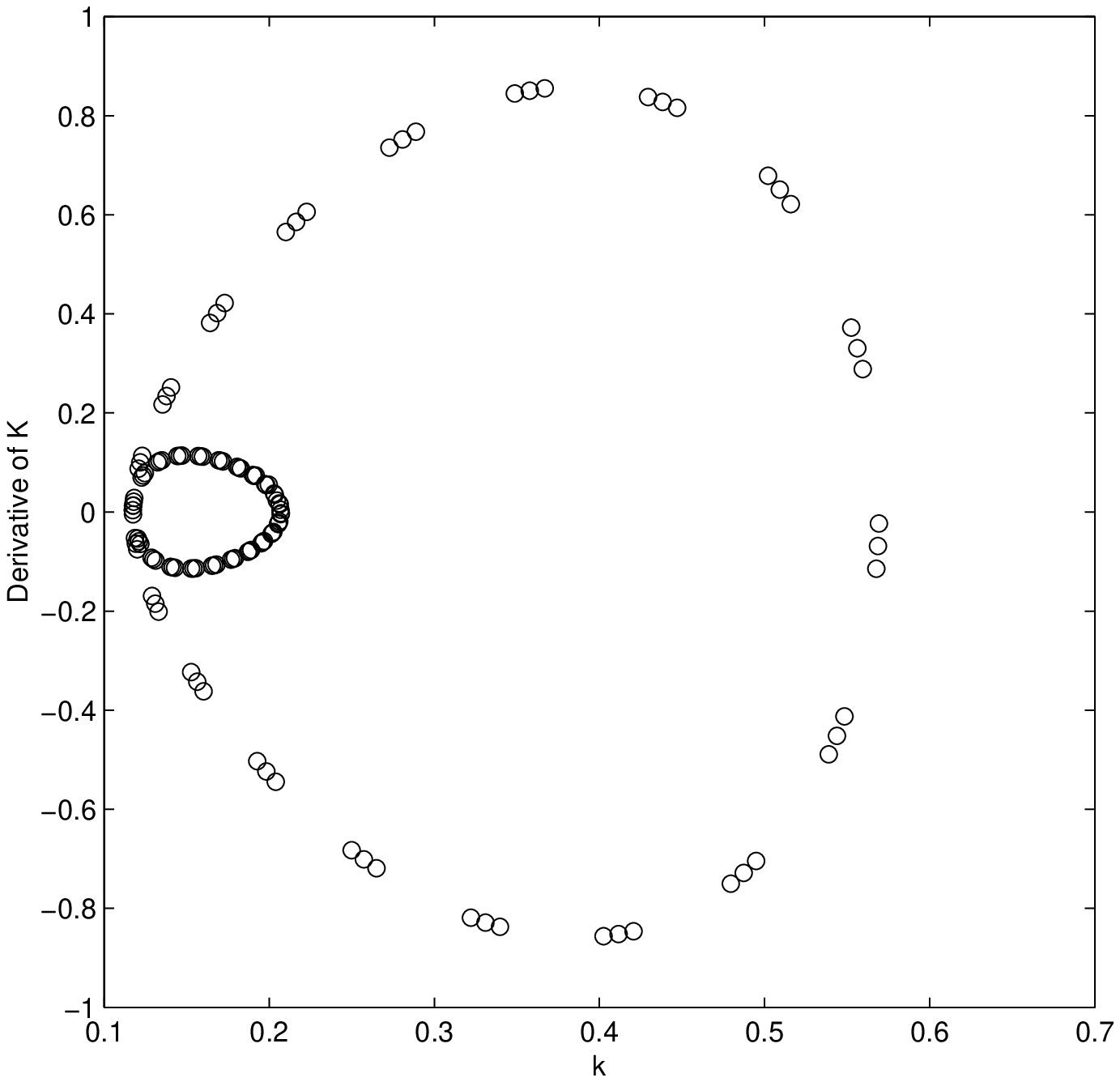}}&

\epsfysize=5cm
\epsfxsize=5cm
{ \epsfbox{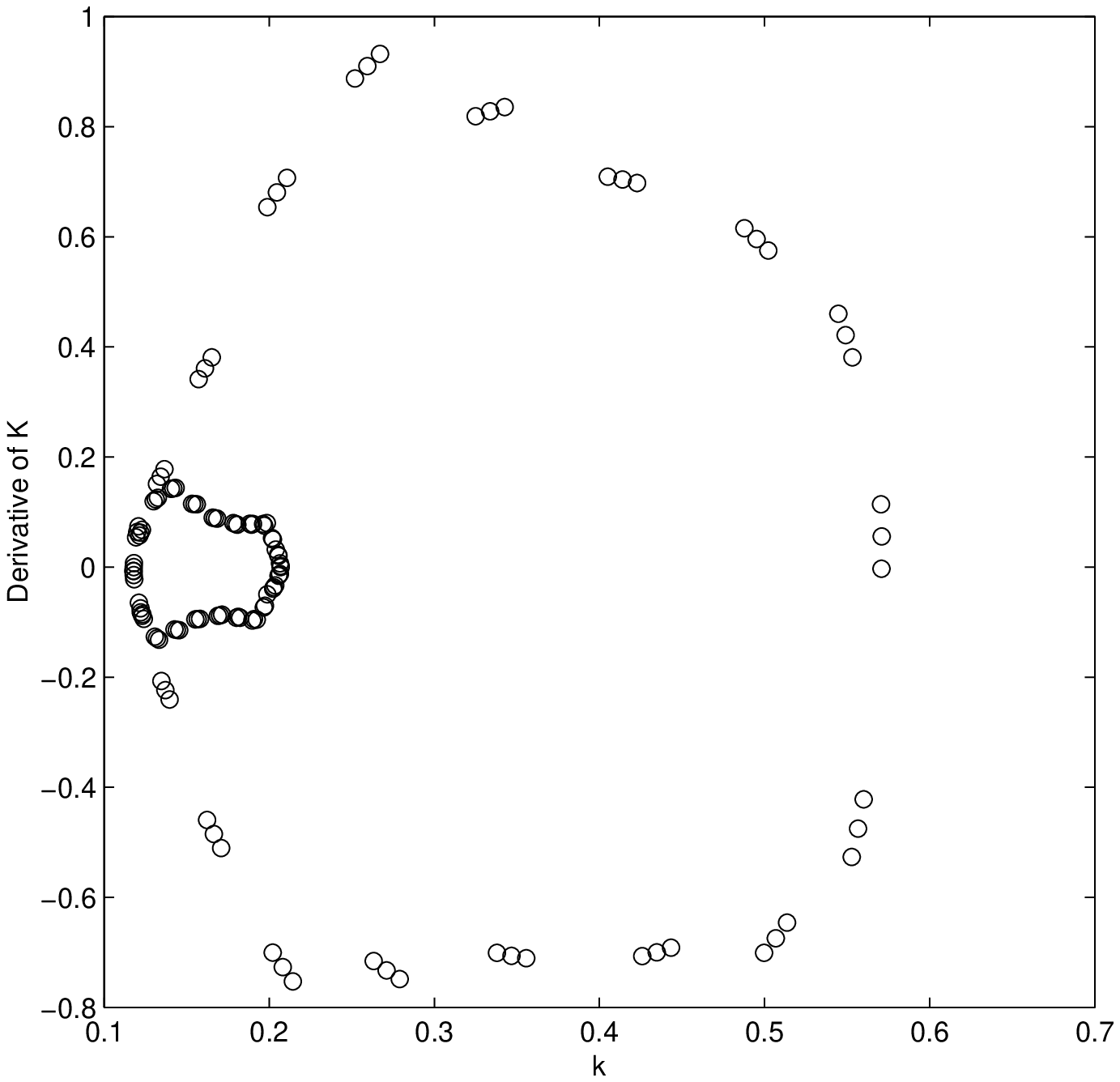}  }&

\epsfysize=5cm
\epsfxsize=5cm
{ \epsfbox{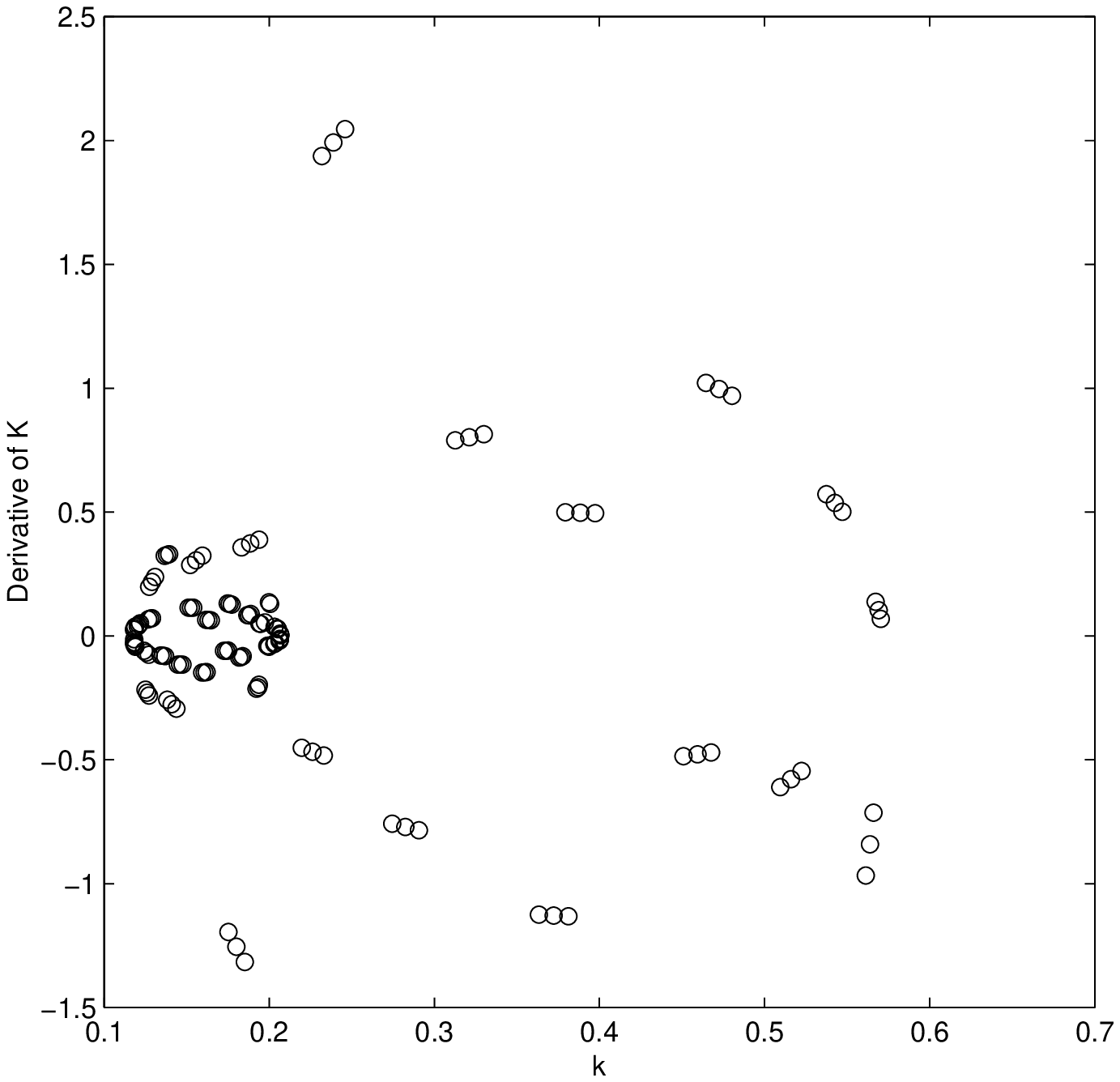}}
\\
\text{\small With partition 10-a.} &\text{\small With partition 10-b.} & \text{\small With partition 10-c.}
\end{array}
$
\end{figure}

\vspace{0.5cm}

\begin{figure}[here]
\caption{Approximations of the Affine Signature Curve Obtained with (\ref{affine1}).}
\vspace{1cm}
$ 
\begin{array}{lll}
\epsfysize=5cm
\epsfxsize=5cm
{ \epsfbox{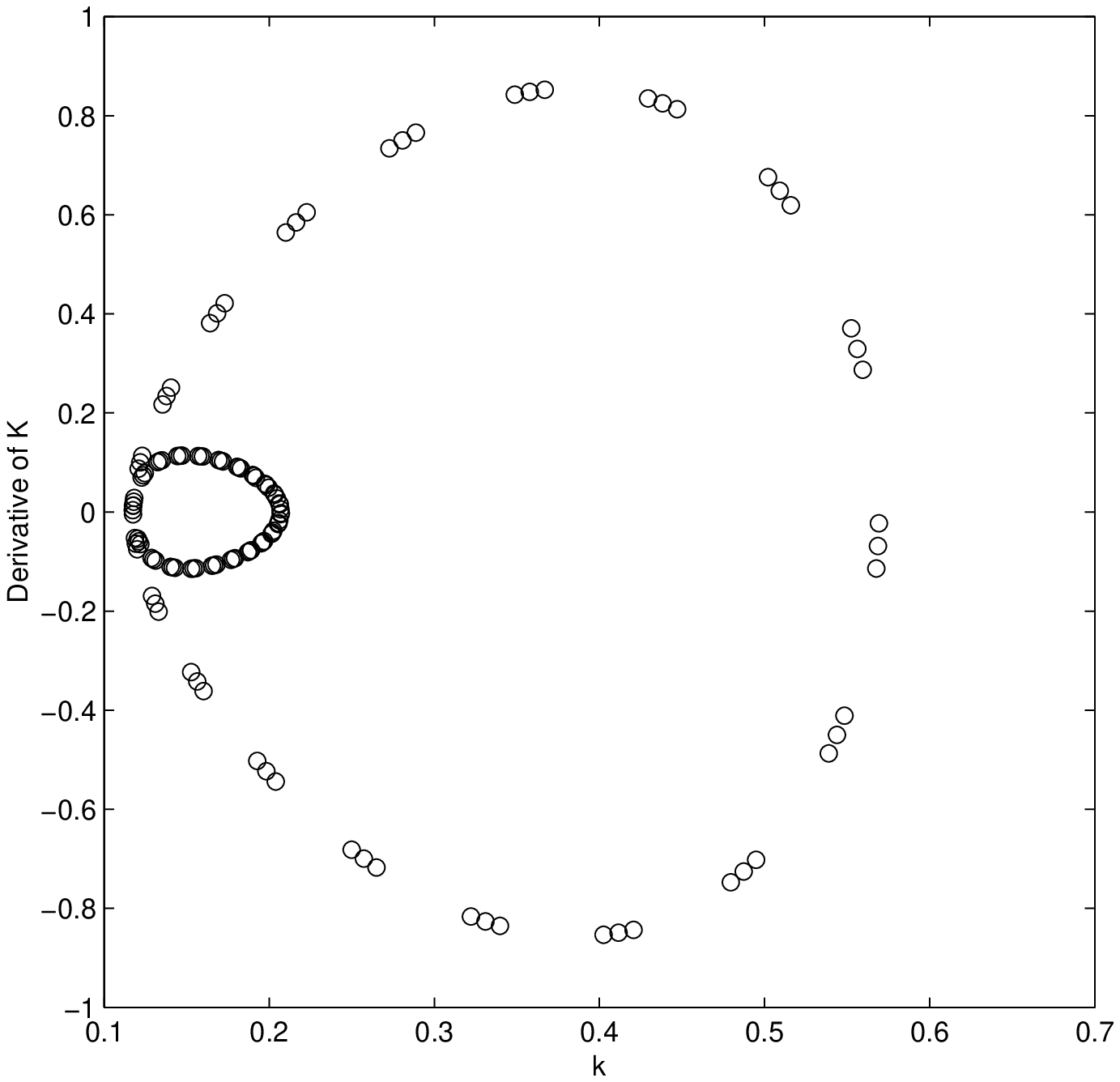}}&

\epsfysize=5cm
\epsfxsize=5cm
{ \epsfbox{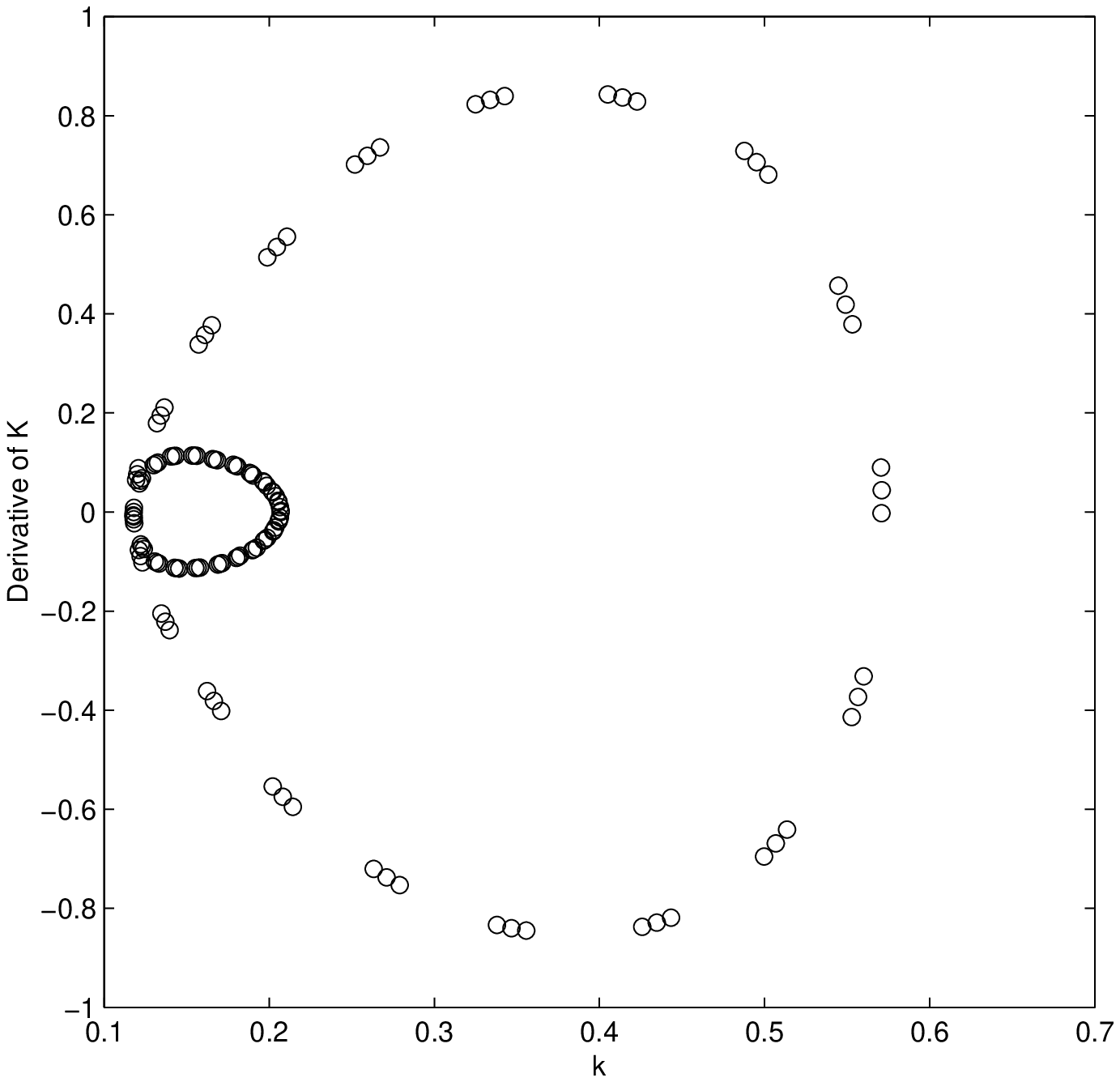}  }&

\epsfysize=5cm
\epsfxsize=5cm
{ \epsfbox{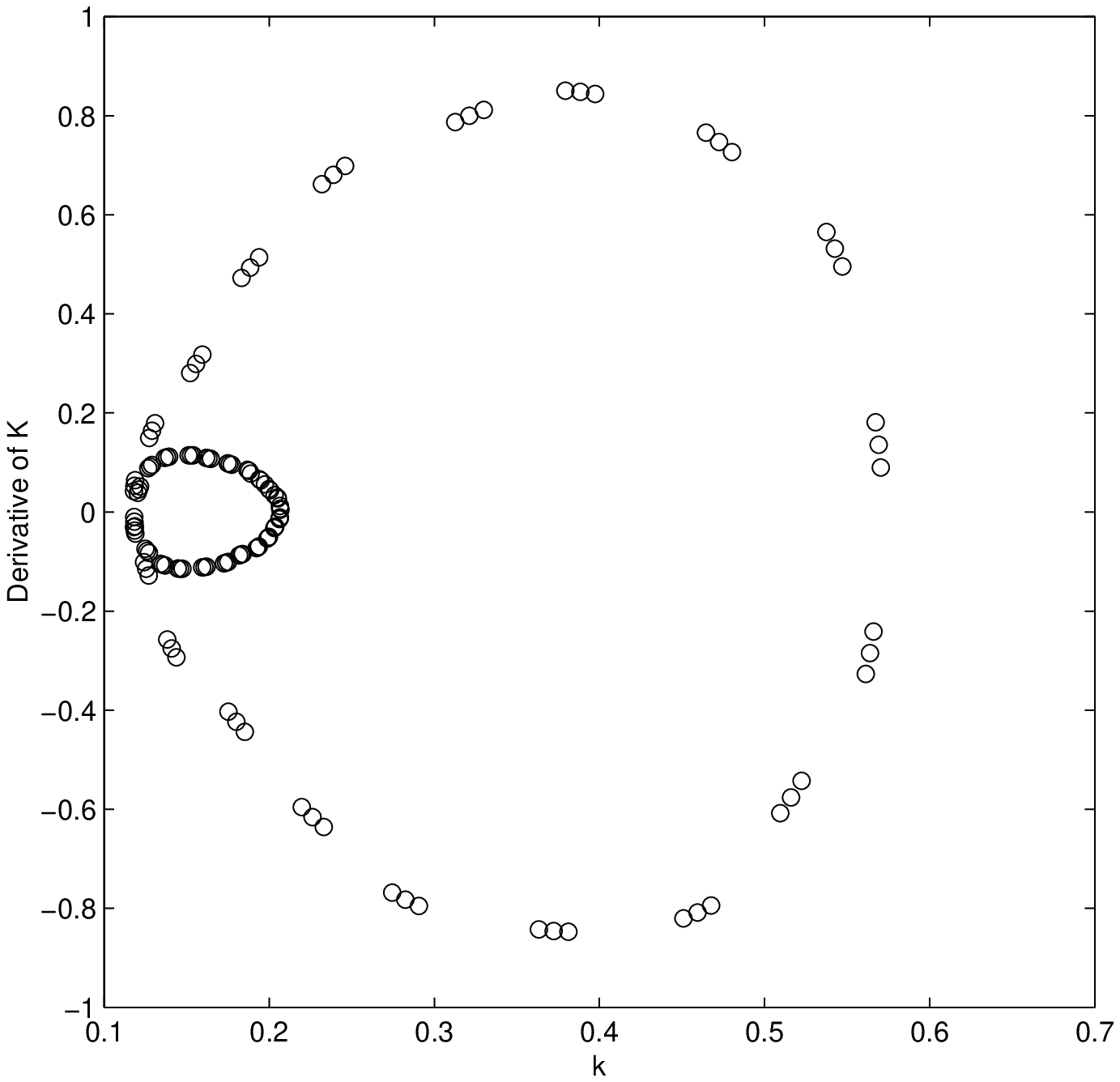}}
\\
\text{\small With Partition 10-a.} &\text{\small With Partition 10-b.} & \text{\small With Partition 10-c.}
\end{array}
$
\end{figure}
\vspace{0.5cm}

\newpage

\begin{figure}[here]
\caption{Curve Used to Test the Formulas Proposed for the Spatial Signature}
\vspace{0.5cm}
\centerline{
\hbox{
\epsfysize=7.0cm
\epsfxsize=7.0cm
{\leavevmode \epsfbox{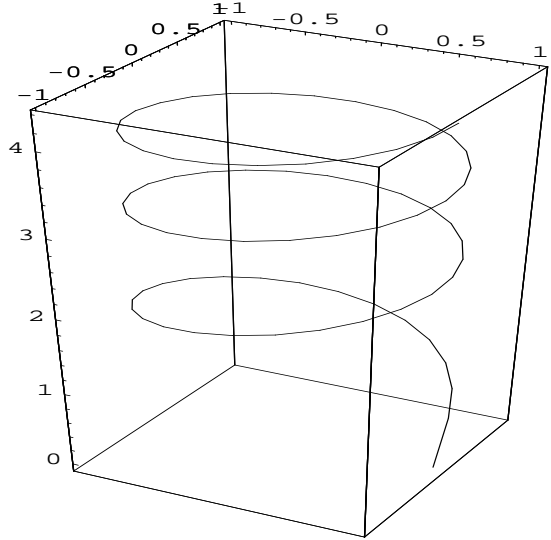}}
}}
\centerline{\text{The Curve $\alpha(t)=(\cos t,\sin t,\sqrt{t})$, for $0\leq t\leq 6\pi$ }}

\end{figure}

\vspace{0.5cm}

\begin{figure}[here]
\caption{Projections of the Exact Signature for the Curve Above}
\vspace{1cm} 
$
\begin{array}{lll}
\epsfysize=5cm
\epsfxsize=5cm
{ \epsfbox{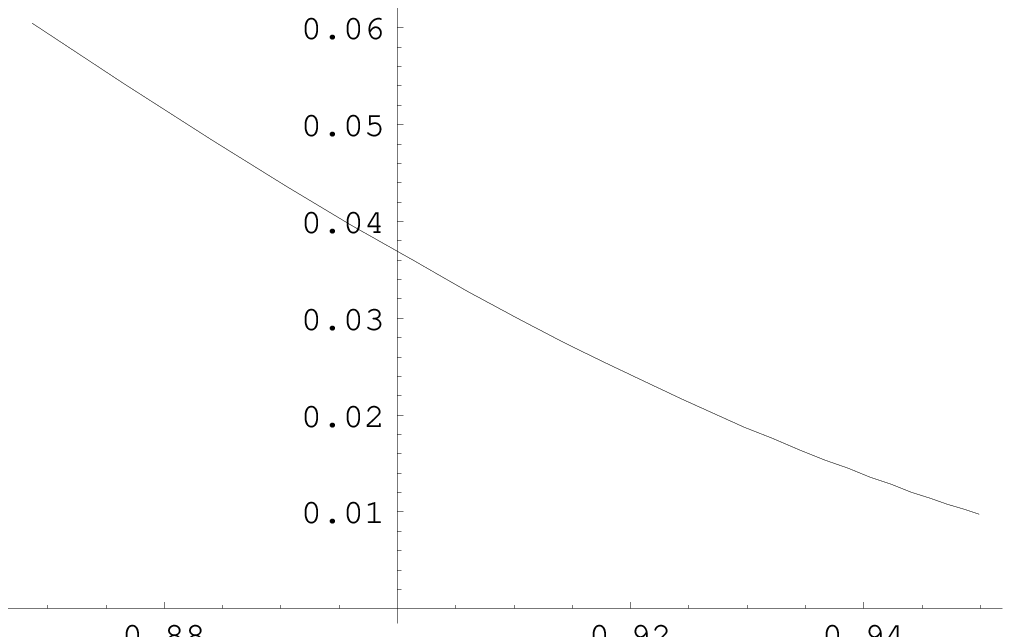}}&\hspace{1.5cm} &

\epsfysize=5cm
\epsfxsize=5cm
{ \epsfbox{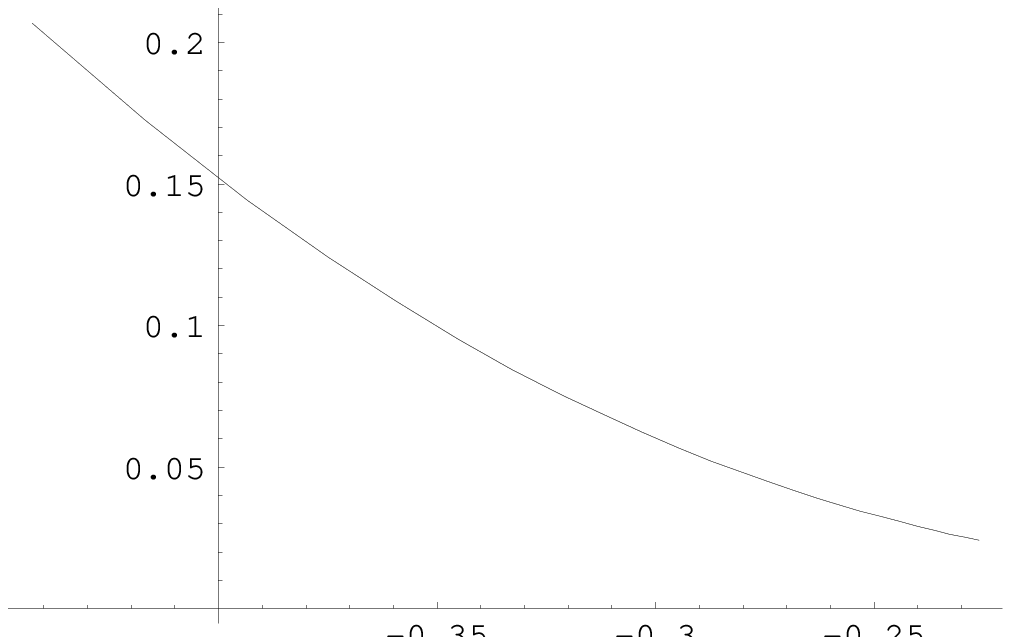}}
\\
\vspace{2cm}
\text{\small Derivative of the Curvature vs Curvature} &\hspace{1.5cm} &\text{\small Derivative of the Torsion vs Torsion }
\end{array}
$
\end{figure}
\vspace{0.5cm}

\newpage

\begin{figure}[here]
\caption{Approximations Obtained with $\Delta t=0.1$}
\vspace{1cm}
$ 
\begin{array}{lll}
\epsfysize=5cm
\epsfxsize=5cm
{ \epsfbox{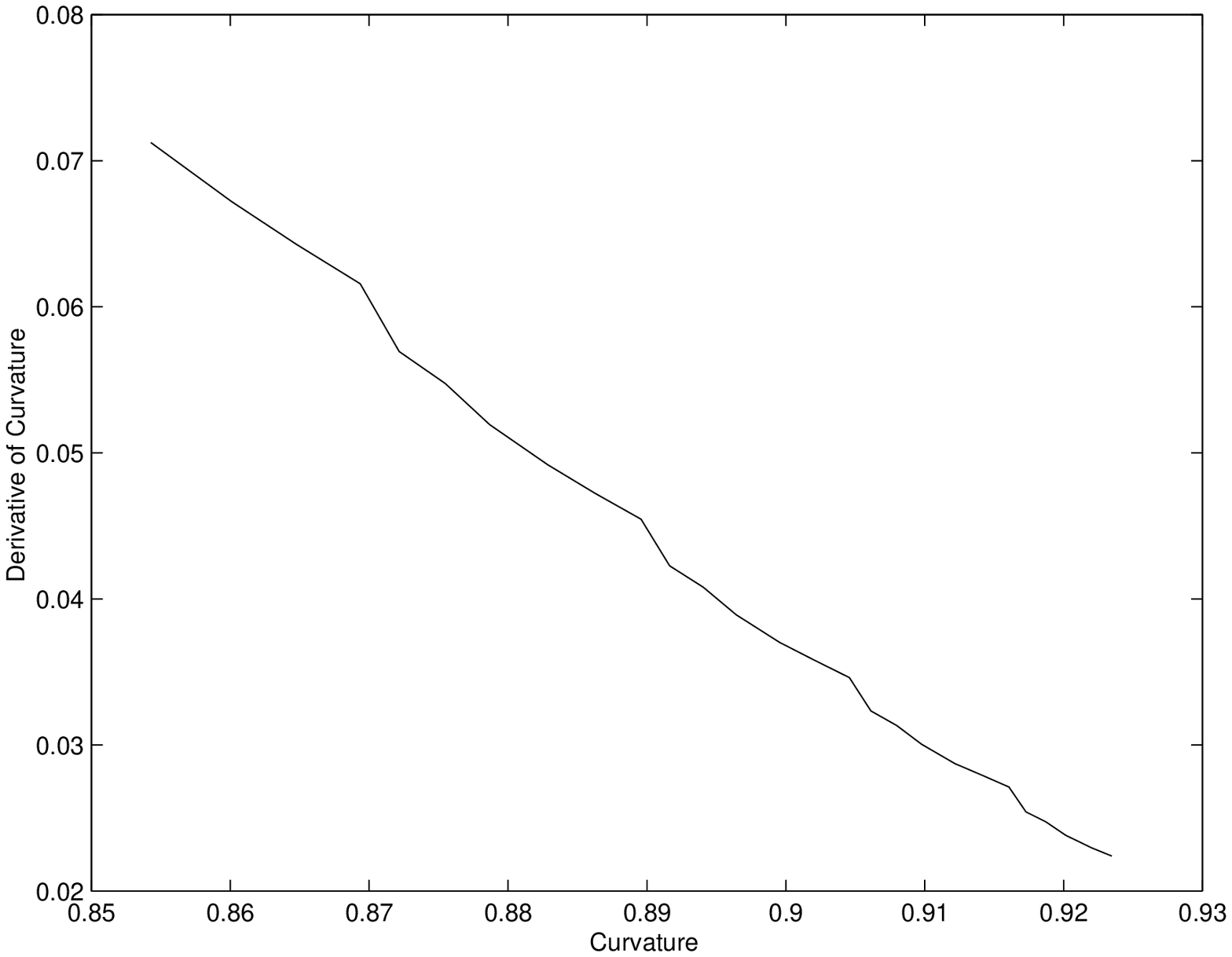}}&

\epsfysize=5cm
\epsfxsize=5cm
{ \epsfbox{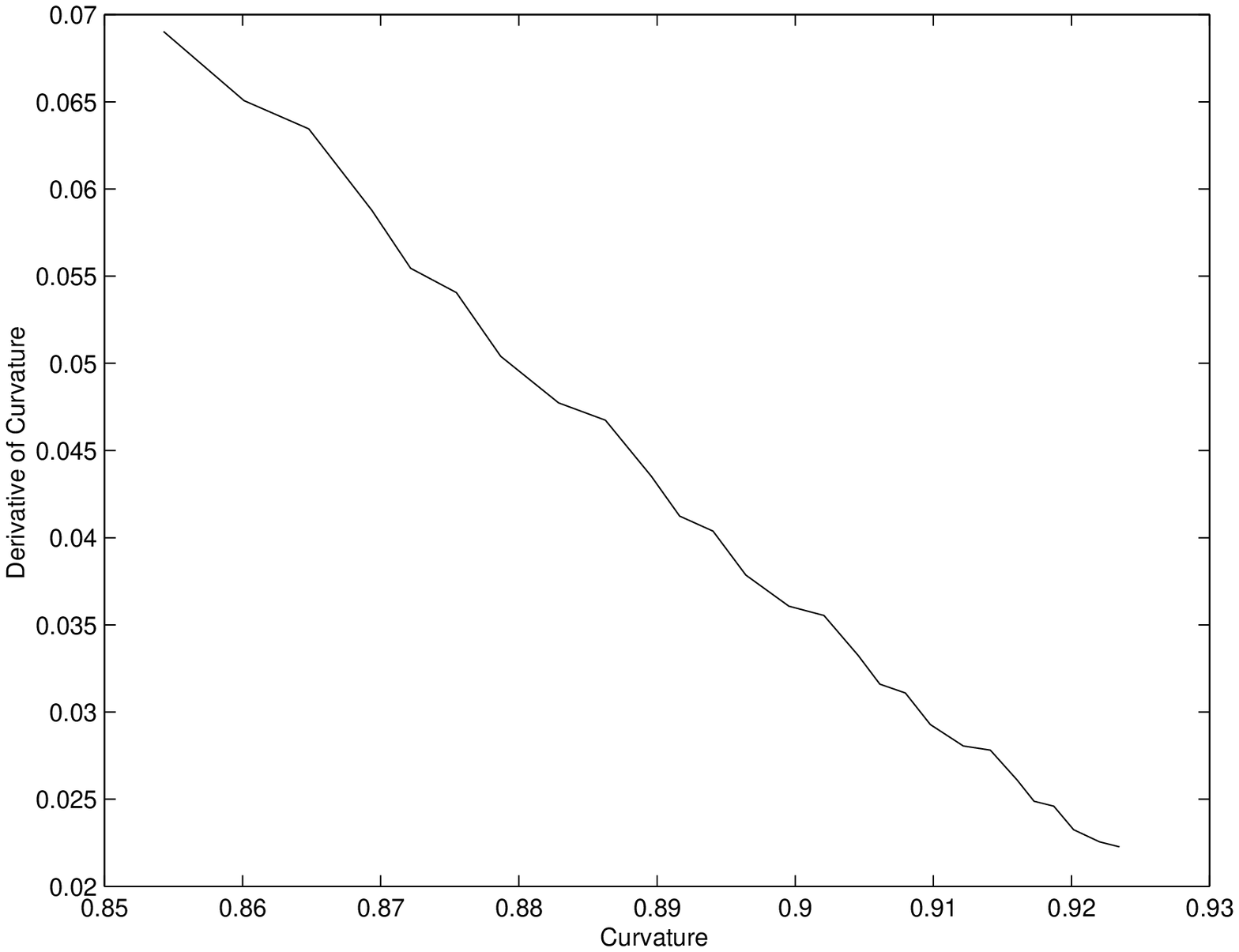}  }&

\epsfysize=5cm
\epsfxsize=5cm
{ \epsfbox{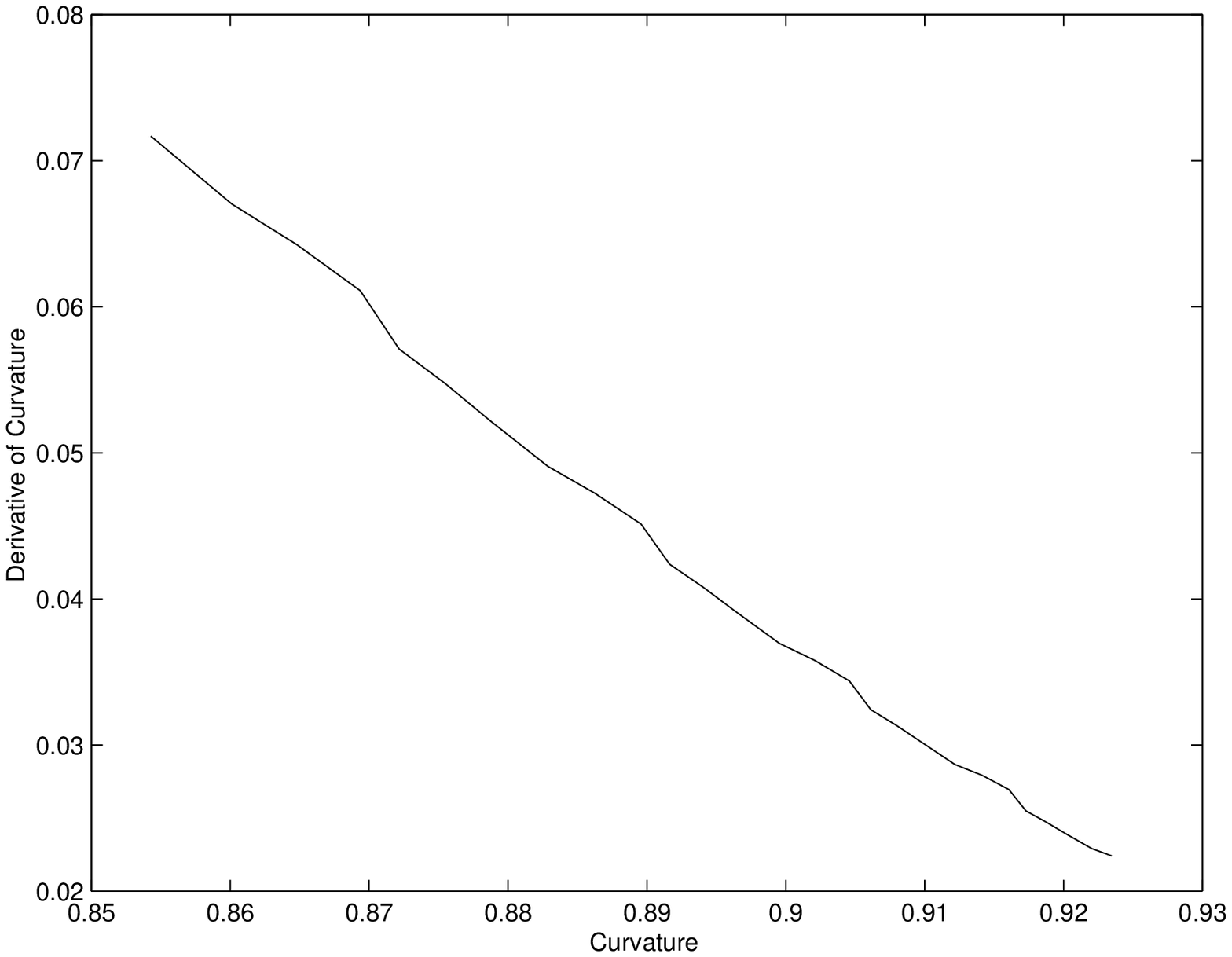}}
\\
\text{\small Using $\tilde{\kappa}$ and $\tilde{\kappa}_{s,5}$} &\text{\small Using $\tilde{\kappa}$ and $\tilde{\kappa}_{s,4}$ } & \text{\small Using $\tilde{\kappa}$ and $\tilde{\kappa}_{s,3}$}
\end{array}
$
\end{figure}
\vspace{0.5cm}

\begin{figure}[here]
\caption{Approximations Obtained with $\Delta t=0.05$}
\vspace{1cm}
$ 
\begin{array}{lll}
\epsfysize=5cm
\epsfxsize=5cm
{ \epsfbox{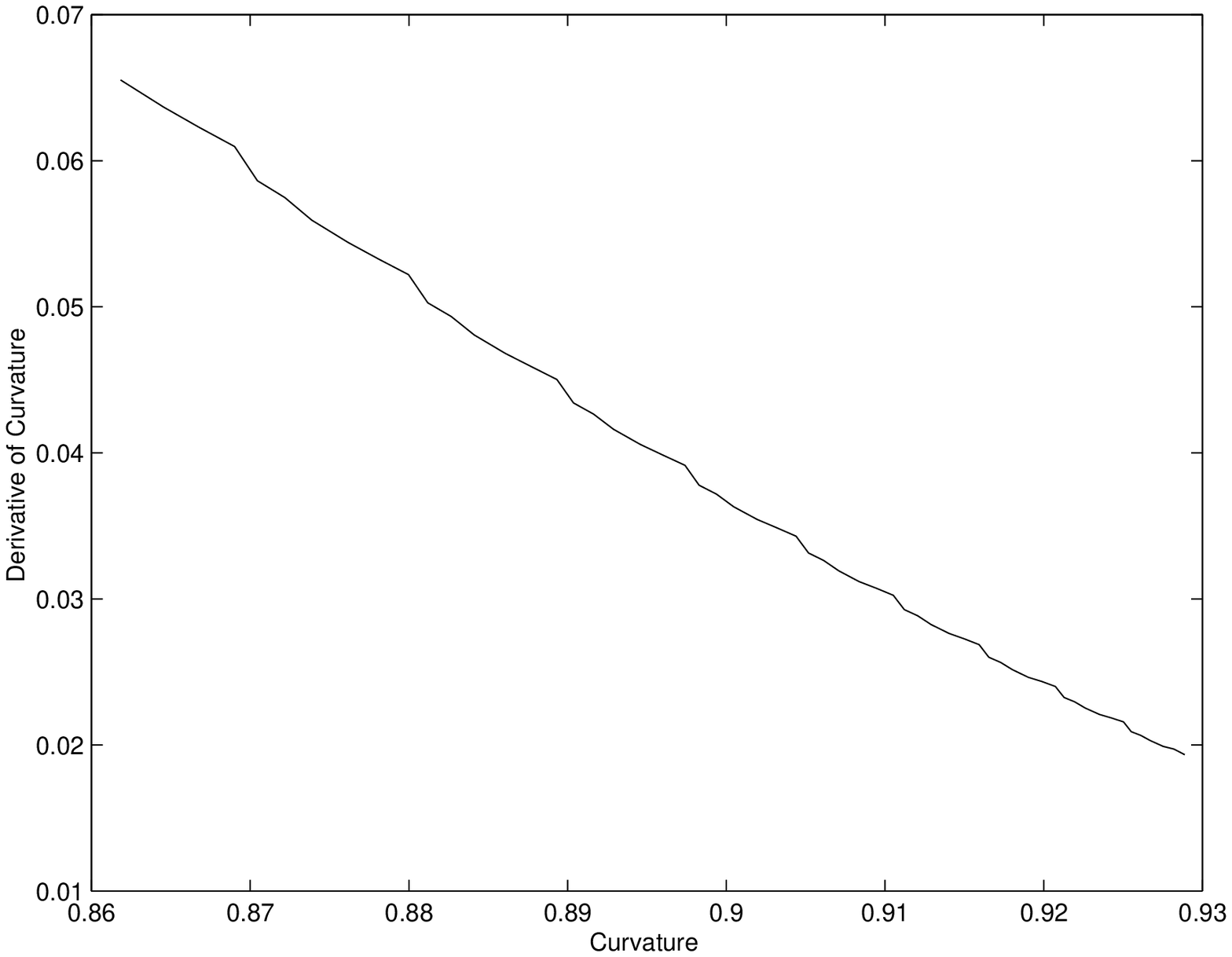}}&

\epsfysize=5cm
\epsfxsize=5cm
{ \epsfbox{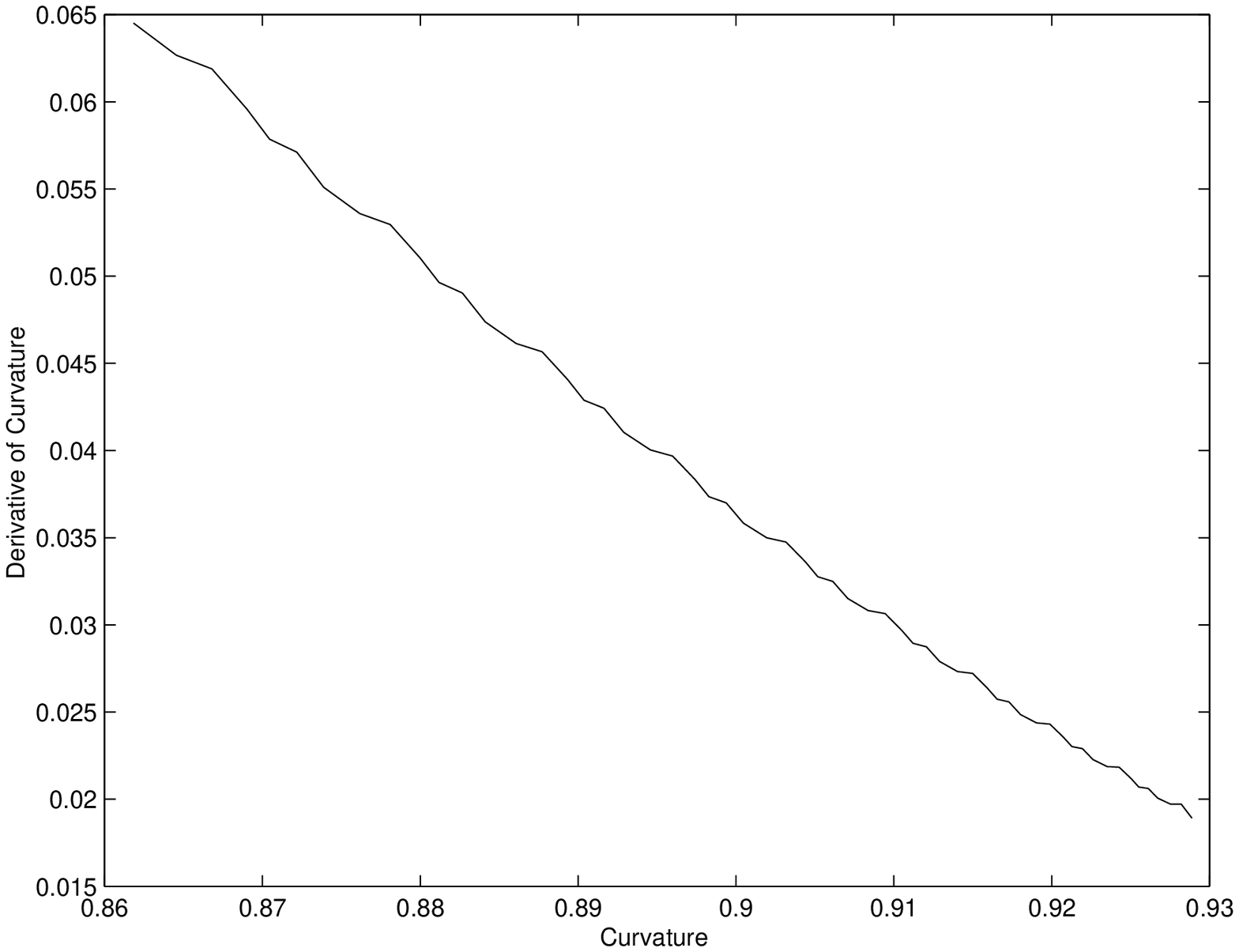}  }&

\epsfysize=5cm
\epsfxsize=5cm
{ \epsfbox{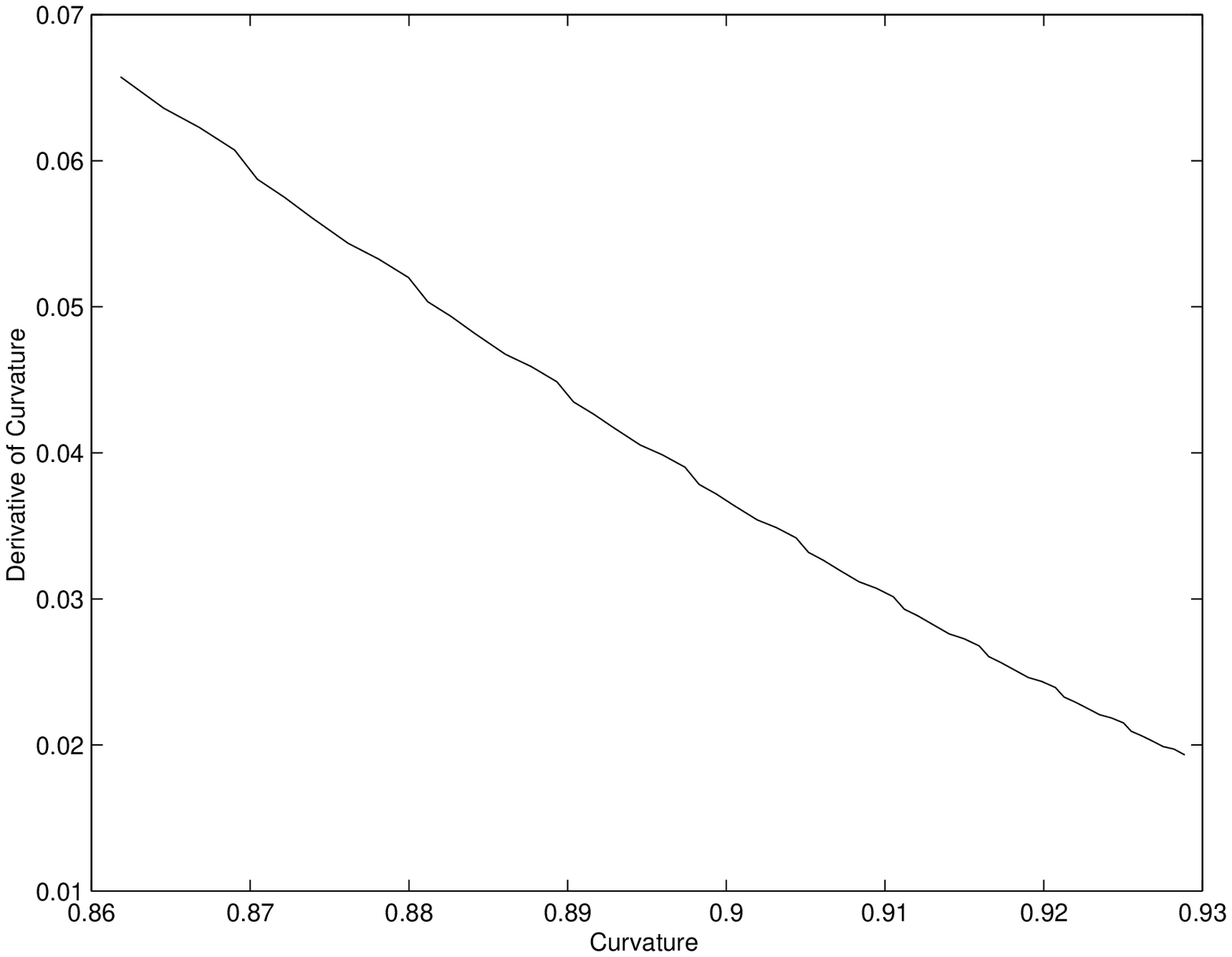}}
\\
\text{\small Using $\tilde{\kappa}$ and $\tilde{\kappa}_{s,5}$} &\text{\small Using $\tilde{\kappa}$ and $\tilde{\kappa}_{s,4}$ } & \text{\small Using $\tilde{\kappa}$ and $\tilde{\kappa}_{s,3}$}
\end{array}
$
\end{figure}
\vspace{0.5cm}

\newpage
\begin{figure}[here]
\caption{Approximations Obtained with $\Delta t=0.025$}
\vspace{1cm} 
$
\begin{array}{lll}
\epsfysize=5cm
\epsfxsize=5cm
{ \epsfbox{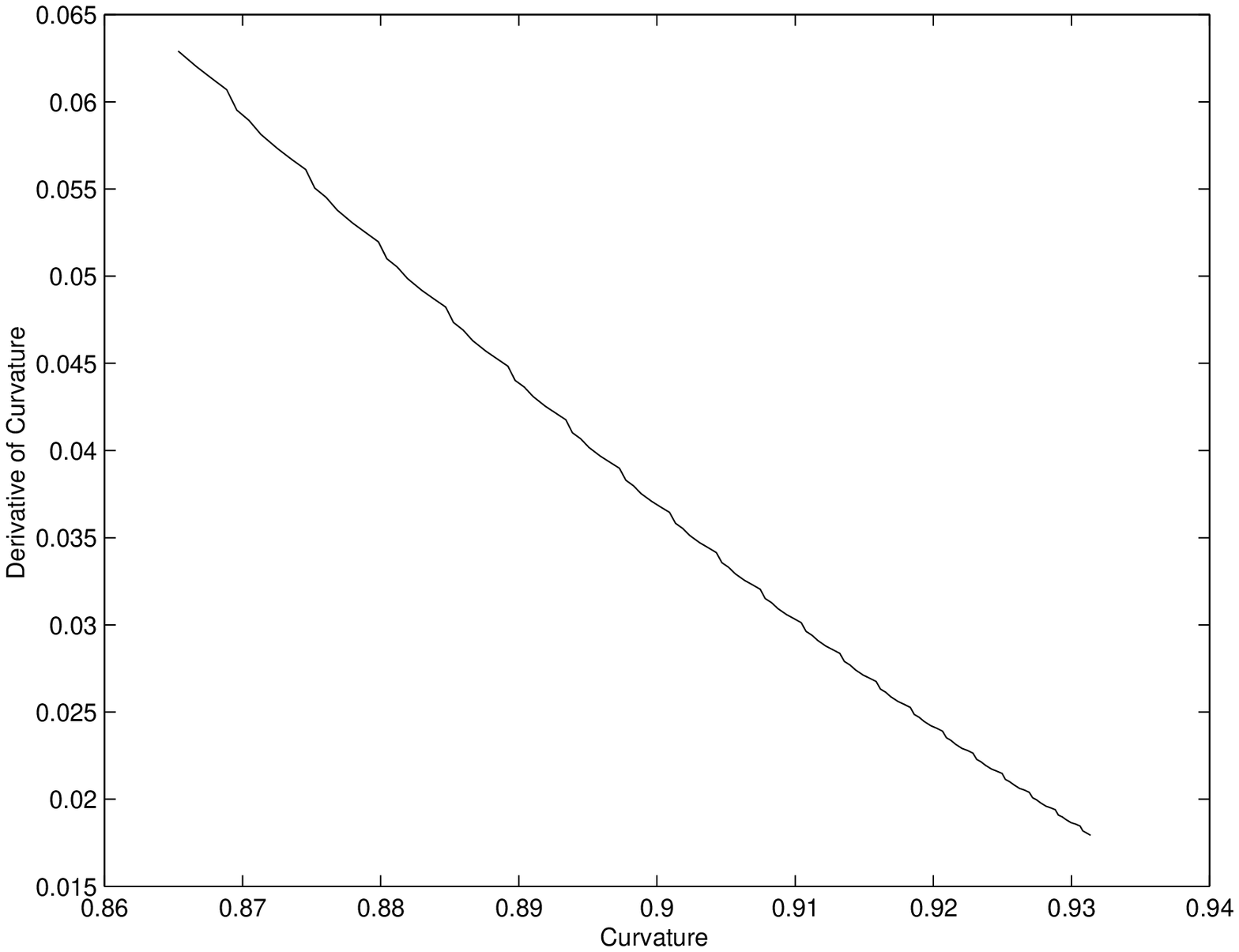}}&

\epsfysize=5cm
\epsfxsize=5cm
{ \epsfbox{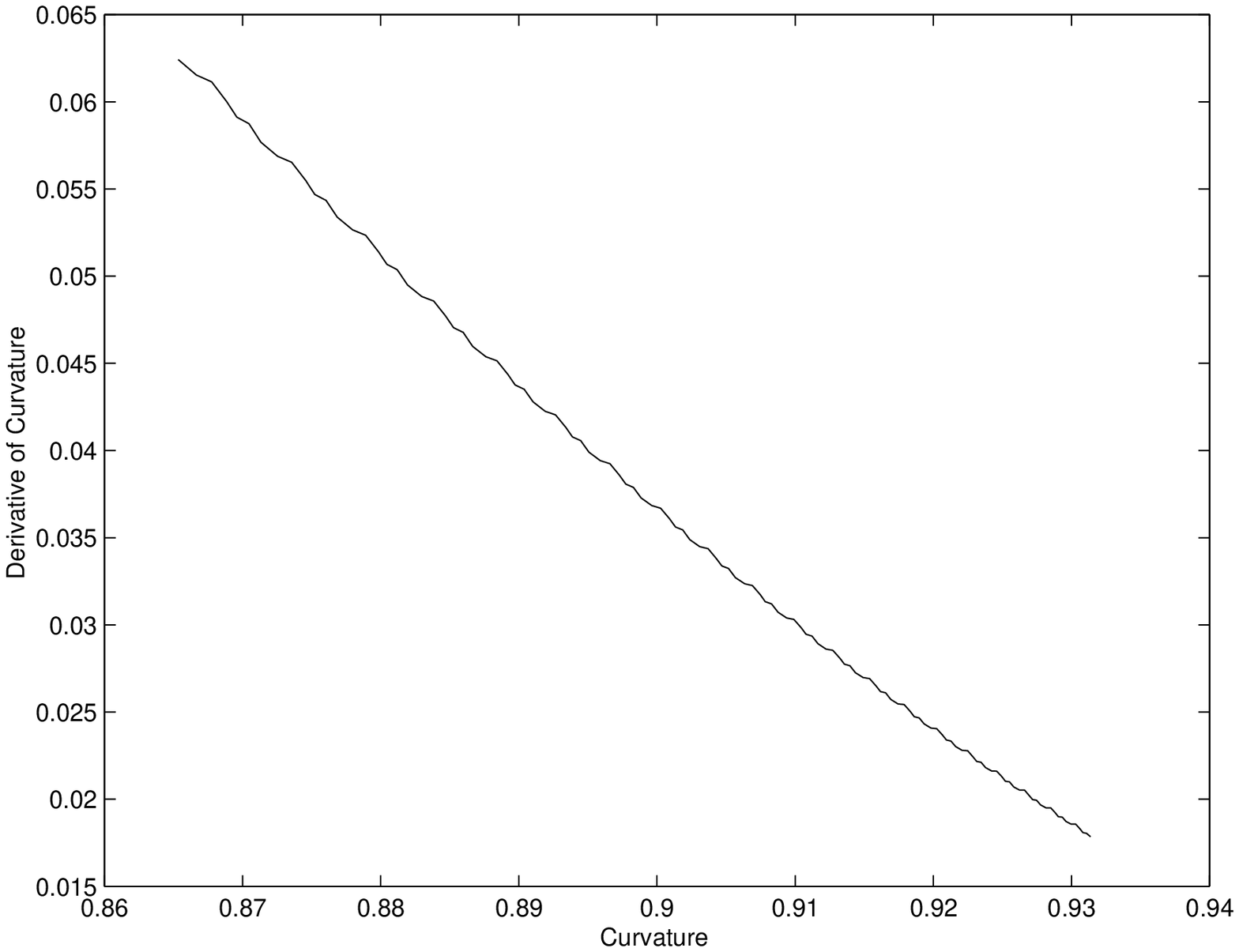}  }&

\epsfysize=5cm
\epsfxsize=5cm
{ \epsfbox{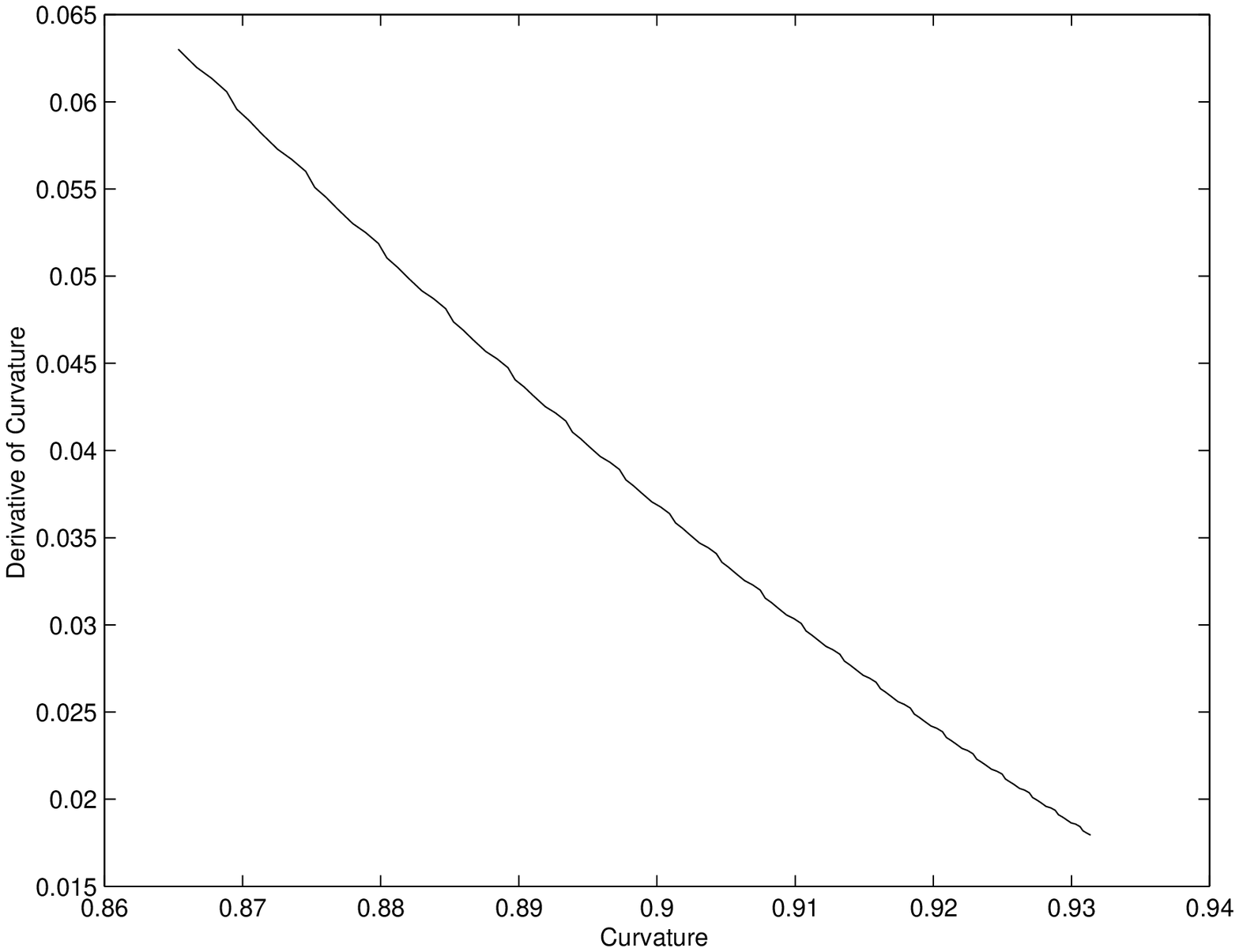}}
\\ \text{\small Using $\tilde{\kappa}$ and $\tilde{\kappa}_{s,5}$} &\text{\small Using $\tilde{\kappa}$ and $\tilde{\kappa}_{s,4}$ } & \text{\small Using $\tilde{\kappa}$ and $\tilde{\kappa}_{s,3}$}
\end{array}$
\end{figure}
\vspace{0.5cm}

\begin{figure}[here]
\caption{Approximations Obtained with $\Delta t=0.0125$}
\vspace{1cm} 
$\begin{array}{lll}
\epsfysize=5cm
\epsfxsize=5cm
{ \epsfbox{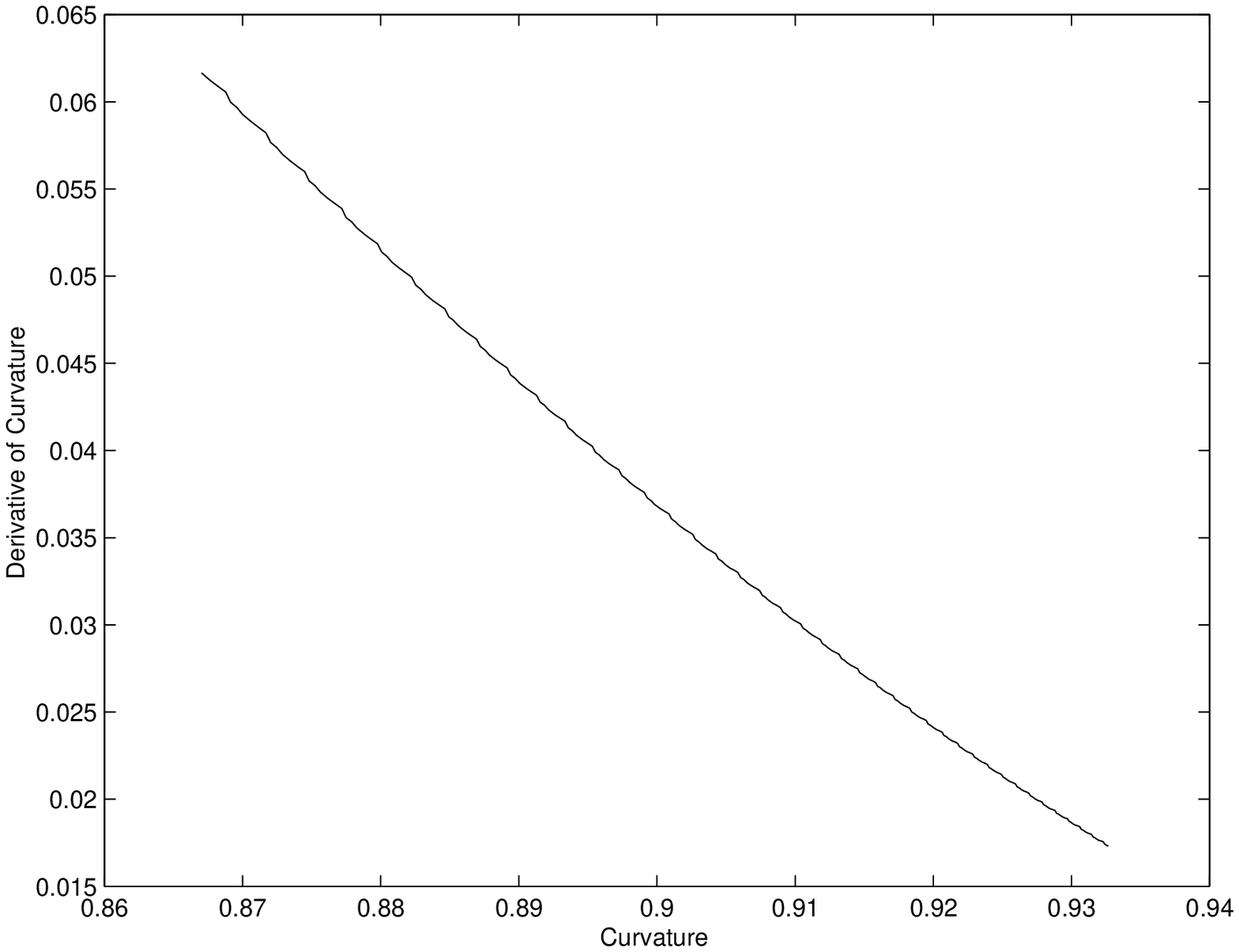}}&

\epsfysize=5cm
\epsfxsize=5cm
{ \epsfbox{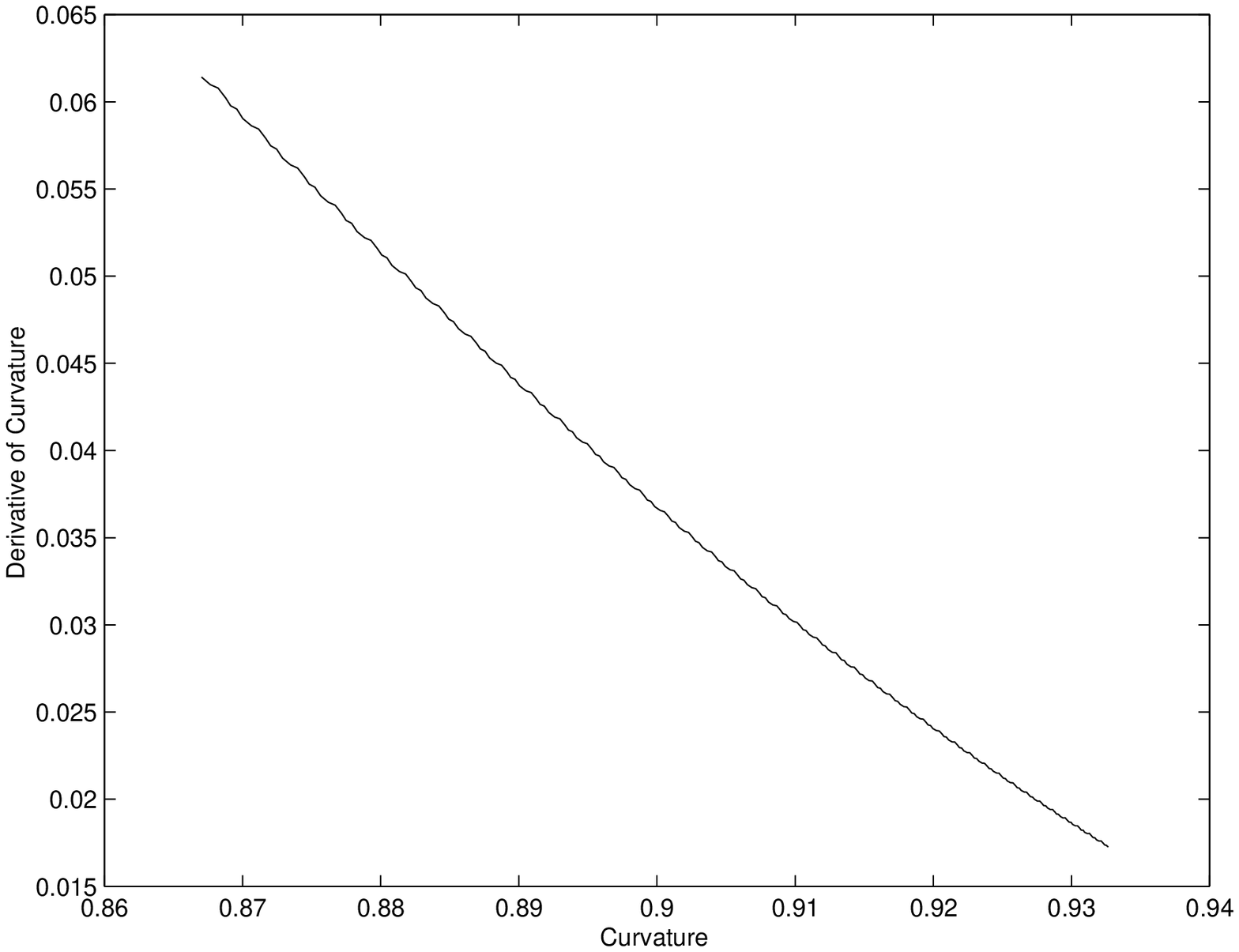}  }&

\epsfysize=5cm
\epsfxsize=5cm
{ \epsfbox{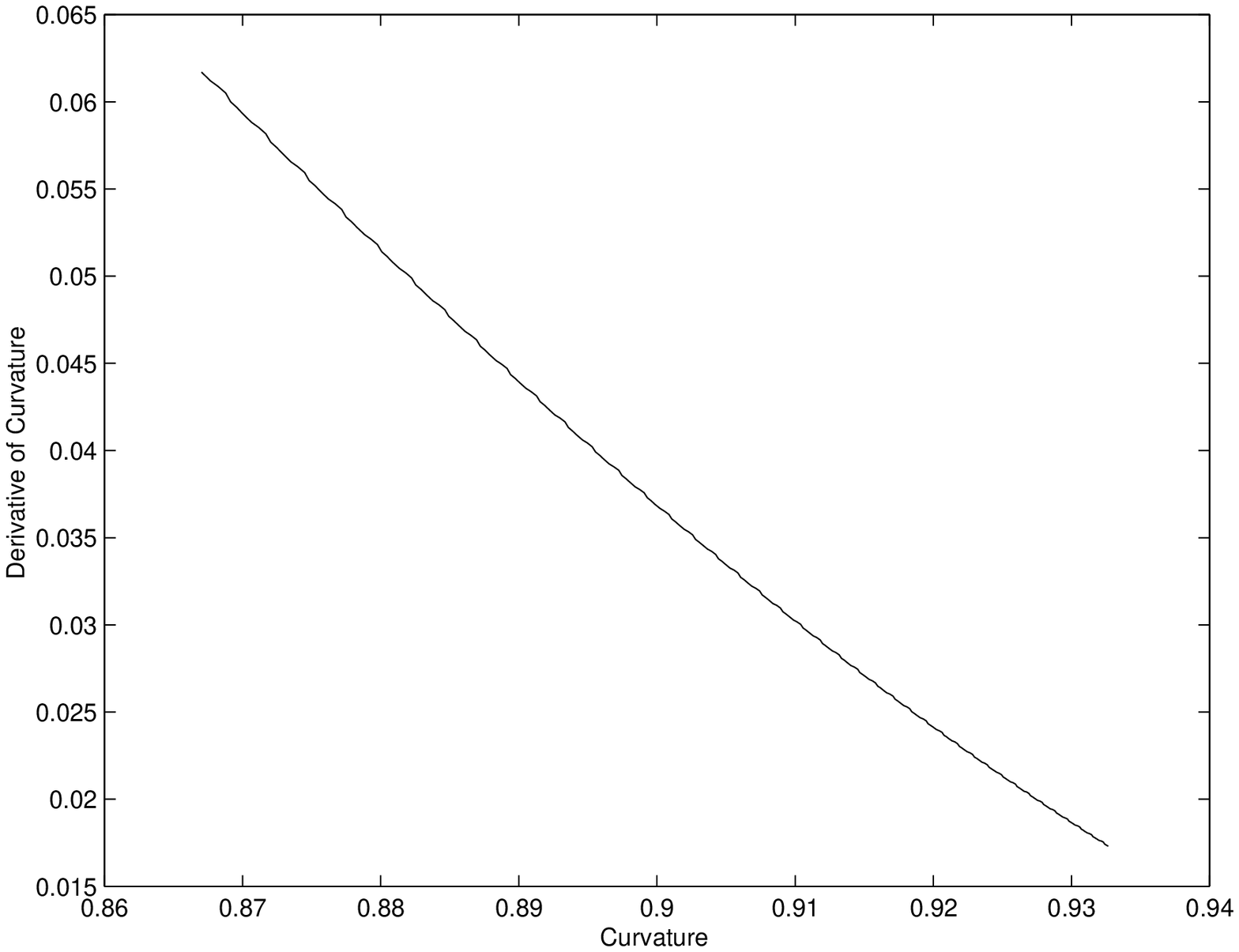}}
\\
\text{\small Using $\tilde{\kappa}$ and $\tilde{\kappa}_{s,5}$} &\text{\small Using $\tilde{\kappa}$ and $\tilde{\kappa}_{s,4}$ } & \text{\small Using $\tilde{\kappa}$ and $\tilde{\kappa}_{s,3}$}

\end{array}$
\end{figure}
\vspace{0.5cm}

\newpage

\begin{figure}[here]
\caption{Approximations Obtained with $\Delta t=0.1$}
\vspace{1cm} 
$\begin{array}{ll}
\epsfysize=5cm
\epsfxsize=5cm
{ \epsfbox{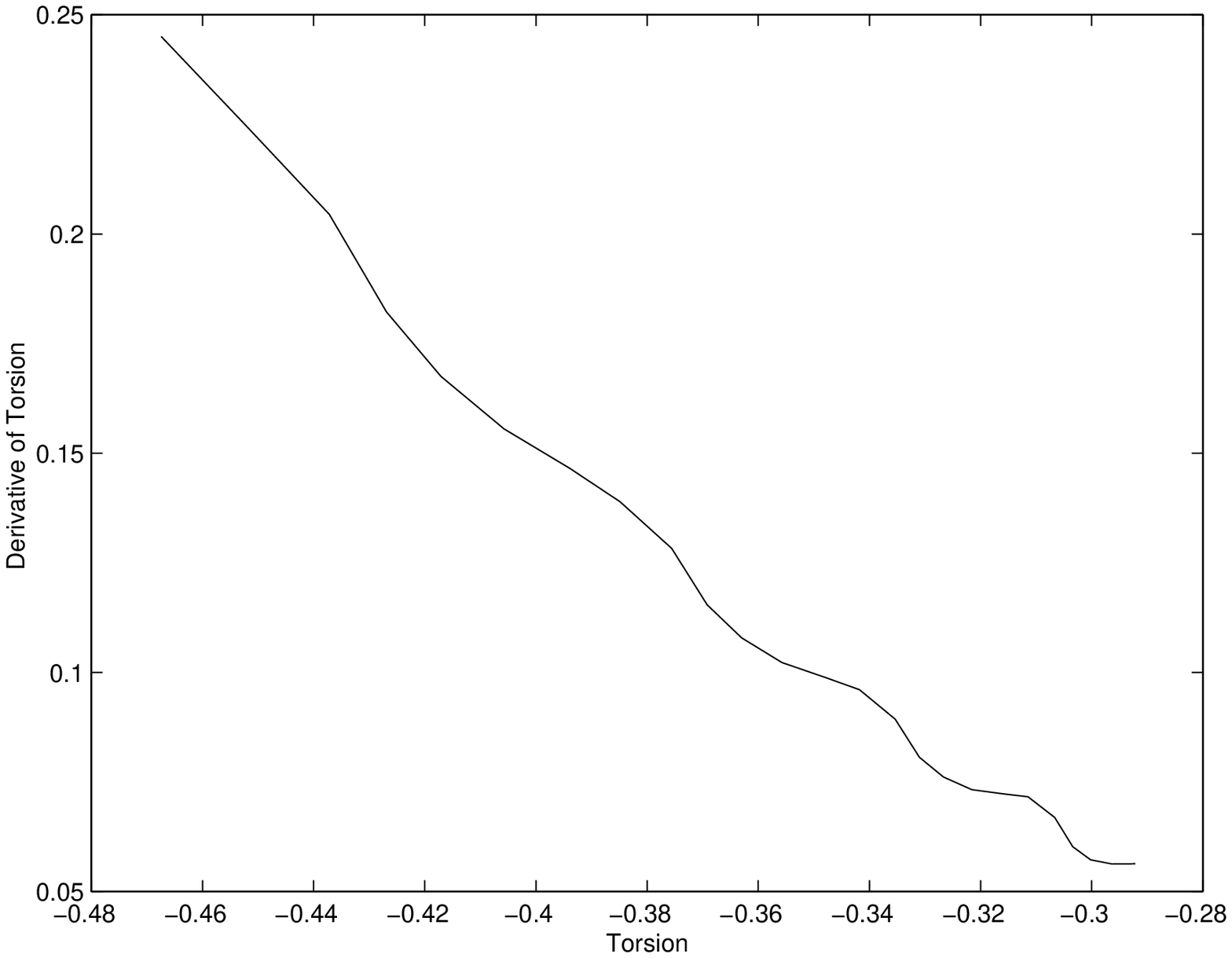}}&

\epsfysize=5cm
\epsfxsize=5cm
{ \epsfbox{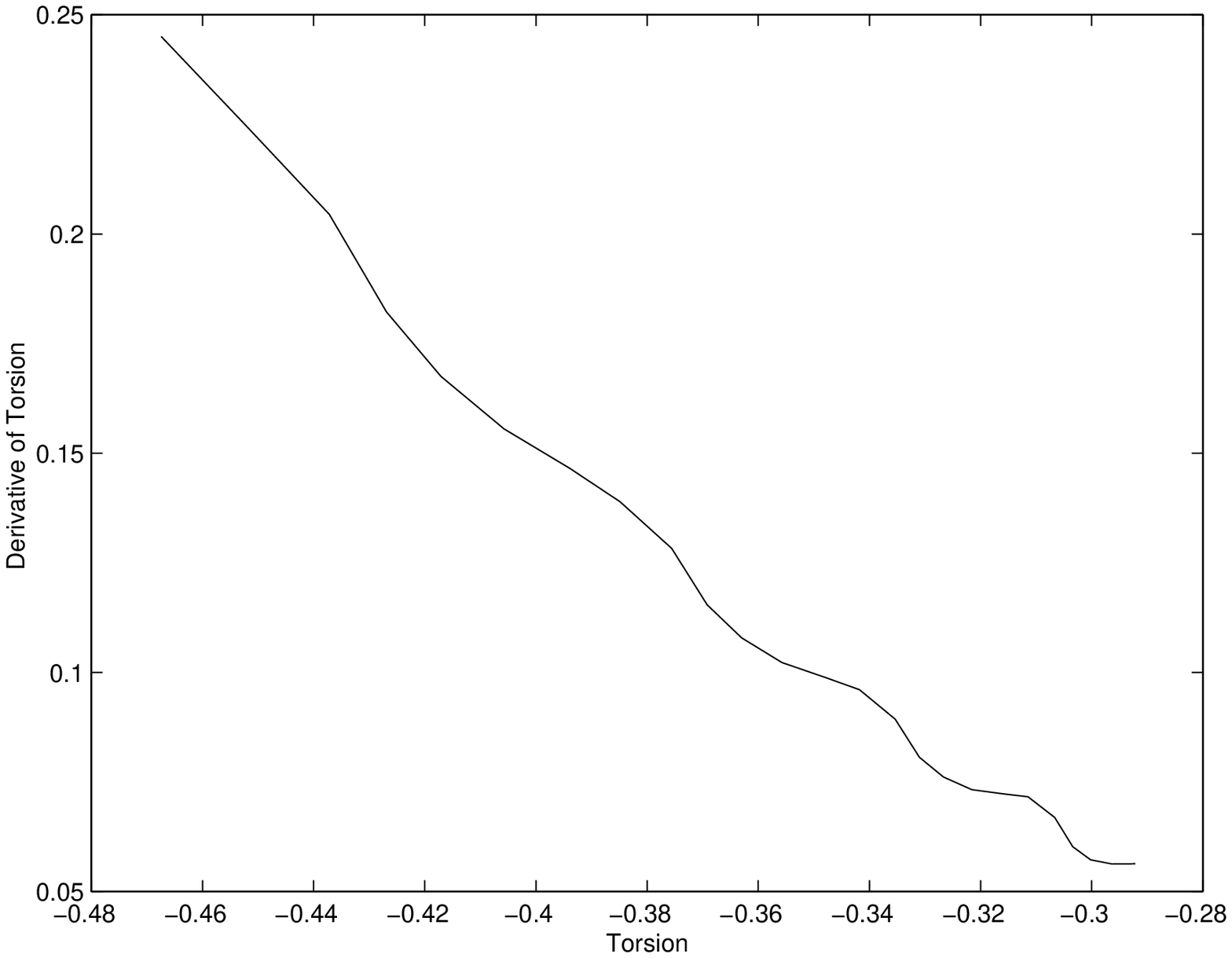}}
\\
\text{\small Using $\tilde{\tau}_1$ and $\tilde{\tau}_s$} &\text{\small Using $\tilde{\tau}_2$ and $\tilde{\tau}_s$} 
\end{array}$
\end{figure}
\vspace{0.5cm}

\begin{figure}[here]
\caption{Approximations Obtained with $\Delta t=0.05$}
\vspace{1cm} 
$\begin{array}{ll}
\epsfysize=5cm
\epsfxsize=5cm
{ \epsfbox{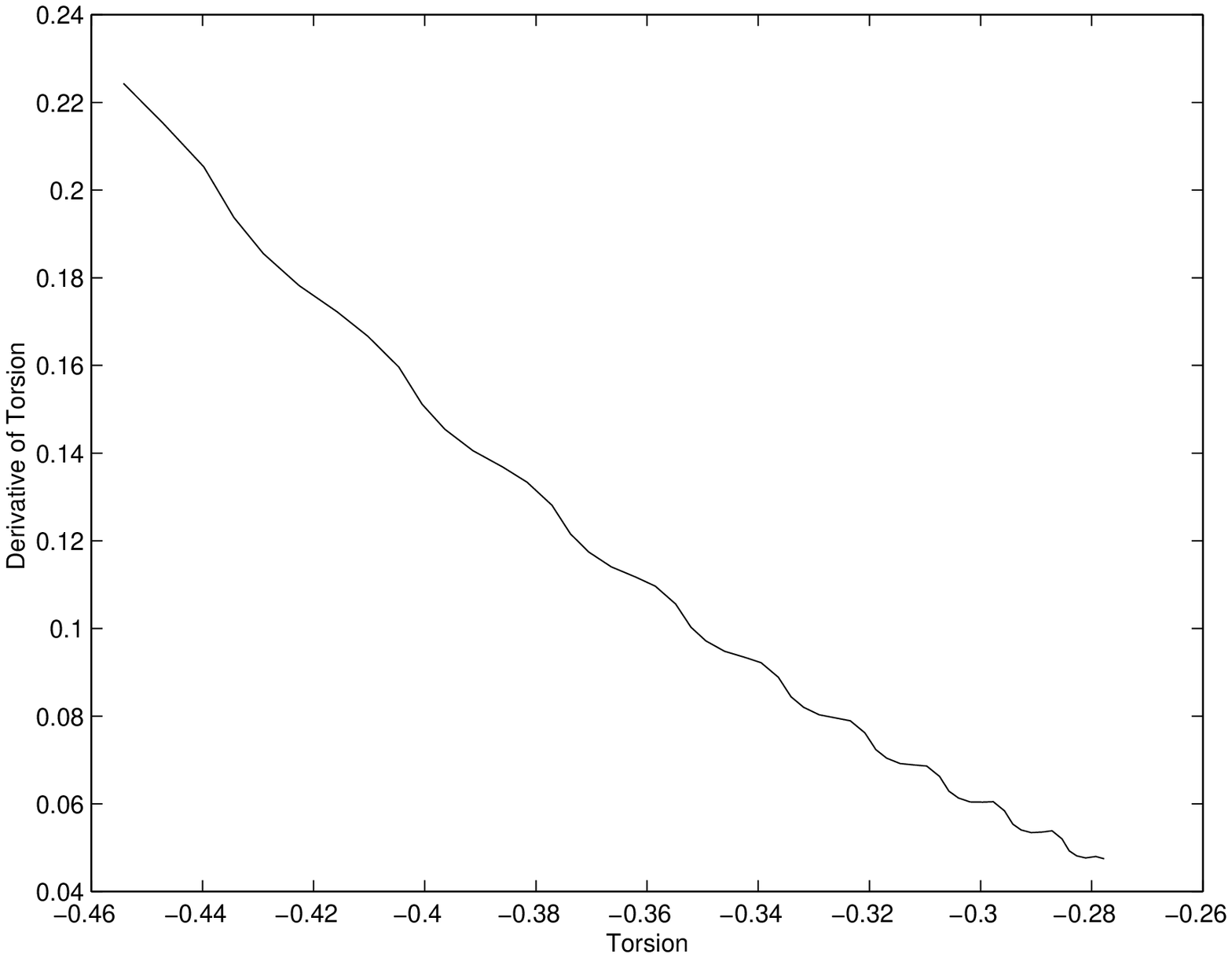}}&

\epsfysize=5cm
\epsfxsize=5cm
{ \epsfbox{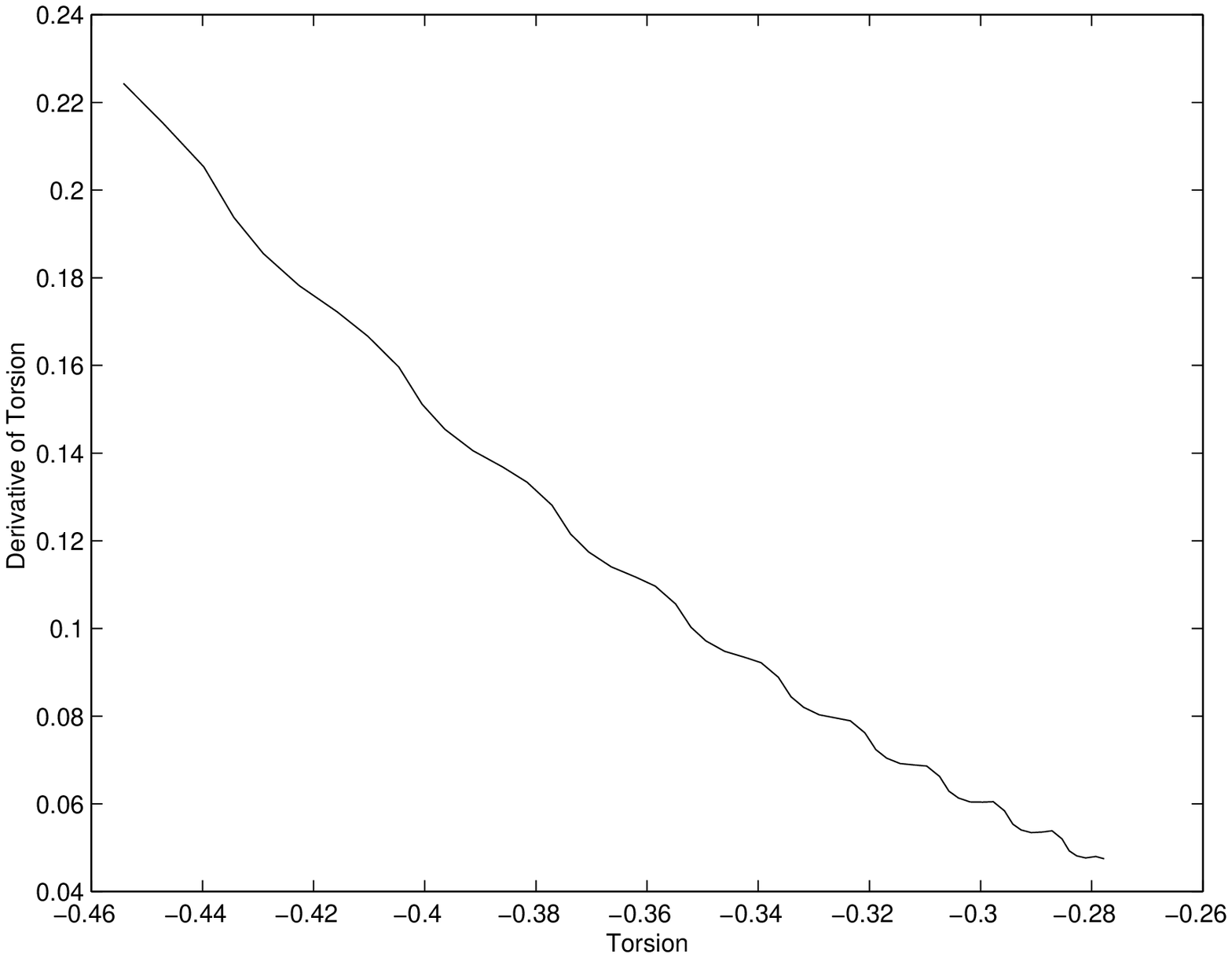}}
\\  \text{\small Using $\tilde{\tau}_1$ and $\tilde{\tau}_s$} &\text{\small Using $\tilde{\tau}_2$ and $\tilde{\tau}_s$}  
\end{array}$
\end{figure}
\vspace{0.5cm}

\newpage

\begin{figure}[here]
\caption{Approximations Obtained with $\Delta t=0.025$}
\vspace{1cm} 
$\begin{array}{ll}
\epsfysize=5cm
\epsfxsize=5cm
{ \epsfbox{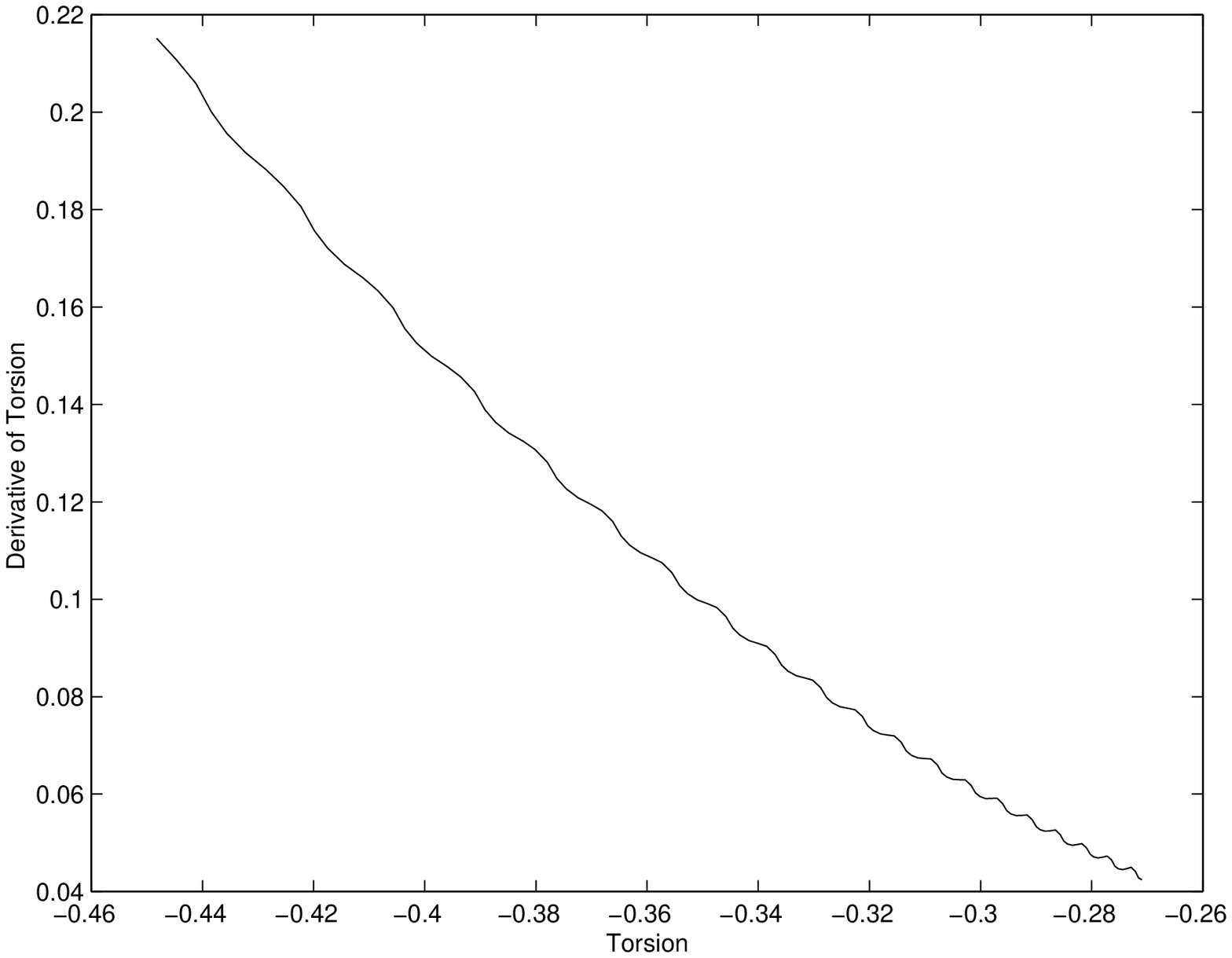}}&

\epsfysize=5cm
\epsfxsize=5cm
{ \epsfbox{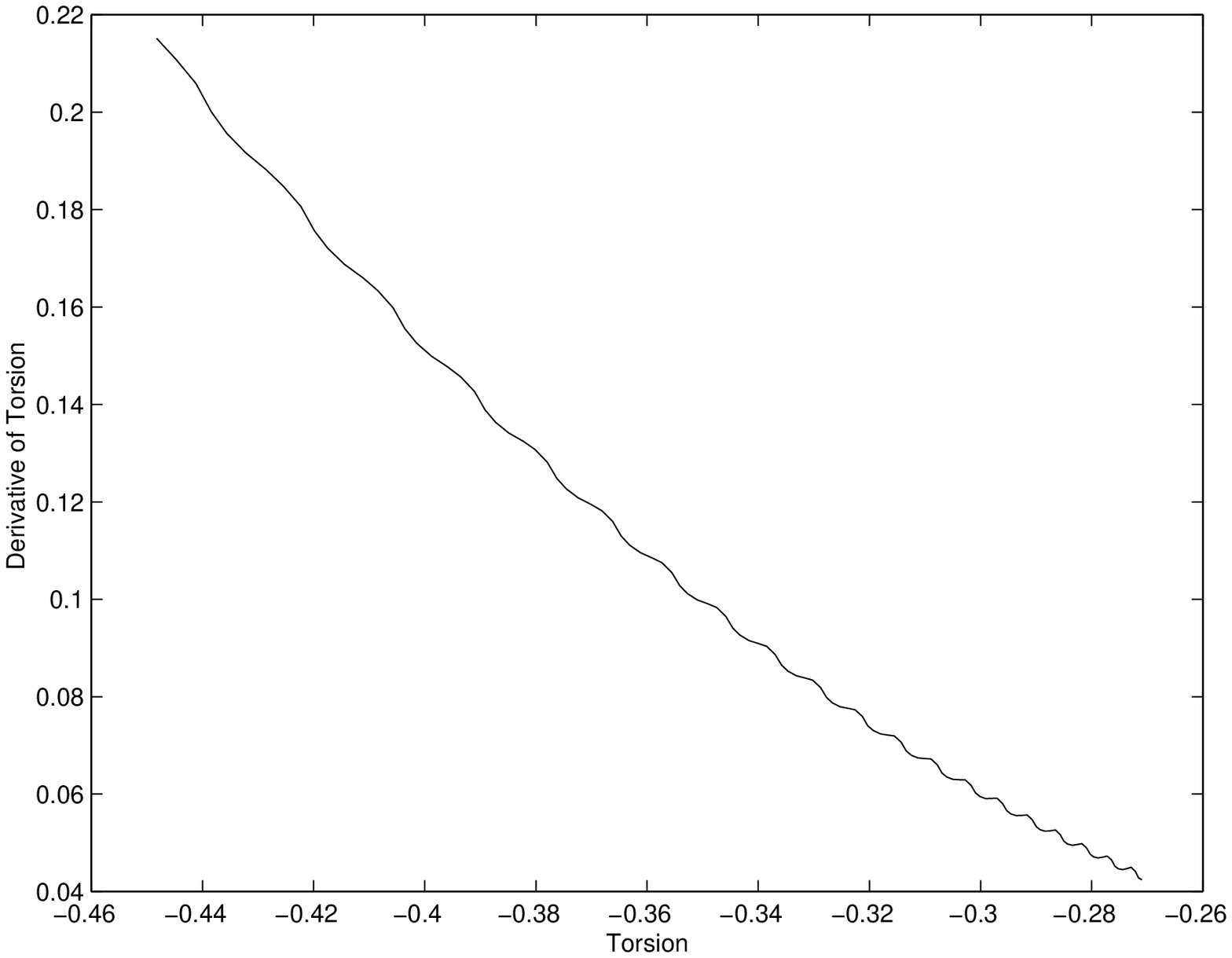}}
\\ \text{\small Using $\tilde{\tau}_1$ and $\tilde{\tau}_s$} &\text{\small Using $\tilde{\tau}_2$ and $\tilde{\tau}_s$} 
 
\end{array}$
\end{figure}
\vspace{0.5cm}

\begin{figure}[here]
\caption{Approximations Obtained with $\Delta t=0.0125$}
\vspace{1cm} 
$\begin{array}{ll}
\epsfysize=5cm
\epsfxsize=5cm
{ \epsfbox{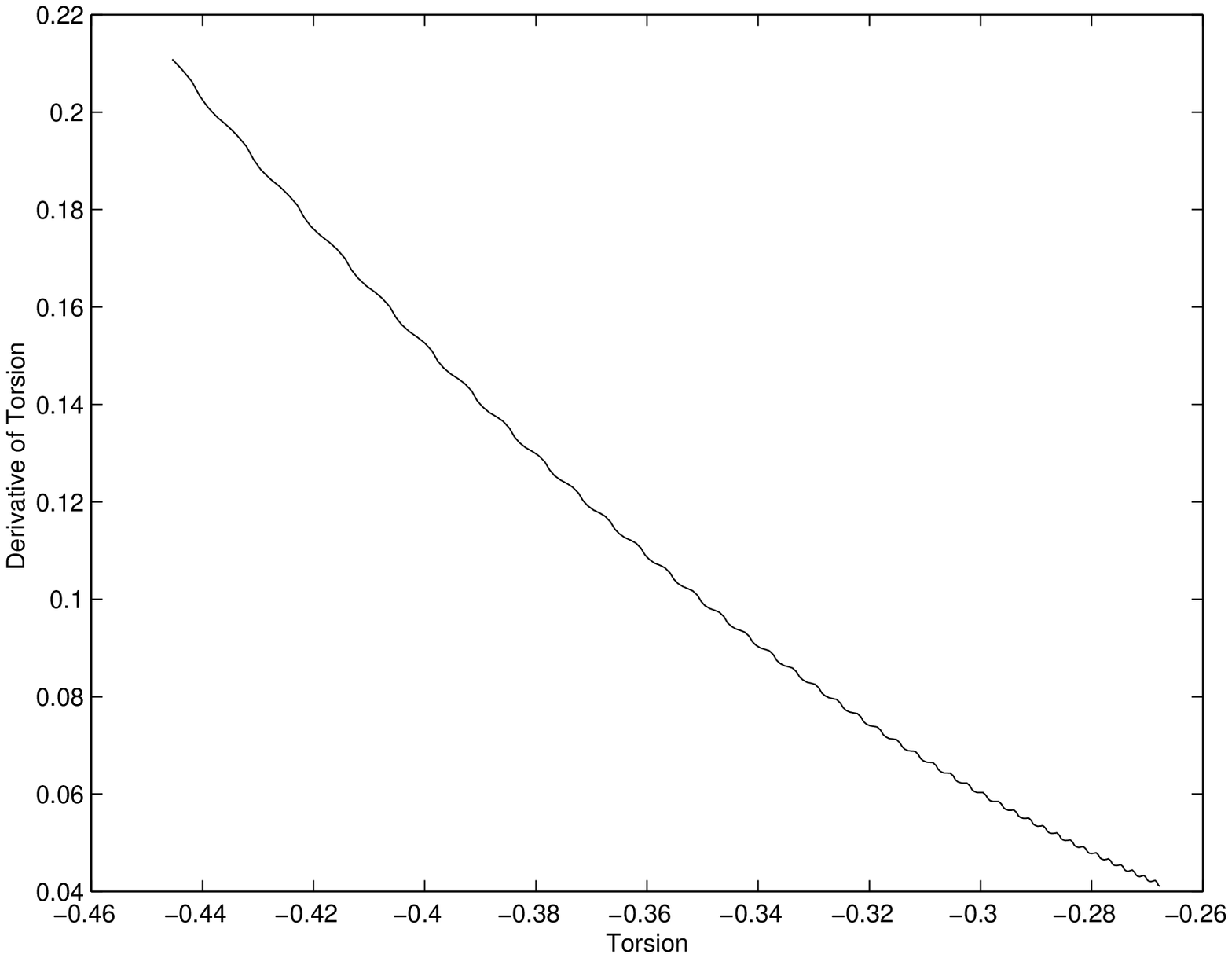}}&

\epsfysize=5cm
\epsfxsize=5cm
{ \epsfbox{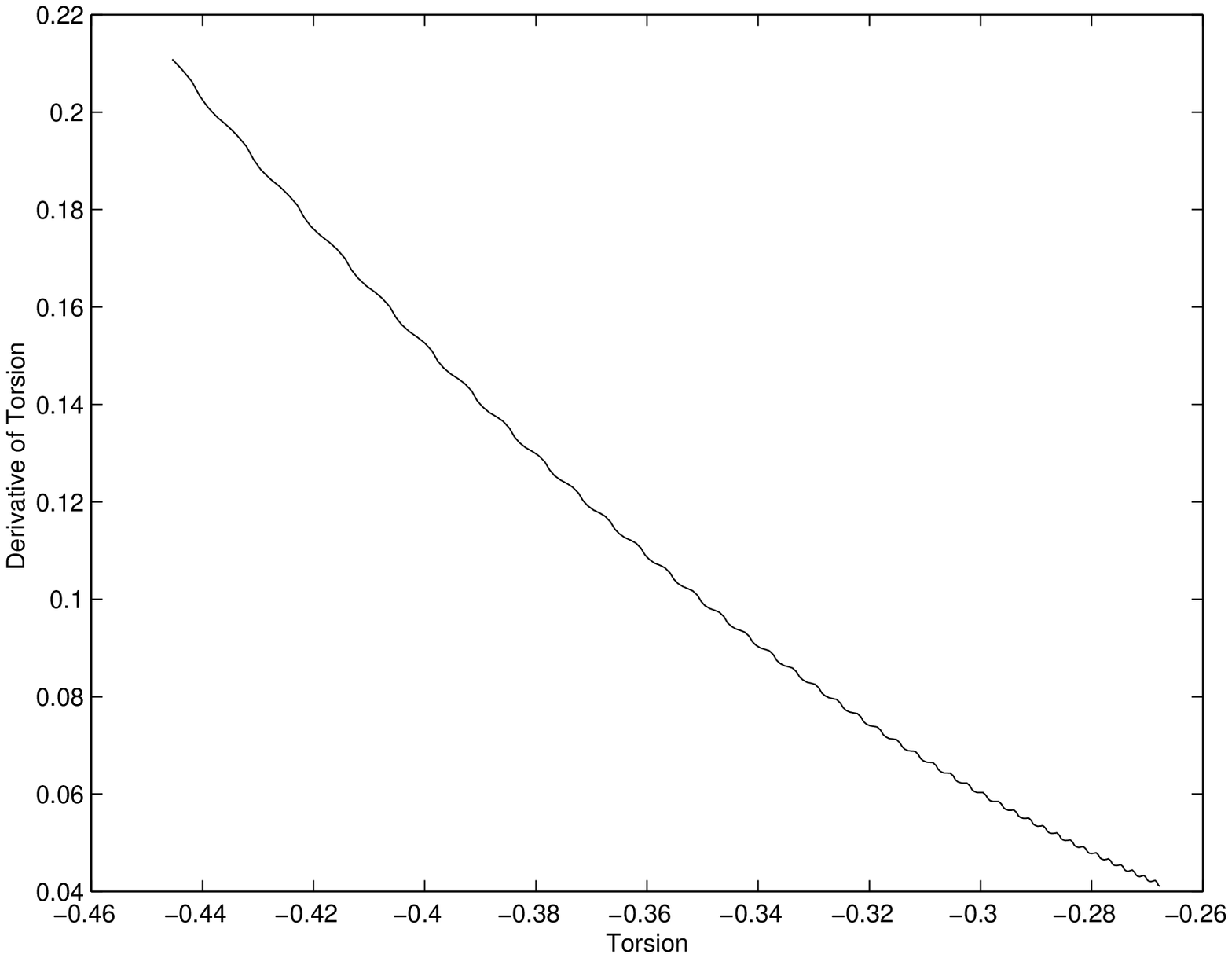}}
\\
\text{\small Using $\tilde{\tau}_1$ and $\tilde{\tau}_s$} &\text{\small Using $\tilde{\tau}_2$ and $\tilde{\tau}_s$}  
\end{array}$
\end{figure}
\vspace{0.5cm}

\newpage
\section*{Acknowledgements}
I want to thank my advisor Peter J.\ Olver for his advice and support. I would also like to thank Allen Tannenbaum for stimulating discussions and Steve Haker for letting me use code of his.
\bibliography{bib}
\bibliographystyle{abbrv}
\end{document}